\documentclass[fleqn,usenatbib]{mnras}

\usepackage{amsmath}
\usepackage{bigints}
\usepackage{graphicx}
\usepackage{float}
\usepackage{caption}

\usepackage{bm}
\usepackage{caption}
\usepackage{dcolumn}
\usepackage{graphicx}
\usepackage{hyperref}
\usepackage[mathscr]{eucal}
\usepackage{subcaption}
\usepackage{stackengine}
\usepackage{setspace}
\usepackage{txfonts}
\usepackage{tikz} 
\usepackage{wasysym}
\usepackage{ulem,xpatch}

\DeclareRobustCommand{\VAN}[3]{#2}
\let\VANthebibliography\thebibliography
\def\thebibliography{\DeclareRobustCommand{\VAN}[3]{##3}\VANthebibliography}

\title[Multimessenger Emission from AIC of WDs]{Multimessenger Emission from the Accretion-Induced Collapse of White Dwarfs}

\author[L. F. L. Micchi et al.]{
Luís Felipe Longo Micchi$^{1,2,3,4}$, David Radice\thanks{Alfred P.~Sloan fellow} $^{2,3,5}$, and Cecilia Chirenti$^{6,7,8,9}$ \\
$^{1}$ Center for Natural and Human Sciences, Universidade Federal do ABC, Santo Andr\'e, SP 09210-170, Brazil \\
$^{2}$ Institute for Gravitation \& the Cosmos, The Pennsylvania State University, University Park, PA 16802, USA \\
$^{3}$ Department of Physics, The Pennsylvania State University, University Park, PA 16802, USA \\
$^{4}$ Theoretisch-Physikalisches Institut, Friedrich-Schiller-Universit\"at Jena, 07743, Jena, Germany \\
$^{5}$ Department of Astronomy \& Astrophysics, The Pennsylvania State University, University Park, PA 16802, USA \\
$^{6}$ Department of Astronomy, University of Maryland, College Park, Maryland 20742, USA \\
$^{7}$ Astroparticle Physics Laboratory NASA/GSFC, Greenbelt, Maryland 20771, USA \\
$^{8}$ Center for Research and Exploration in Space Science and Technology \\
$^{9}$ Center for Mathematics, Computation and Cognition, UFABC, Santo Andr\'e-SP, 09210-170, Brazil
}

\date{Submitted to Monthly Notices of the Royal Astronomical Society on June 14th, 2023}

\pubyear{2022}

\begin{document}

\newcommand{\blue}[1]{\textcolor{blue}{#1}}
\newcommand{\red}[1]{\textcolor{red}{#1}}
\newcommand{\dr}[1]{\textcolor{purple}{[DR: #1]}}

\label{firstpage}
\pagerange{\pageref{firstpage}--\pageref{lastpage}}
\maketitle
\vspace{-1.9cm}
\begin{abstract}

We present fully general-relativistic three-dimensional numerical simulations of accretion-induced collapse (AIC) of white dwarfs (WDs). We evolve three different WD models (nonrotating, rotating at 80\% and 99\% of the Keplerian mass shedding limit) that collapse due to electron capture. For each of these models, we provide a detailed analysis of their gravitational waves (GWs), neutrinos and electromagnetic counterpart and discuss their detectability. Our results suggest that fast rotating AICs could be detectable up to a distance of 8 Mpc with third-generation GW observatories, and up to 1 Mpc with LIGO. AIC progenitors are expected to have large angular momentum due to their accretion history, which is a determining factor for their stronger GW emission compared to core-collapse supernovae (CCSNe). Regarding neutrino emission, we found no significant difference between AICs and CCSNe. In the electromagnetic spectrum, we find that AICs are two orders of magnitude fainter than type Ia supernovae. Our work places AICs as realistic targets for future multimessenger searches with third generation ground-based GW detectors.
\end{abstract}

\begin{keywords}
White Dwarf  -- Gravitational Collapse -- Gravitational Waves --  Neutrino -- Multi-messenger Astronomy 
\end{keywords}



\newpage

\section{Introduction}

Since the first detection of gravitational waves (GWs) was performed by the LIGO collaboration \citep{GW150914_dection}, we have faced a great expansion of multimessenger astronomy (MMA). Commonly discussed sources for MMA are compact binary mergers and events involving stellar collapse. Among the stellar collapse events, perhaps the most studied cases in the literature are type Ia supernovae (SNeIa) and core collapse supernovae (CCSNe). Far from being the only types of stellar explosion, other physical processes can be behind the explosion mechanism of a collapsing star. In this paper, we focus on accretion-induced collapse (AIC) of white dwarfs (WDs).

\begin{table*}
        \centering
        \begin{tabular}{cccccc}
          \hline
          Simulation &  $\rho_{0}$ [g  ${\rm cm}^{-3}$]  &  Temp. [MeV] &  $M$ [$M_{\odot} $]  & $M_{{\rm bar.}}$ [ $M_{\odot}$]   &  $R$ [km] \\
         \hline \hline
         \texttt{rot} &  9.95 $\times 10^{9}$  &   0.01  &   1.52 & 1.53 & $1.93 \times 10^{3}$  \\
         \texttt{rot\_ar075} & 9.95 $\times 10^{9}$  &   0.01 &    1.51 &     1.53 &   $1.71\times10^{3}$ \\
         \texttt{nrot} & 9.95 $\times 10^{9}$  &   0.01  &    1.45 &   1.47  &  $1.33\times10^{3}$  \\
         \hline 
         \hline
         Simulations  &  $J$ [g cm$^{2}$ s${}^{-1}$] &  $T$/$W$  &  $\Omega$ [Hz]  &  $\Omega_{\rm Kepler}$ [Hz]   &  $a_{r}$ \\
         \hline 
         \texttt{rot} & 3.26 $\times 10^{49}$  &  1.49 $\times 10^{-2}$  & 5.28  & 5.30 & 0.66   \\
        \texttt{rot\_ar075} & 3.08 $\times 10^{49}$ &   1.37 $\times 10^{-2}$ &  5.09  & 6.38 &  0.75\\
        \texttt{nrot} &  0  &  0 &  0 &   9.03  &  1.00 \\
        \hline
        \end{tabular}
        \caption{Initial conditions for the three simulations performed during this work. The parameters are as follows: $\rho_{0}$: central baryonic density, Temp.: central temperature, $M_{{\rm bar.}}$: baryonic mass, $M$: gravitational mass, $R$: stellar equatorial radius, $J$: total angular momentum, $\frac{T}{\rm{W}}$: the ratio between rotational and gravitational energy, $\Omega$: rotational frequency, $\Omega_{\rm Kepler}$: the Keplerian angular velocity at the equator on the surface of the star,
        $a_{r}$: ratio between the polar radius and the equatorial radius of the star. The \texttt{rot\_ar075} model spins at $0.8\, \Omega_{\rm Kepler}$ and the \texttt{rot} model spins close to the break-up limit at $0.99\, \Omega_{\rm Kepler}$.}
        \label{tableID}
\end{table*}

WDs are compact stars supported by electron degeneracy pressure \citep{Chandra_deg_e_gas}. In order for gravitational collapse to take place, the total gravitational mass of the WD has to exceed the Chandrasekhar mass, $M_{\rm{Chand.}}\approx 1.44 M_{\odot}$  \citep{Chandra_deg_e_gas,Chandra_mass_lim}. \citet{Nomoto1986} studied the final fate of accreting WDs based on their accretion rate, initial total mass, and composition. If a low-mass ($M\lesssim1.2M_{\odot}$) C-O WD faces an intense accretion history ($10^{-7} M_{\odot} \rm{yr}^{-1}\lesssim\dot{M}\lesssim10^{-5} M_{\odot} \rm{yr}^{-1}$), then the star is likely to undergo a carbon deflagration and it will detonate as an SNIa, leaving no remnant behind \citep[see also the work of][]{Shen_2009,Shen_2009_2,Moore_2013}. The AIC happens if a high mass ($M \gtrsim 1.2 M_{\odot}$) O+Ne+Mg or Si+O WD accretes slowly ($\dot{M}\lesssim10^{-9} M_{\odot} \rm{yr}^{-1}$), leading the star to collapse and form a neutron star (NS). The different outcomes, SNIa explosion or AIC, depend on the WD composition since lower-Z compositions release a larger amount of energy during burning.

\begin{figure}
    \includegraphics[width=0.5\textwidth]{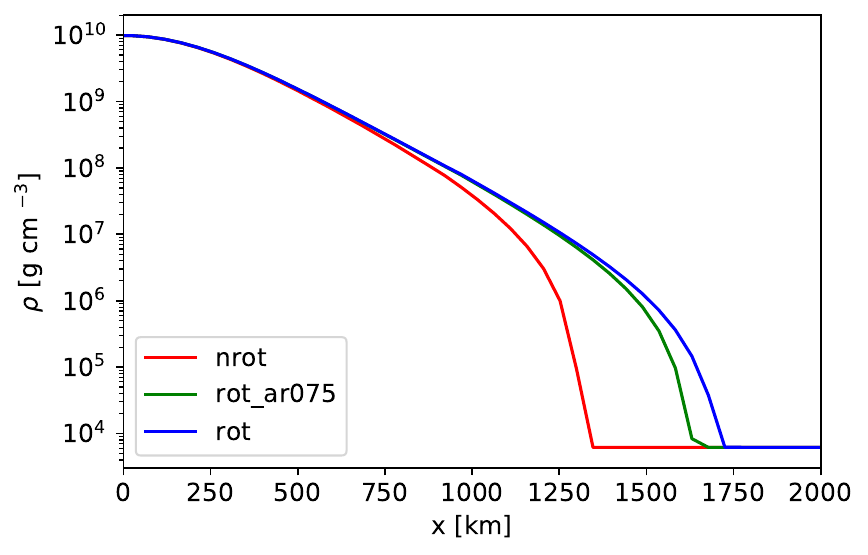}
    \caption{Initial density profile in the equatorial plane. For fixed density $\rho_0$, rotating stars have  larger radii and higher total masses.}
    \label{fig:IDmass}
\end{figure}

AIC events have been connected to the nucleosynthesis of very rare neutron-rich heavy elements, such as ${}^{62}$Ni, ${}^{66}$Zn,${}^{68}$Zn,${}^{87}$Rb and ${}^{88}$Sr. \citet{Fryer_1999} studied the r-process in AICs to constrain their event rate given the cosmological abundances of these elements and estimated between 1 and 200 events per $10^{6}$ years per galaxy.

AICs have also been invoked to explain a plethora of different astrophysical phenomena. AICs are expected to have highly rotating progenitors due to their accretion history \citep{Uenishi2003,Piersanti2003a,Piersanti2003b,Saio2004,Yoon_e_capture,Yoon2005,Kashyap_2018}, as they could form either in a double degenerate case (WD-WD merger) or a single degenerate case (a WD accreting from a companion star that overflows its Roche lobe).\footnote{A third formation channel for AICs was proposed by \citet{CMIC_Ablimit}, in which the WD enters a common envelope phase with its companion (core-merger-induced collapse).} Therefore angular momentum is expected to play a major role in their GW emission \citep{Dimmelmeier:2008iq, CCSNerot2}. Additionally, AICs could form magnetars as their  remnants, possibly powering gamma-ray bursts \citep{Yi_GRB,AICasSGRBEE,Perley2009}. Continued accretion after the formation of the NS remnant from the AIC could lead to the formation of a millisecond pulsar or low-mass X-ray binary \citep{MSP_Hurley_2010, MSP, SyXB_Ablimit, WangMSP}.

Fast radio bursts (FRBs) have also been tentatively connected with AICs. According to \citet{Waxman_FRB} and \citet{Margalit:2019hke}, the properties of FRB 121102 and FRB 180924 are consistent with those of a magnetized NS formed in a weak stellar explosion, which could for example be an AIC or a binary WD merger.

AIC simulations have been performed by different groups.
\citet{Fryer_1999} employed 1D and 2D simulations to determine the composition of the ejected material and concluded that an AIC does not produce enough nickel to explain the formation of a SNIa. \citet{Dessart_2006} performed studies in 2D Newtonian gravity and showed that AIC events with no magnetic fields should display low ejecta mass ($M_{\textrm{ej.}}\sim 10^{-3} M_{\odot}$) with an entropy ranging between 20 and 50 $k_{\rm{B}}$/baryon and explosion energy of $\lesssim 10^{50}$ erg. When magnetic fields are included, \citet{Dessart_2007} reported, among other differences, an increase in ejecta mass by a factor of 100 ($M_{\textrm{ej.}}\sim 0.1 M_{\odot}$). More recent 2D simulations including general relativity (GR) effects were performed by \citet{Abdikamalov2010}, who predicted that AICs are subjected to low $\frac{T}{|W|}$-type dynamical instabilities \citep{Shibata:2000jt,Low1,Low2,Low3,Low4,Low5}, due to the existence of a corotating point inside the star \citep{lowb_condition1,lowb_condition2}. Finally, GW signals generated by the AIC of dark matter admixed WDs have been studied by \citet{Chan_2023}, who find that dark matter accumulated in the WD can have an impact on the GW emission.

\begin{figure}

\begin{subfigure}{0.5\textwidth}
\includegraphics[width=\textwidth]{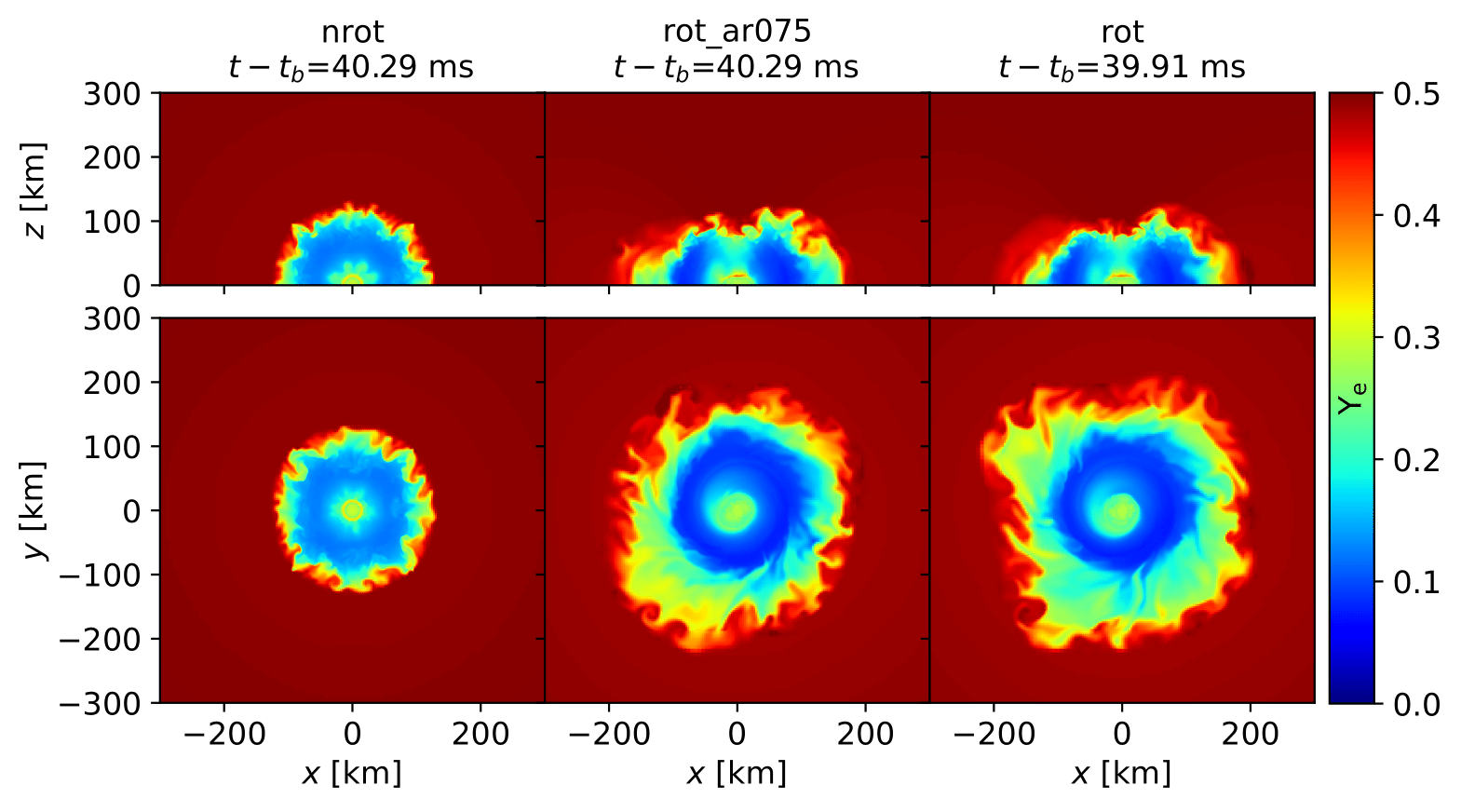}
\end{subfigure}
     
\begin{subfigure}{0.5\textwidth}
\includegraphics[width=\textwidth]{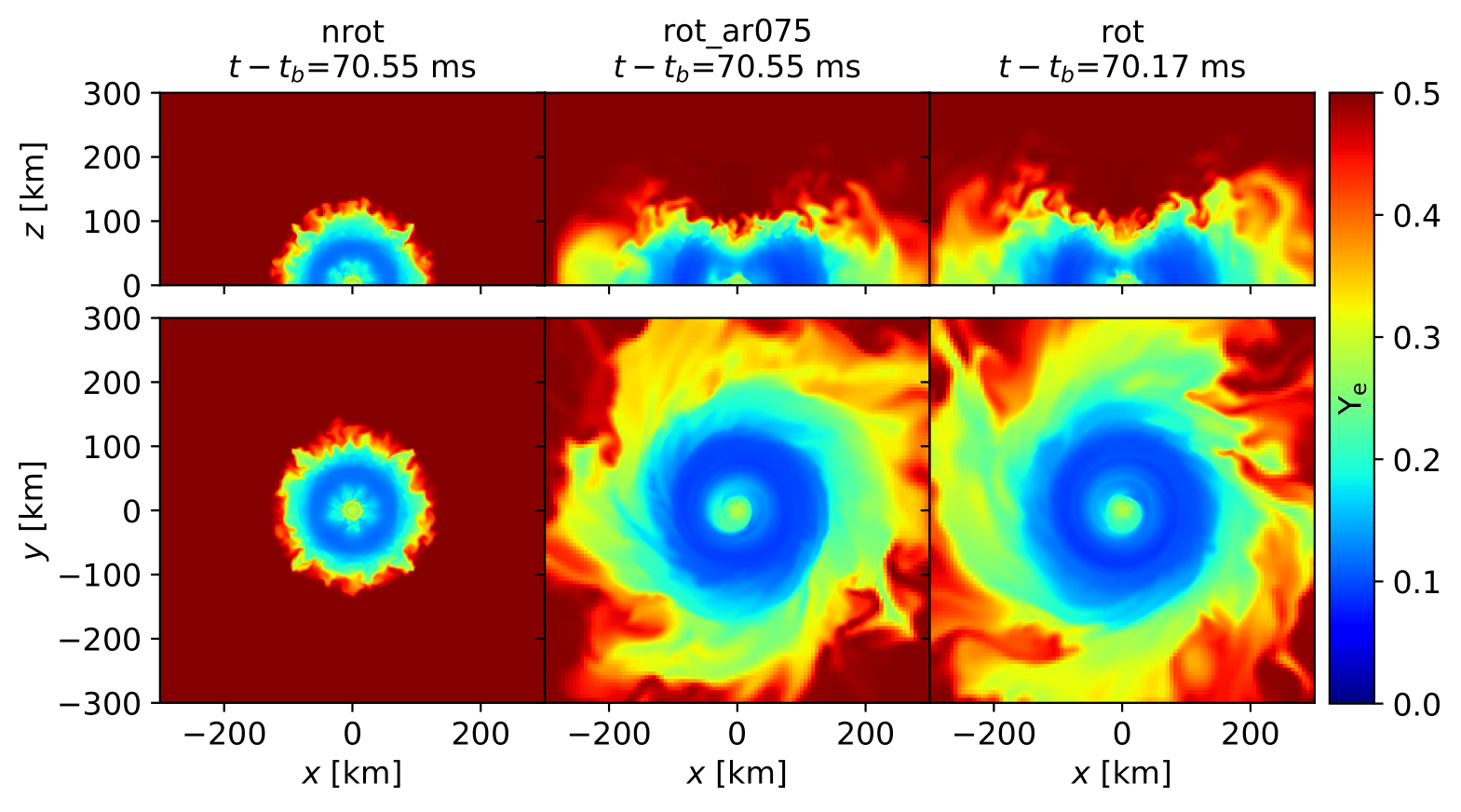}
\end{subfigure}

\begin{subfigure}{0.5\textwidth}
\includegraphics[width=\textwidth]{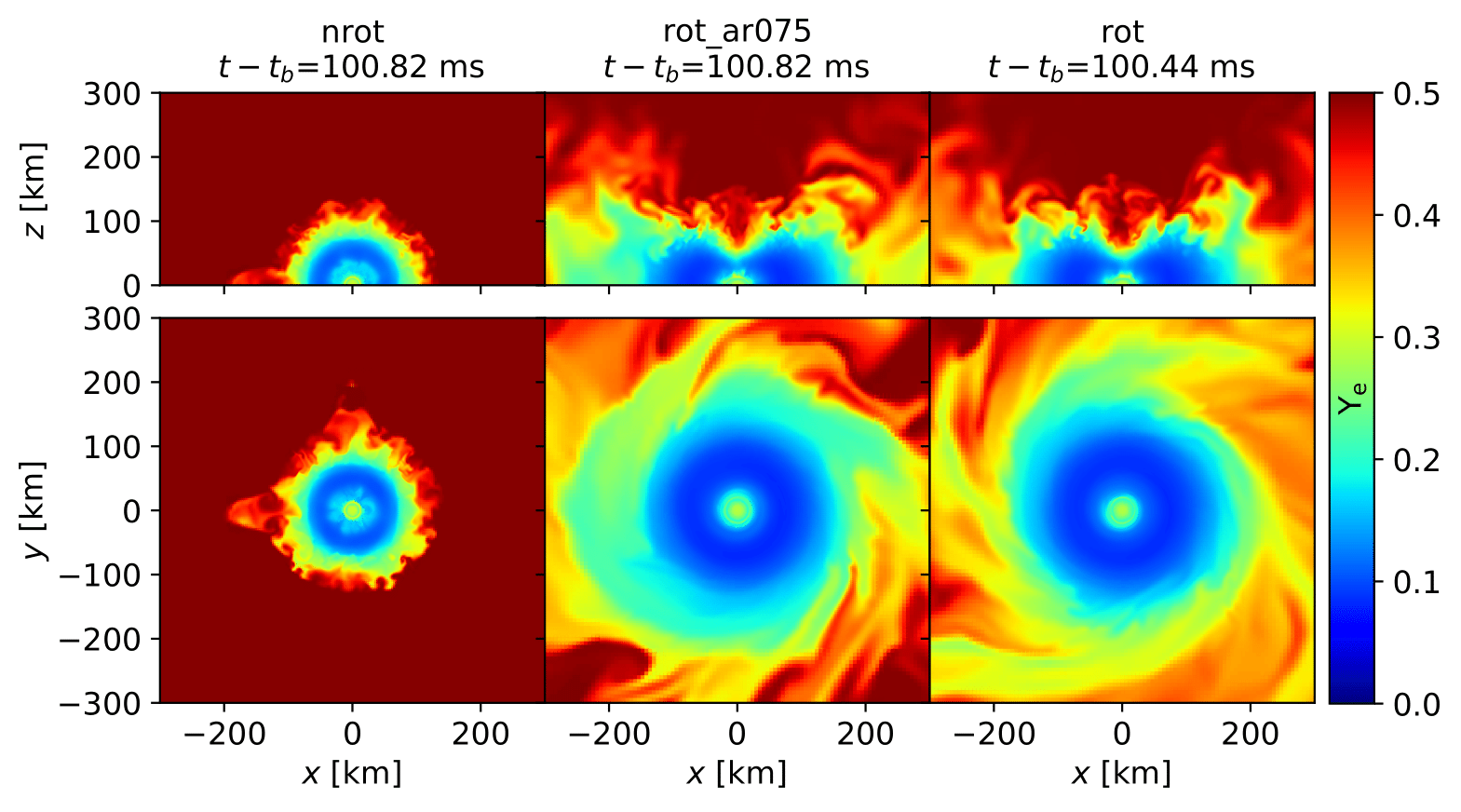}
\end{subfigure}

\begin{subfigure}{0.5\textwidth}
\includegraphics[width=\textwidth]{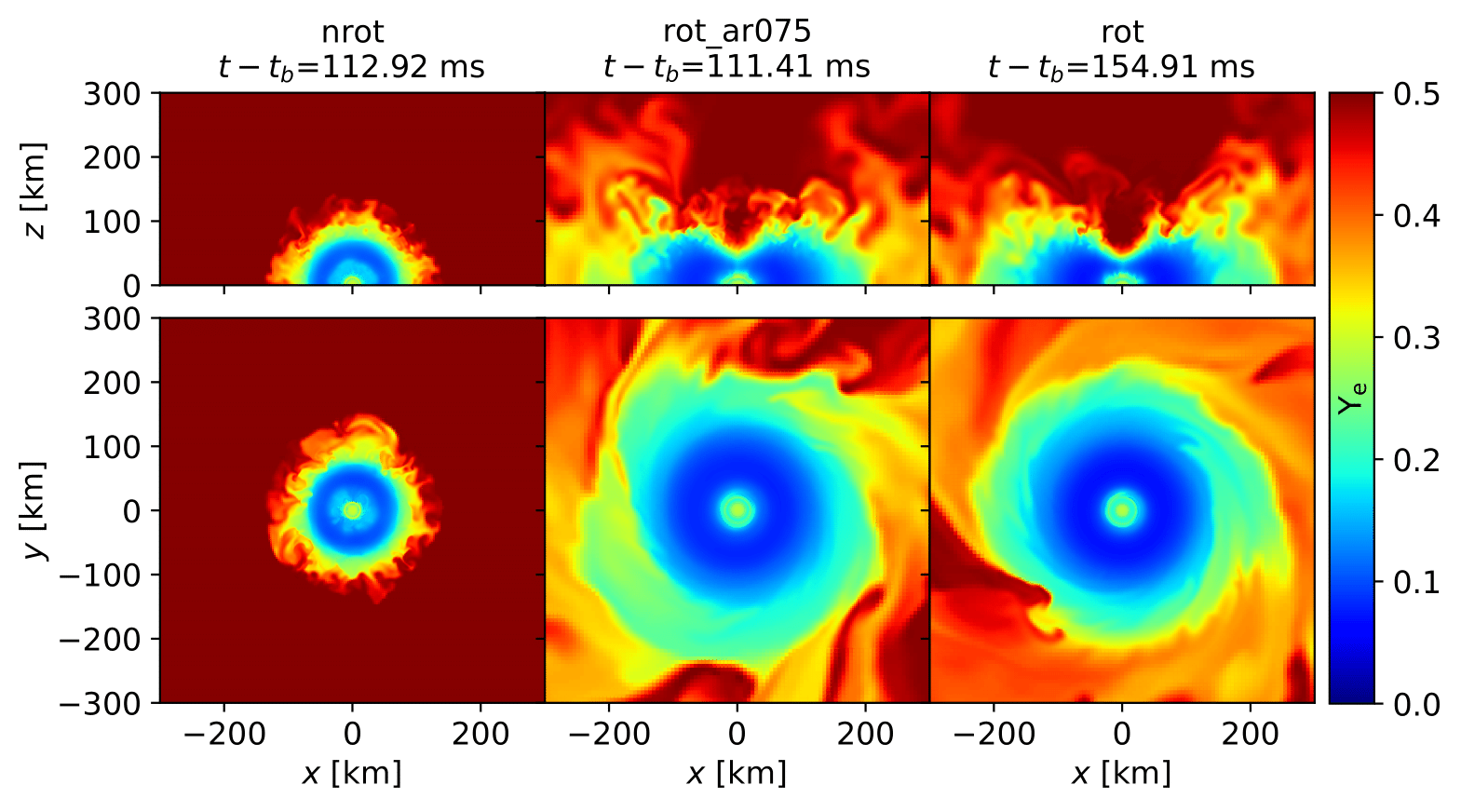}
\end{subfigure}

\caption{Electron fraction $Y_{e}$ in the equatorial and meridional planes. Most of the low $Y_{e}$ material ($Y_e \lesssim 0.2$) remains bound to the central region with $r<100$ km. As the dense spiral arms travel outwards, the material becomes electron rich again ($Y_{e}\gtrsim 0.3$). The region with low electron fraction is larger for the rotating cases. Both \texttt{rot} and \texttt{rot\_ar075} display a torus-like morphology, while \texttt{nrot} is closer to spherical symmetry. The slight increase in $Y_{e}$ towards the center of the star is due to the fact that electrons captures reaction are suppressed by Pauli blocking.}
    \label{fig:Ye}
\end{figure}

In this paper, we perform fully general relativistic 3D simulations of AIC of rotating WDs. Our simulations include an M1 grey neutrino radiation transport scheme via the \texttt{WhiskyTHC} code \citep{M1}, which is implemented within the \texttt{Einstein Toolkit} \citep{Einstein_Toolkit}. This paper is organized as follows. The numerical setup and initial data are presented in Section \ref{methods}. Our results are discussed in the following Sections: collapse and post-bounce dynamics of the AIC (Section \ref{delep}), expected gravitational radiation emission of an AIC event (Section \ref{sec:GW}), effects of the weak (im)balance of the nucleons and how it gives rise to the neutrino flux (Section \ref{chem}) and properties of the ejected material and estimates for the electromagnetic emission (Section \ref{EMcounter}). In Section \ref{conclusions} we present our conclusions.

\section{Numerical Methods}\label{methods}

\begin{figure}
    \centering
    \includegraphics[width=0.45\textwidth]{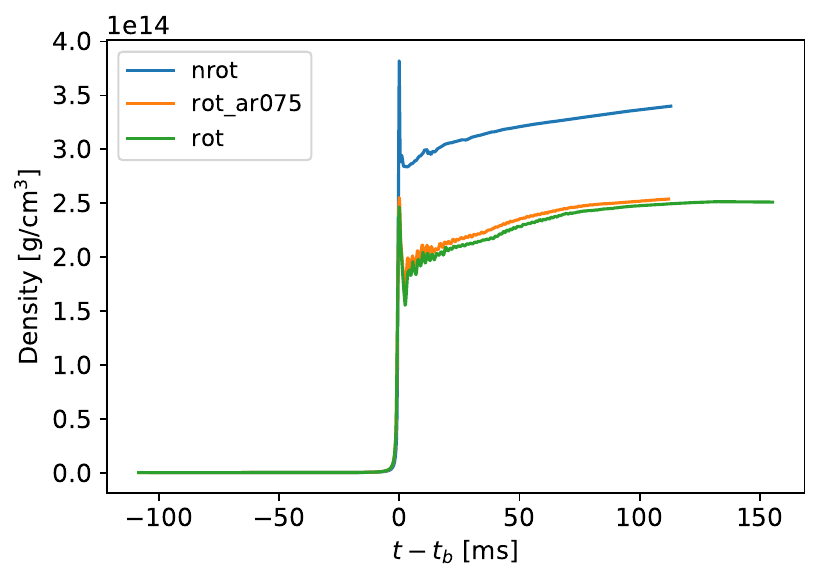}
    \caption{Time evolution of the central density of the star. The time of bounce ($t_{b}$) is marked by the first local maximum of $\rho_{0}(t)$. After the bounce, the central density increases due to the collapse of the outer layers. The oscillations visible at early times after the bounce represent perturbation modes of the remnant PNS \citep{Fuller2015,Rodriguez2023}. Note the clear effect of rotation in decreasing the central density, as the centrifugal force tends to spread the material to larger radius, see also Fig.~\ref{fig:IDmass}.}
    \label{fig:Central_rho_evolve}
\end{figure}

\begin{figure*}
\begin{minipage}{0.8\textwidth}
\includegraphics[width=\textwidth]{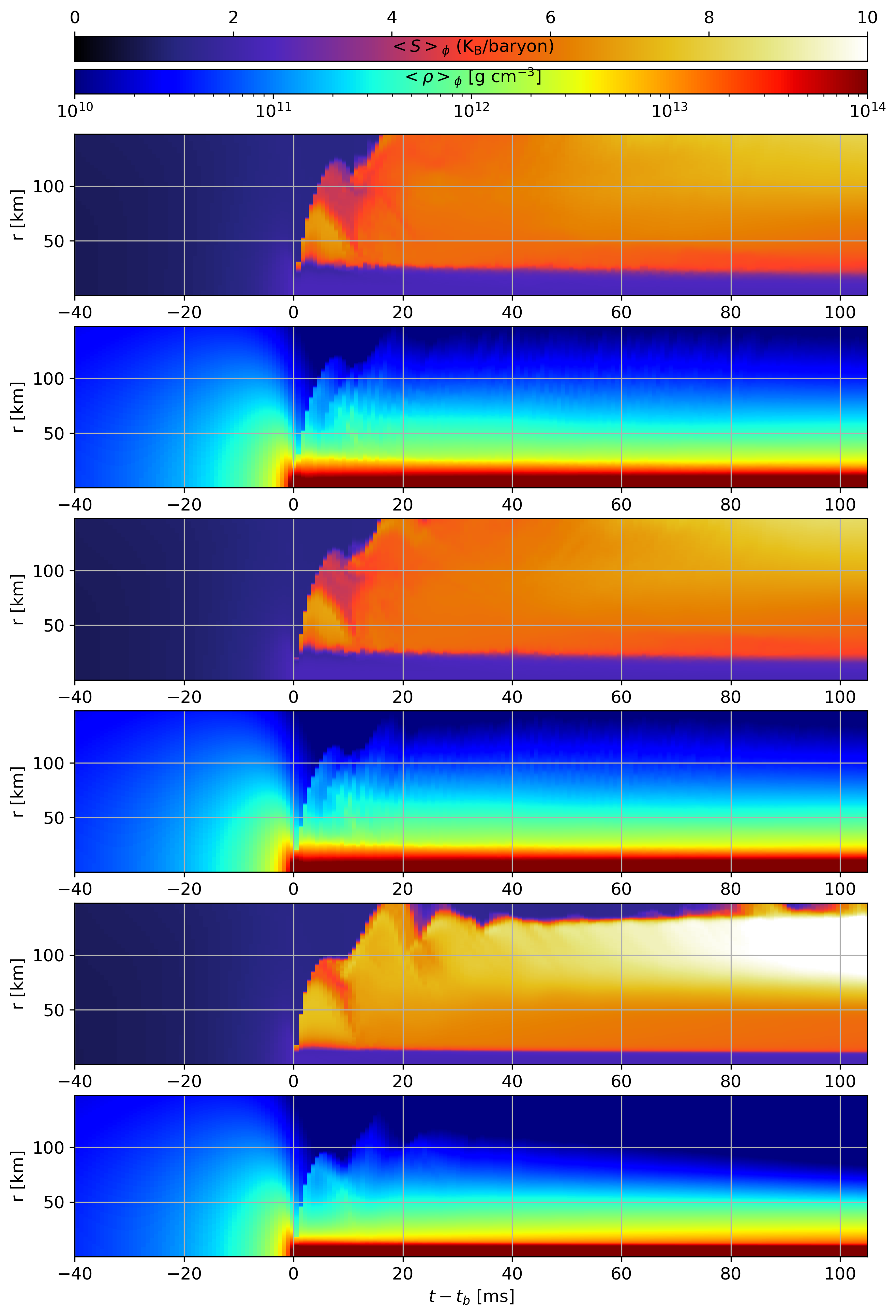}

\caption{Angular average of the equatorial entropy $S$ and density $\rho$ profiles throughout the evolution of the AIC. The regions with low density and large entropy are correlated. The high-entropy material  pushes the fluid outwards (expanding to lower densities) due to the convective instability. The large increase of entropy occurs shortly after the bounce, as the shock starts to move away from the center of the star. The large density region ($r\lesssim 25$ km) is the PNS core. The stalled shock is seen in the \texttt{nrot} model as the large entropy material ceases to expand at later times ($t-t_b \gtrsim 40$ ms). }
\label{fig:densityaverage}
        
\end{minipage}
\begin{minipage}{0.1\textwidth}

\vspace{-0.5cm}
\begin{tikzpicture}

\draw (-2.0,8.0) -- (-2.0,14.0);
\draw (-2.2,14.0) -- (-2.0,14.0);
\draw (-2.2,8.0) -- (-2.0,8.0);
\node [draw, rotate=-90, font=\Large] at (-1.5,11) {\texttt{rot}};

\draw (-2.0,1.8) -- (-2.0,7.8);
\draw (-2.2,+7.8) -- (-2.0,7.8);
\draw (-2.2,1.8) -- (-2.0,1.8);
\node [draw, rotate=-90, font=\Large] at (-1.5,5.0) {\texttt{rot\_ar075}};

\draw (-2.0,-4.4) -- (-2.0,1.6);
\draw (-2.2,1.6) -- (-2.0,1.6);
\draw (-2.2,-4.4) -- (-2.0,-4.4);
\node [draw, rotate=-90, font=\Large] at (-1.5,-1.3) {\texttt{nrot}};

\end{tikzpicture}

\end{minipage}

\end{figure*}
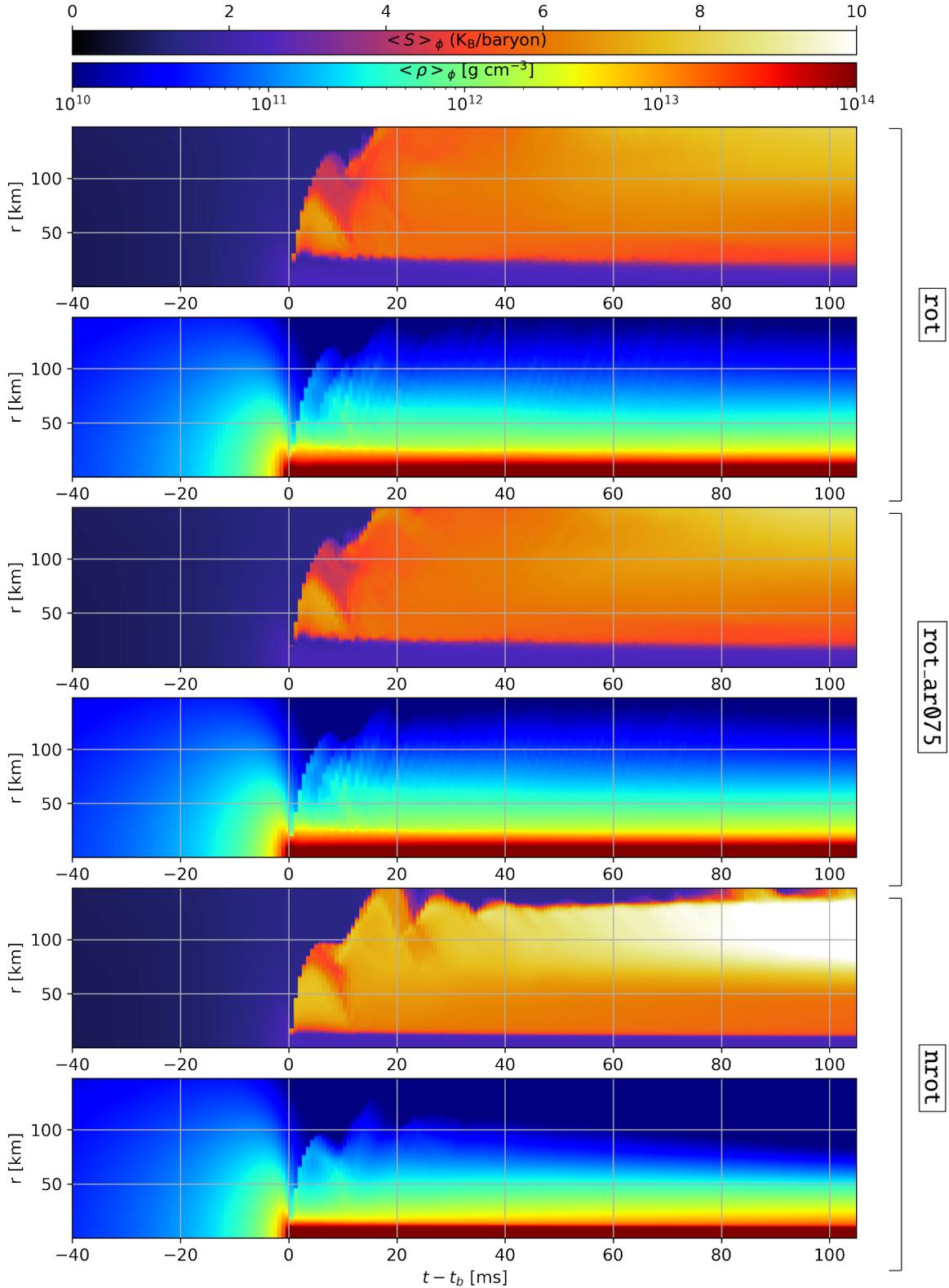

\begin{figure}

\begin{subfigure}{0.5\textwidth}
\includegraphics[width=\textwidth]{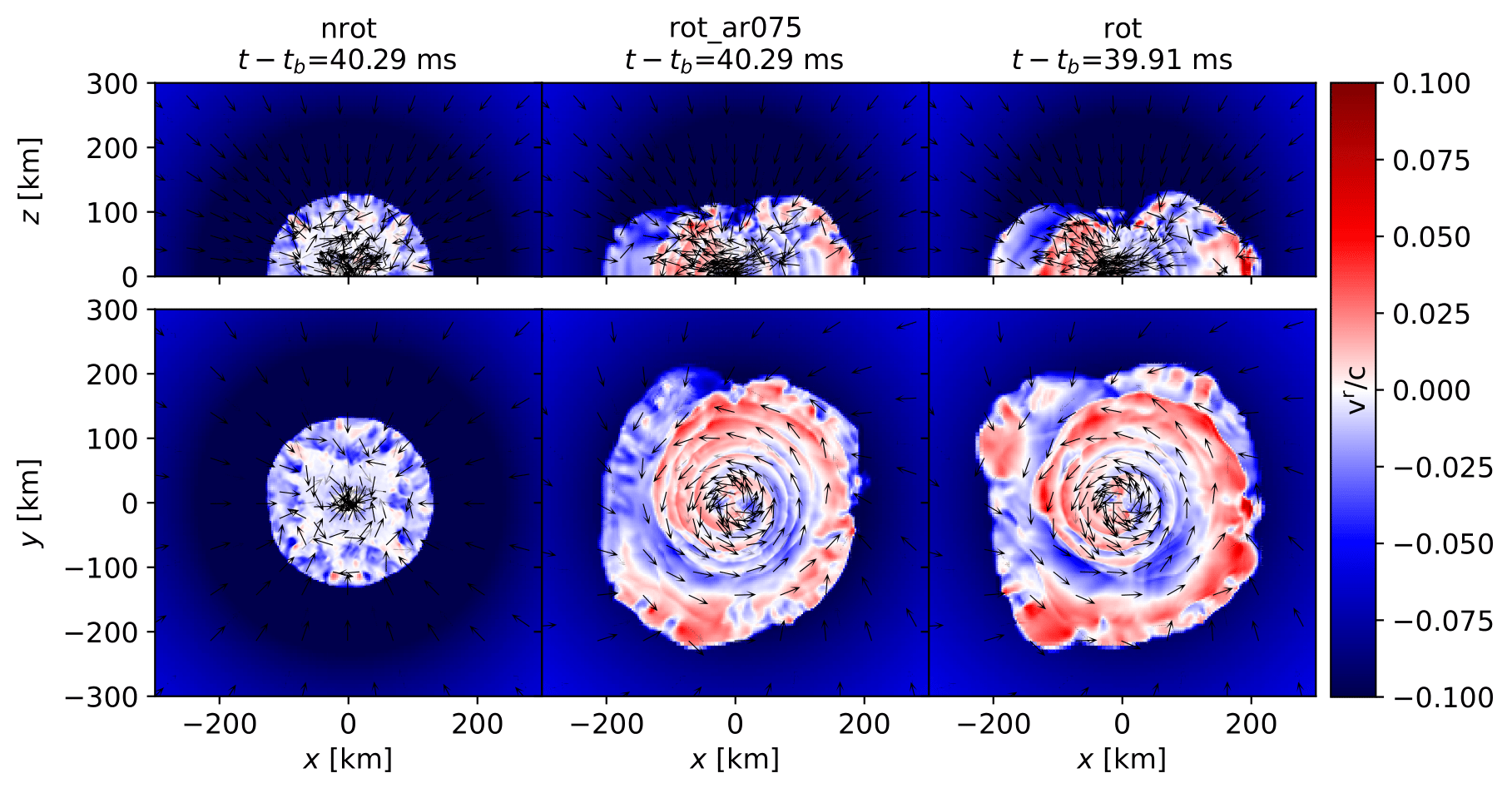}
\end{subfigure}
     
\begin{subfigure}{0.5\textwidth}
\includegraphics[width=\textwidth]{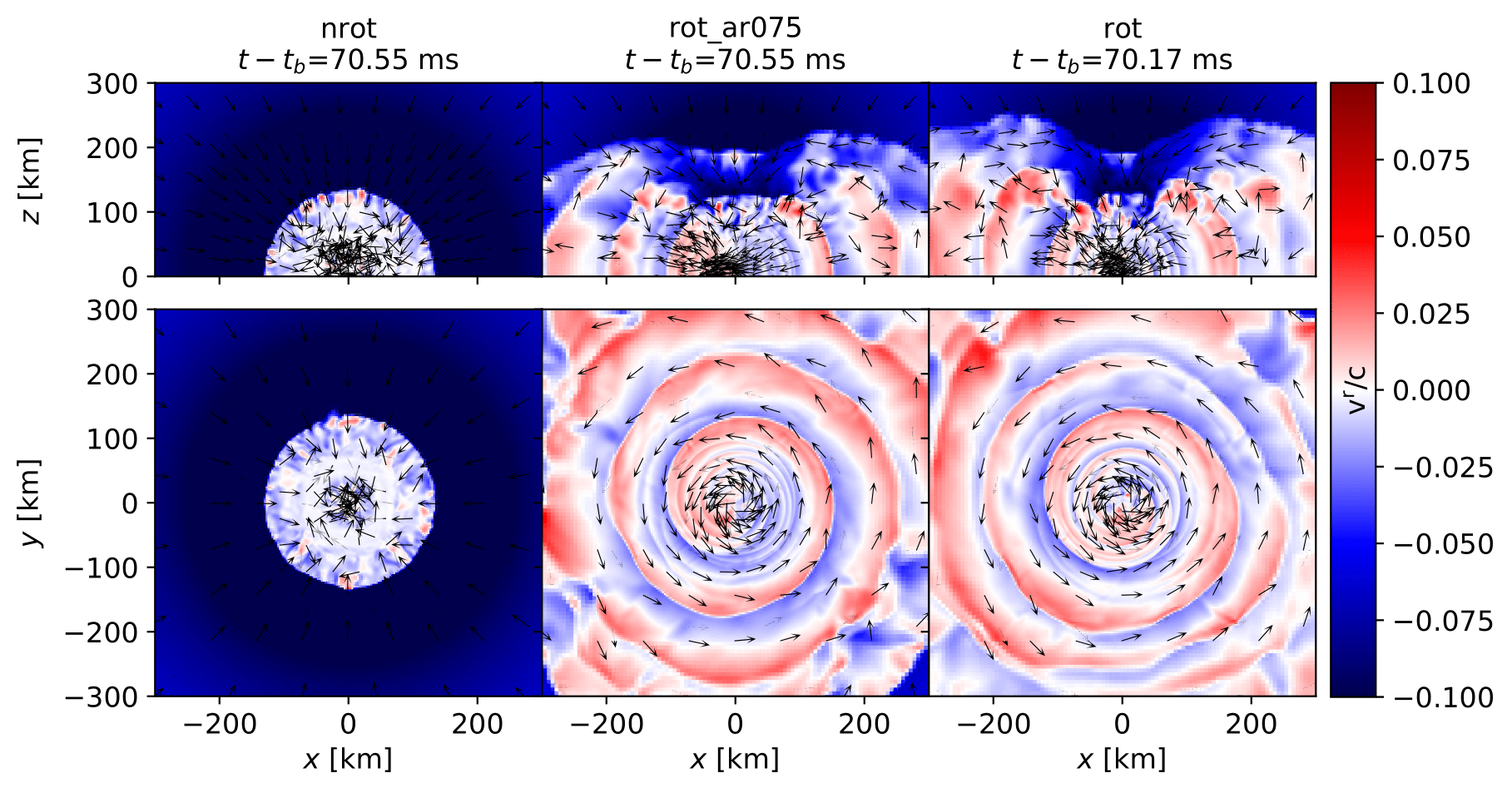}
\end{subfigure}

\begin{subfigure}{0.5\textwidth}
\includegraphics[width=\textwidth]{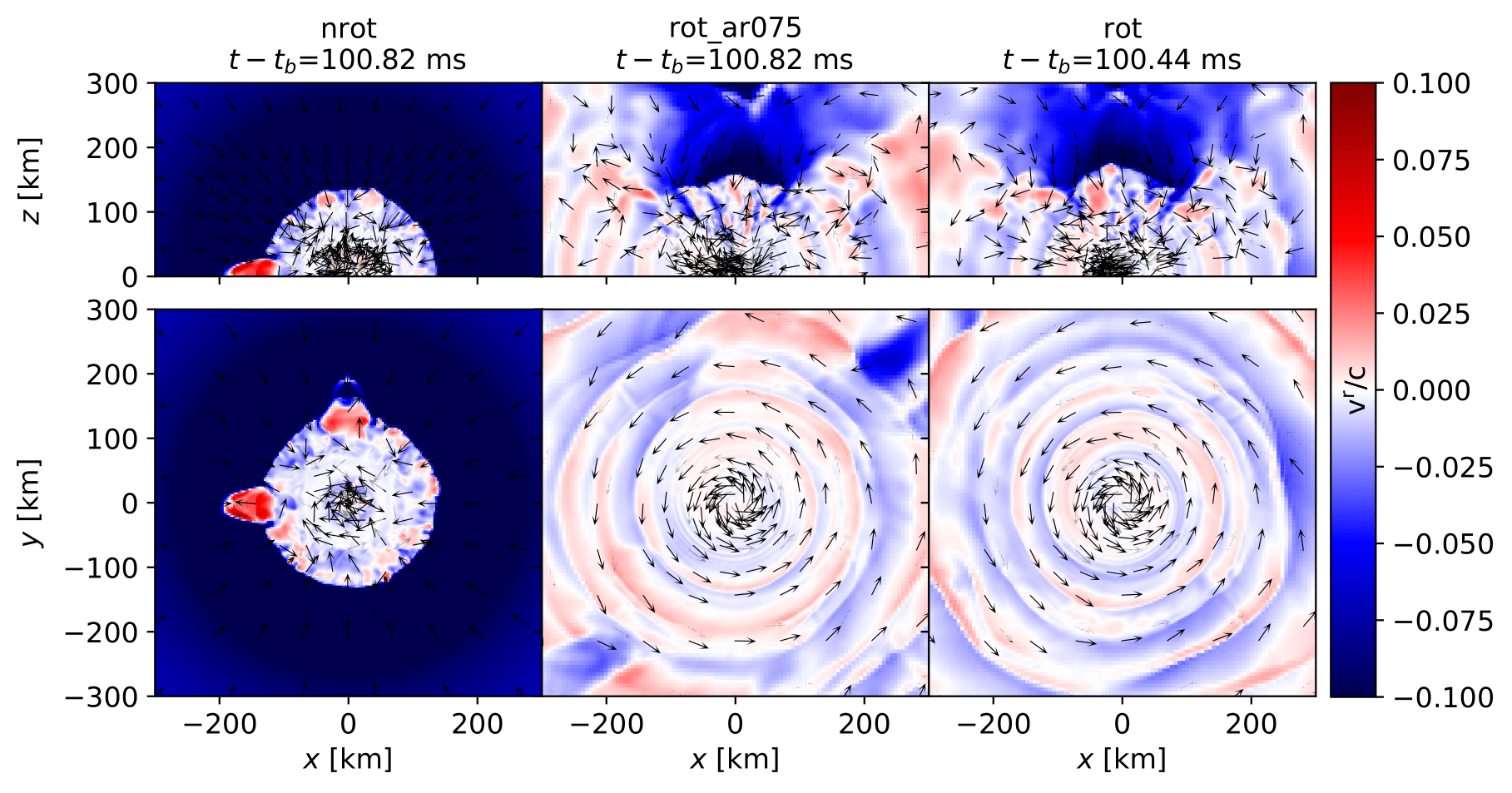}
\end{subfigure}

\begin{subfigure}{0.5\textwidth}
\includegraphics[width=\textwidth]{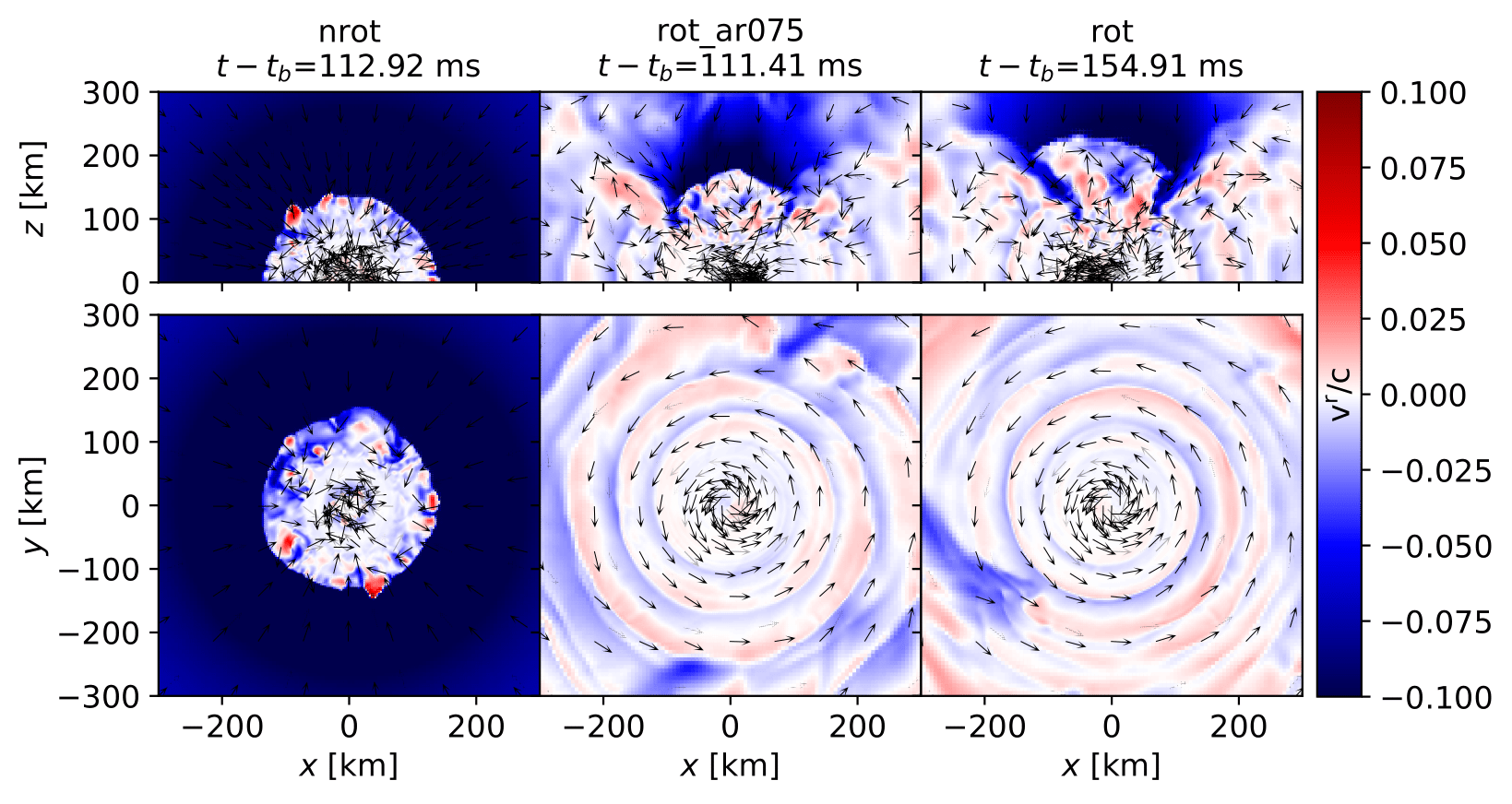}
\end{subfigure}

\caption{Velocity profile of the fluid. The color scheme shows the radial velocity, while the arrows display the direction of the motion. Rotation plays a key role in the explosion, as the shock in the nonrotating case seems to stall very early in the evolution.}
    \label{fig:vel}
\end{figure}

\subsection{Numerical Setup}\label{nsetup}

In our numerical setup, we ignore the path that brought the WD to the brink of collapse and consider only an isolated WD, with initial conditions set to induce the AIC (see Sec. \ref{initialdata}). 

Our models are evolved with the \texttt{WhiskyTHC} code \citep{THC1,THC2,THC3,THC4,THC5}, which uses the \texttt{Einstein Toolkit} framework \citep{Einstein_Toolkit}. More specifically, the space-time evolution is executed via the \texttt{CTGamma} thorn \citep{CTGAMMA1,CTGAMMA2}, which solves the Einstein equations in the Z4c formulation. The hydrodynamical quantities are calculated with \texttt{WhiskyTHC}, which implements high-order convergence schemes for general relativistic hydrodynamics and contains a moment-based neutrino transport scheme \citep{M1}. For the time evolution, we use a finite difference Runge-Kutta scheme with (spatially-dependent) fixed time steps set by a CFL factor of 0.15, which defines the ratio between the temporal and spatial resolution of the grid. The spatial grid is constructed by the \texttt{Carpet} thorn \citep{CARPETfix}. An \texttt{AMR} grid is constructed by requiring 10 logarithmically (base 2) evenly spaced resolution levels. The outermost grid has its $x$ and $y$ coordinates within $\pm 6284.9$km, while $0<z<6284.9$km due to the imposed symmetry with respect to the $xy$ plane. All refinement levels respect this reflection requirement and contain $128\times128\times64$ grid points. The grid resolution in the finest refinement level covering the protoneutron star (PNS) remnant is 0.369 km. The GWs presented here were obtained via the \texttt{WeylScal4} thorn, which evolves and extracts the Weyl scalar at a finite radius. We cross-validated the GW strain found with \texttt{WeylScal4} against the results achieved by the \texttt{ZelmaniQuadWaveExtract} thorn, which computes the GW in the quadrupole approximation at radial infinity.

\subsection{Initial Data} \label{initialdata}

Different evolutionary paths to an AIC could affect, for example, the spin of the WD. We label our simulations with $a_{r}$, the ratio between the polar and equatorial radii of the progenitor WDs. Our models are \texttt{nrot}, \texttt{rot\_ar075}, and \texttt{rot} for the non-rotating, fast rotating, and maximally rotating progenitors, corresponding to values of $a_{r}$ = 1, 0.75, and 0.66, respectively. All of the models are rigidly rotating at the beginning of the simulations. More information about their initial properties is in Table \ref{tableID}. The maximally rotating case is very close to the maximum angular momentum supported by the initial WD before being disrupted by the centrifugal force and spins at $99\%$ of the Keplerian mass-shedding frequency. 

All of the three progenitors have the same initial baryonic central density ($\rho_{0} = 9.95 \times 10^{9}$ g / ${\rm cm}^{3}$ because this value is argued by \citet{Abdikamalov2010}, \citet{Nomoto1991}, \citet{Central_dens0}, and \citet{Dessart_2006} to be representative of a WD in the verge of collapse) and temperature ($T_{0}=0.01$ MeV). Initial data were generated with the \texttt{RNSID} thorn \citep{RNSID}. The equation of state used in this work is the LS220 EOS proposed by \citet{LS220_eos}, which assumes nuclear statistical equilibrium and relies on a compressible liquid-drop model for the nuclei, including effects of interactions and degeneracy of the nucleons outside nuclei. 
The initial central values ($\rho_{0}, T_{0}$) are chosen to emulate configurations on the verge of collapse, which requires that the initial mass of the WD should be greater than the Chandrasekhar limit. The slightly different values of initial mass are a consequence of the extra support provided by the initial rotation and the initial density profiles are shown in Fig.~\ref{fig:IDmass}.

The maximum mass that can be supported in a cold\footnote{The temperature of the star is well below the Fermi energy of its ultra-relativistic electrons. As the star is typically at $\sim 10^{6}-10^{7}$K and  $E_{\rm{F}} \sim 10^{9}$K the $T = 0$ approximation is usually very accurate.}, non-rotating WD composed of a Fermi gas of protons, neutrons, and ultra-relativistic electrons is given by \citep{Glen_book}

\begin{equation}\label{eq:Massmax}
    M_{{\rm{max}}} \approx 5.87 \times Y_{e}^{2} \times M_{\odot}, \quad {\rm{where}} \quad  Y_{e} \equiv \dfrac{{\rm{n}}_{e}}{{\rm{n}}_{p} + {\rm{n}}_{n}}
\end{equation}
is the electron fraction. The number densities of electrons, protons, and neutrons are represented by ${\rm{n}}_{e}$, ${\rm{n}}_{p}$, and ${\rm{n}}_{n}$, respectively. Assuming charge neutrality (${\rm{n}}_{e} = {\rm{n}}_{p}$) and a C+O or O+Ne+Mg composition ($ Y_{e} = 0.5)$, we find $M_{{\rm{max}}} \sim 1.46 M_{\odot}$, consistent with our non-rotating model in \ref{tableID}.

Numerically, it is necessary to impose a low-density atmosphere around the star. In practice, we do this by setting low values for the density, neutrino number density, and neutrino energy density (for all neutrino flavors)  outside of the star: $\rho_{\rm atm} = 6.18\times 10^{3} \rm{g}/\rm{cm}^{3}$, $n_{\nu, {\rm atm}} = 3.10 \times 10^{-31} \rm{cm}^{-3}$, and $E_{\nu, {\rm atm}} = 5.55\times10^{8} \rm{erg}/\rm{cm}^{3}$, respectively.

\section{Collapse Dynamics}\label{delep}

\begin{figure}

\begin{subfigure}{0.5\textwidth}
\includegraphics[width=\textwidth]{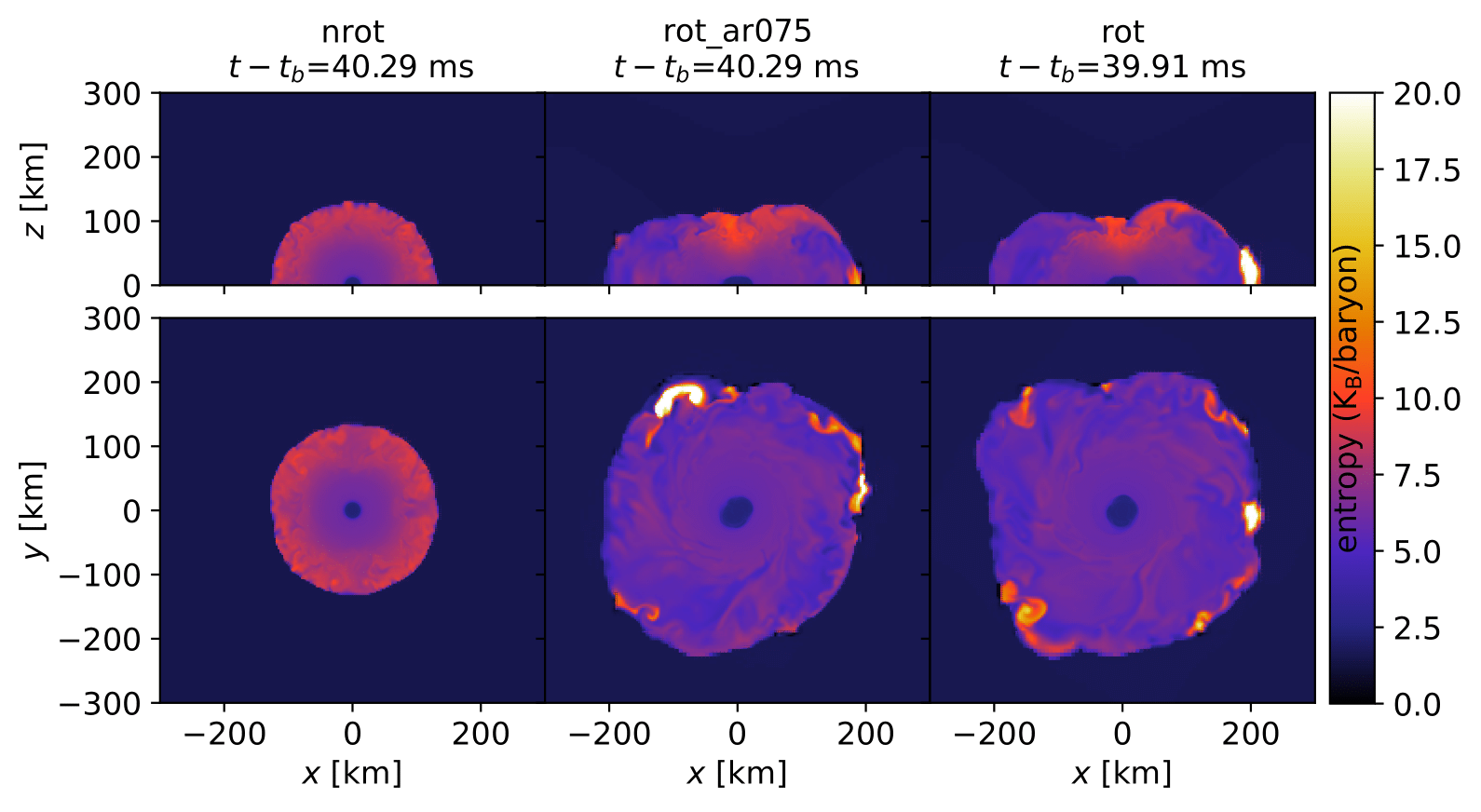}
\end{subfigure}     
\begin{subfigure}{0.5\textwidth}
\includegraphics[width=\textwidth]{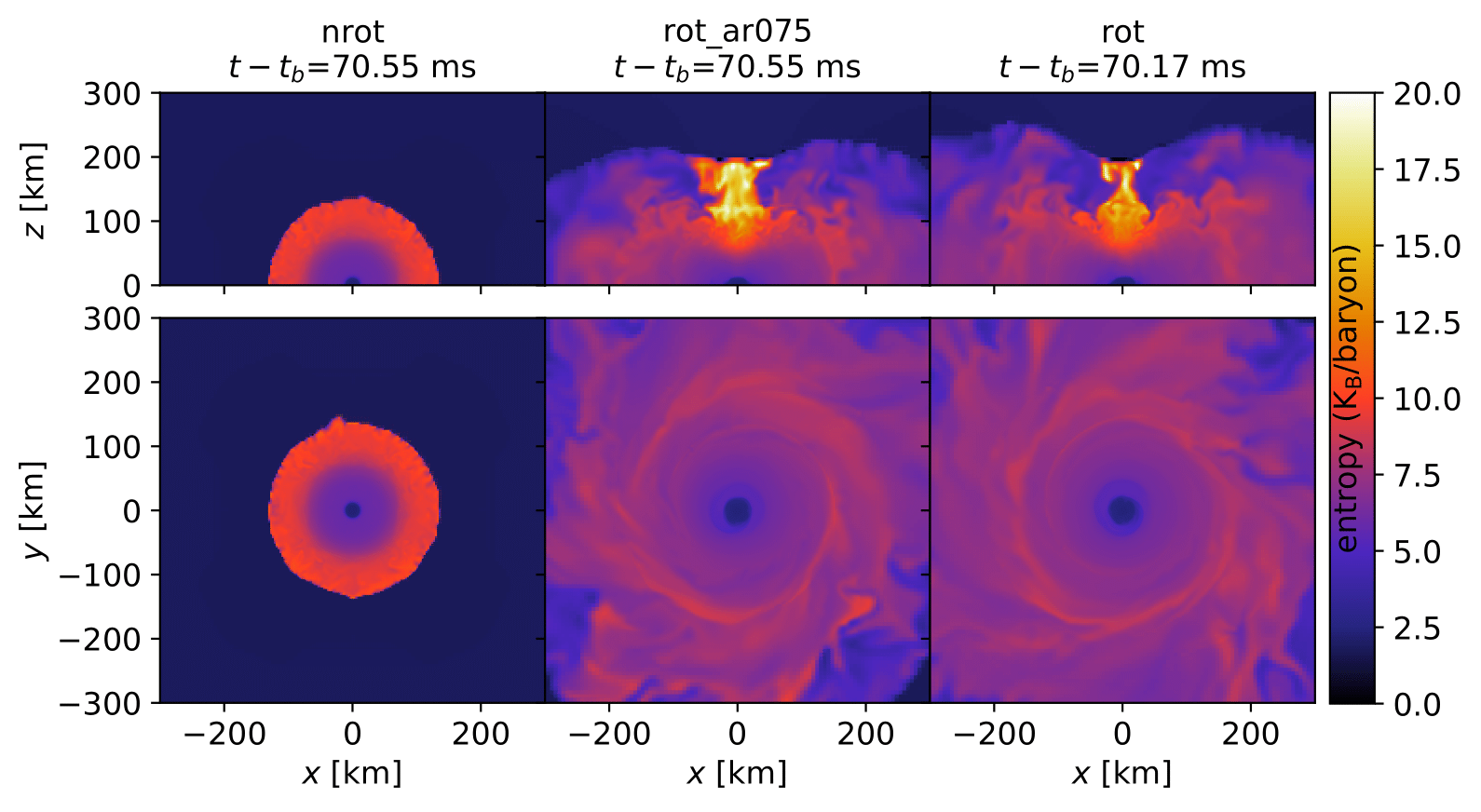}
\end{subfigure}
\begin{subfigure}{0.5\textwidth}
\includegraphics[width=\textwidth]{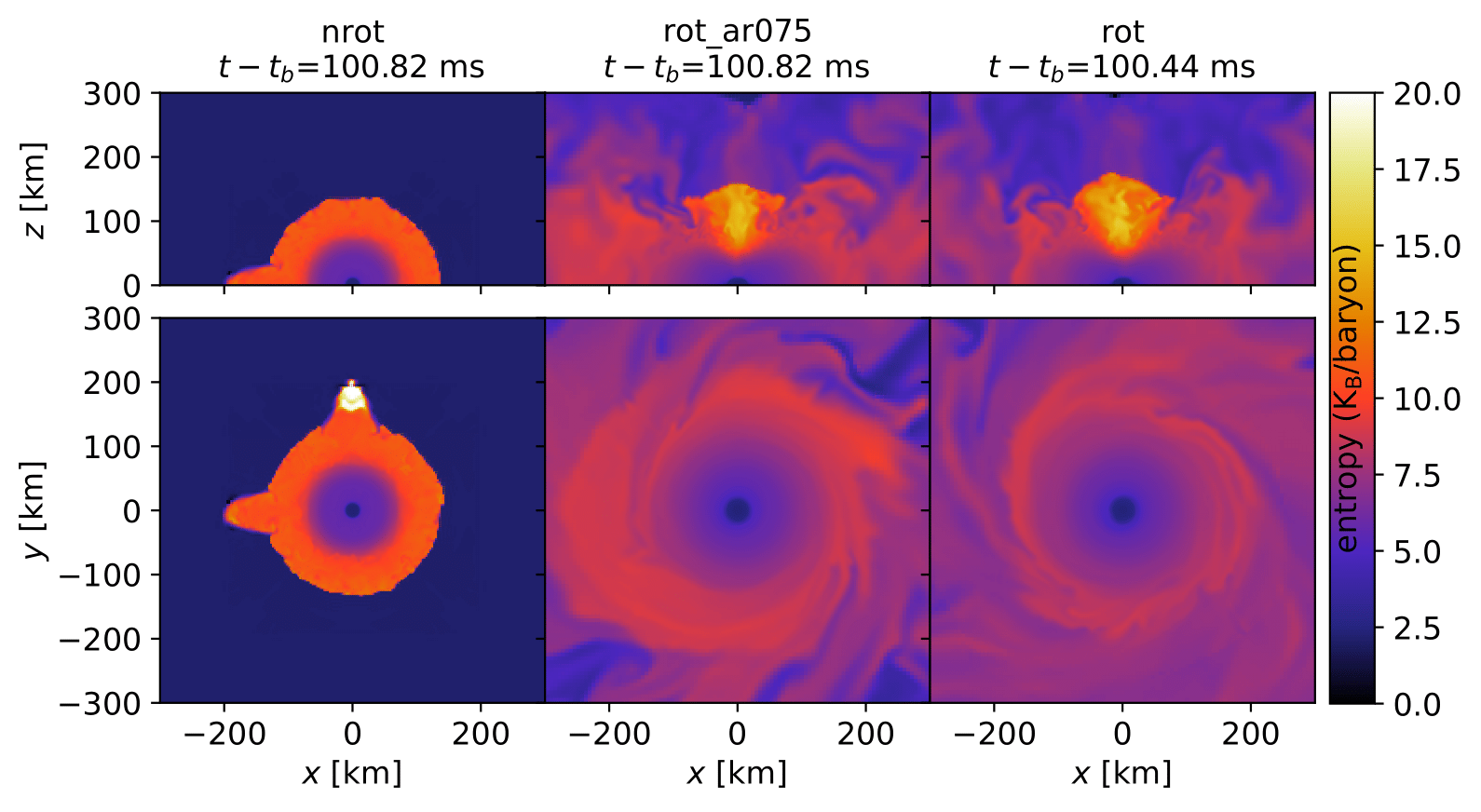}
\end{subfigure}
\begin{subfigure}{0.5\textwidth}
\includegraphics[width=\textwidth]{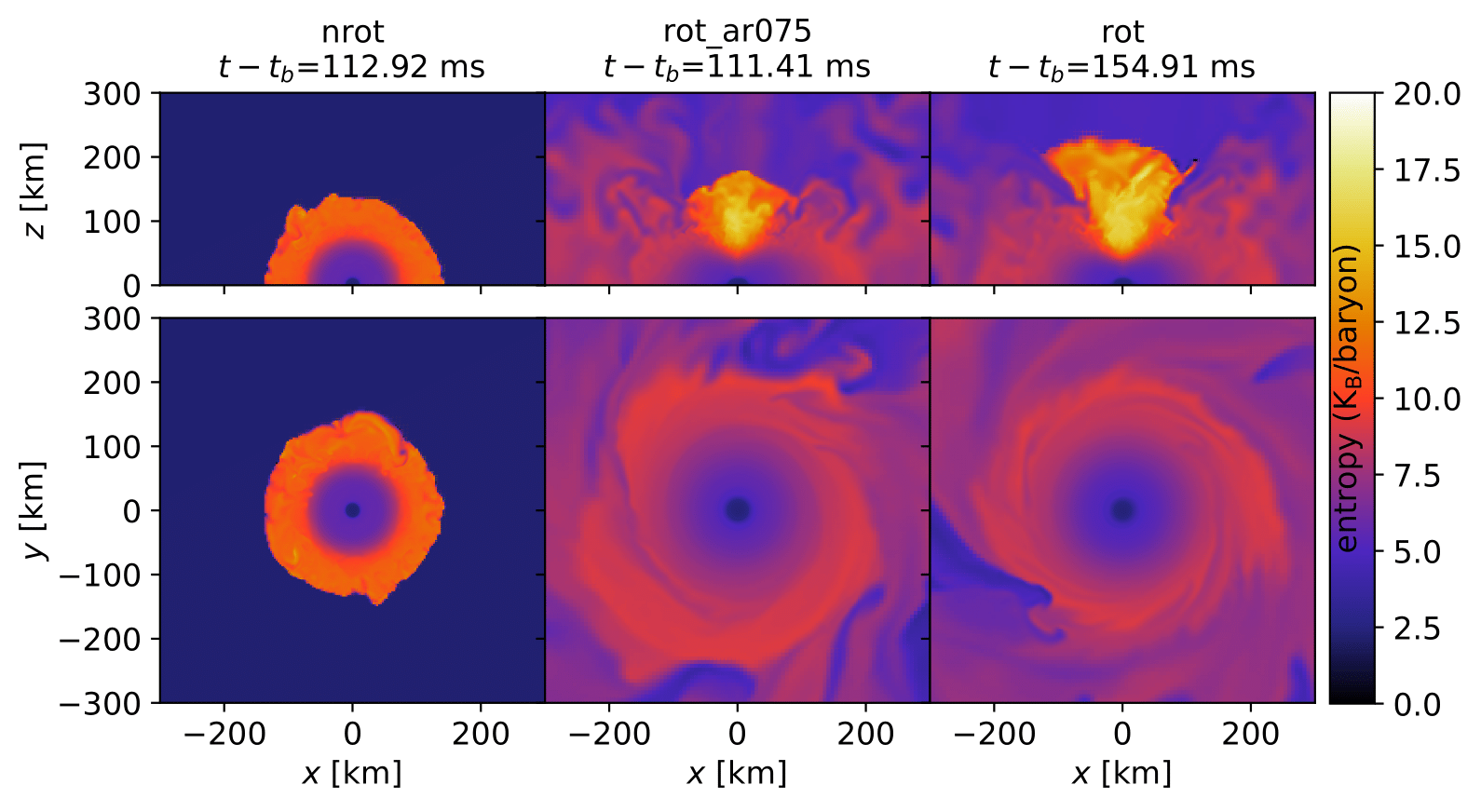}
\end{subfigure}
\caption{Entropy profile of the central part of the collapsing WD. The high entropy material is related to the shock front, while the center of the PNS remains with low entropy, see also Fig.~\ref{fig:vel}. In the \texttt{nrot} model, which  presents a stalled shock, the high entropy material remains contained to $70\ {\rm km} < r < 150\ {\rm km}$ and displays systematically higher entropy than the cases with rotating progenitors. In the  \texttt{rot\_ar075} and \texttt{rot} models, the high entropy material keeps expanding and there is ejection of high entropy material from the pole, see also Fig.~\ref{fig:vel}.}
    \label{fig:entropy}
\end{figure}

\begin{figure}
\begin{center}
\begin{subfigure}{\textwidth}
\includegraphics[width=0.5\textwidth]{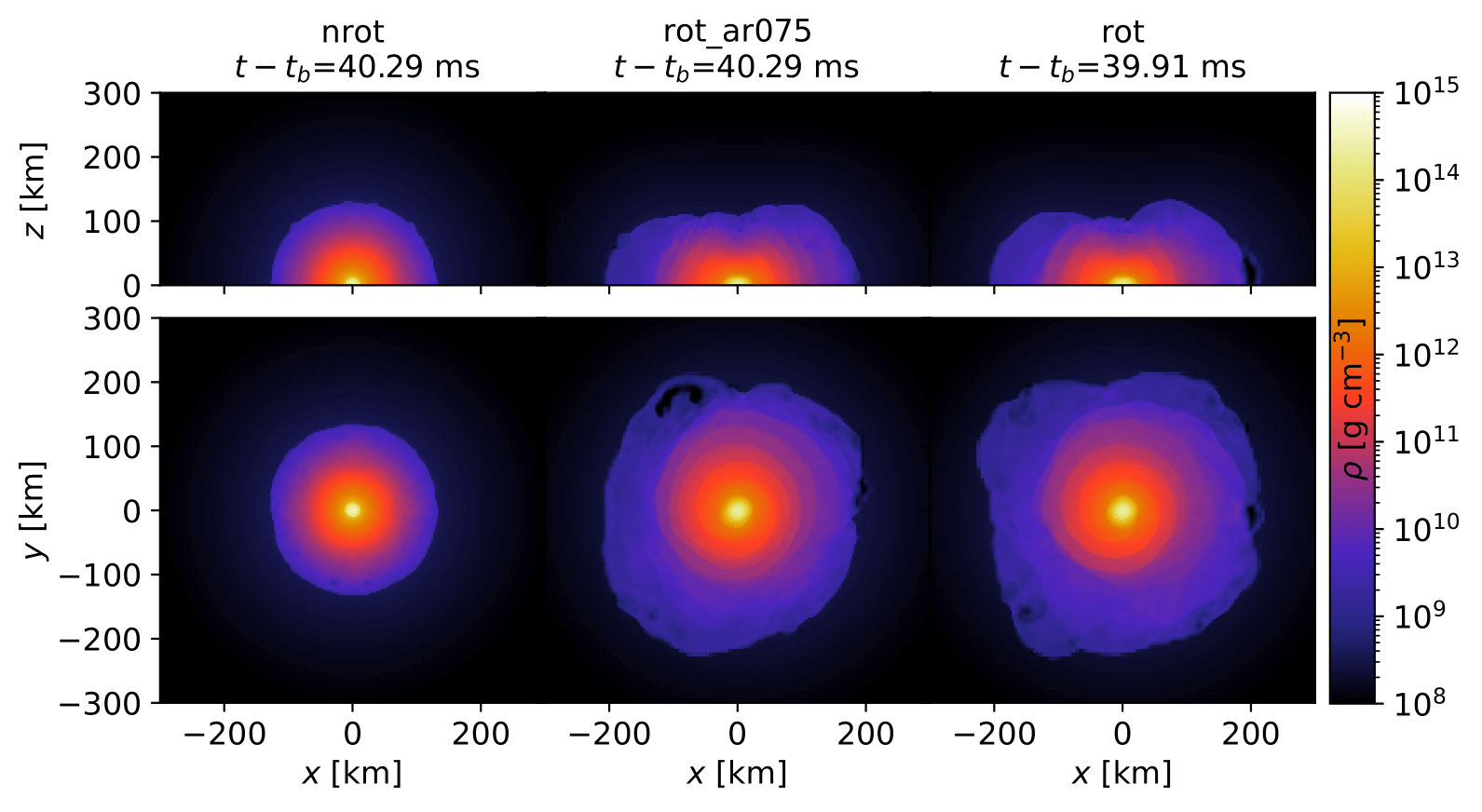}
\end{subfigure}
\end{center}
\begin{center}
\begin{subfigure}{\textwidth}
\includegraphics[width=0.5\textwidth]{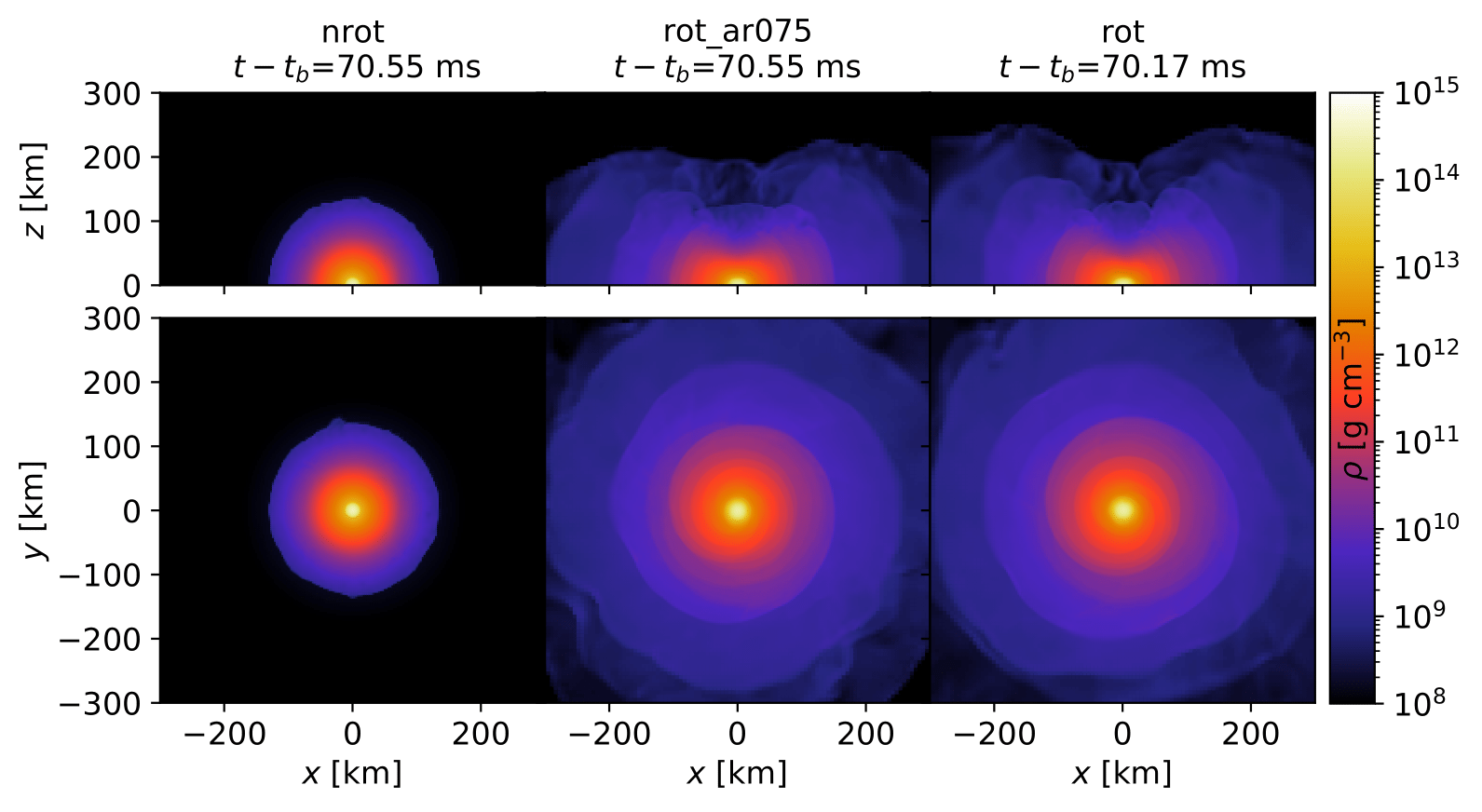}
\end{subfigure}
\end{center}
\begin{center}
\begin{subfigure}{\textwidth}
\includegraphics[width=0.5\textwidth]{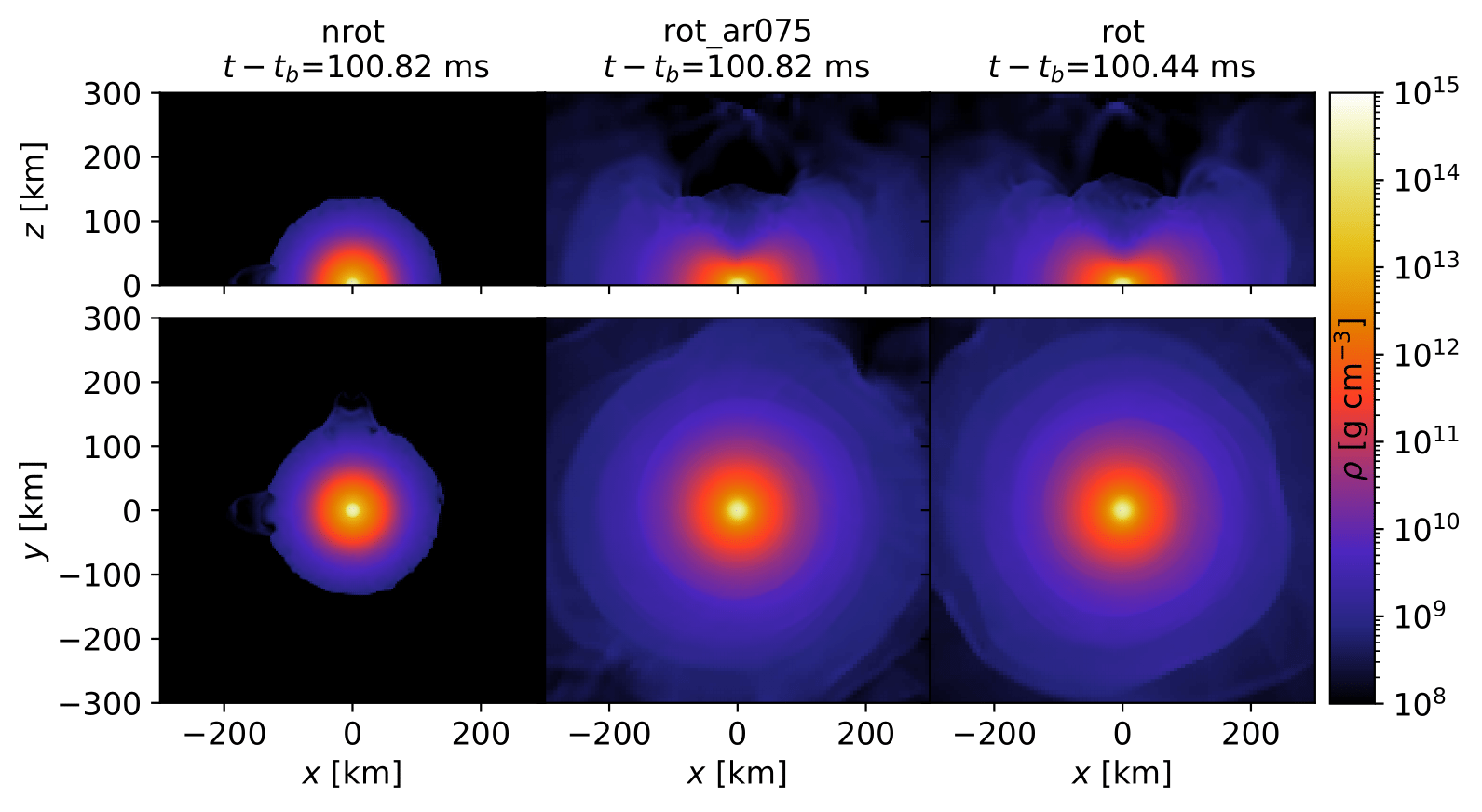}
\end{subfigure}
\end{center}
\begin{center}
\begin{subfigure}{\textwidth}
\includegraphics[width=0.5\textwidth]{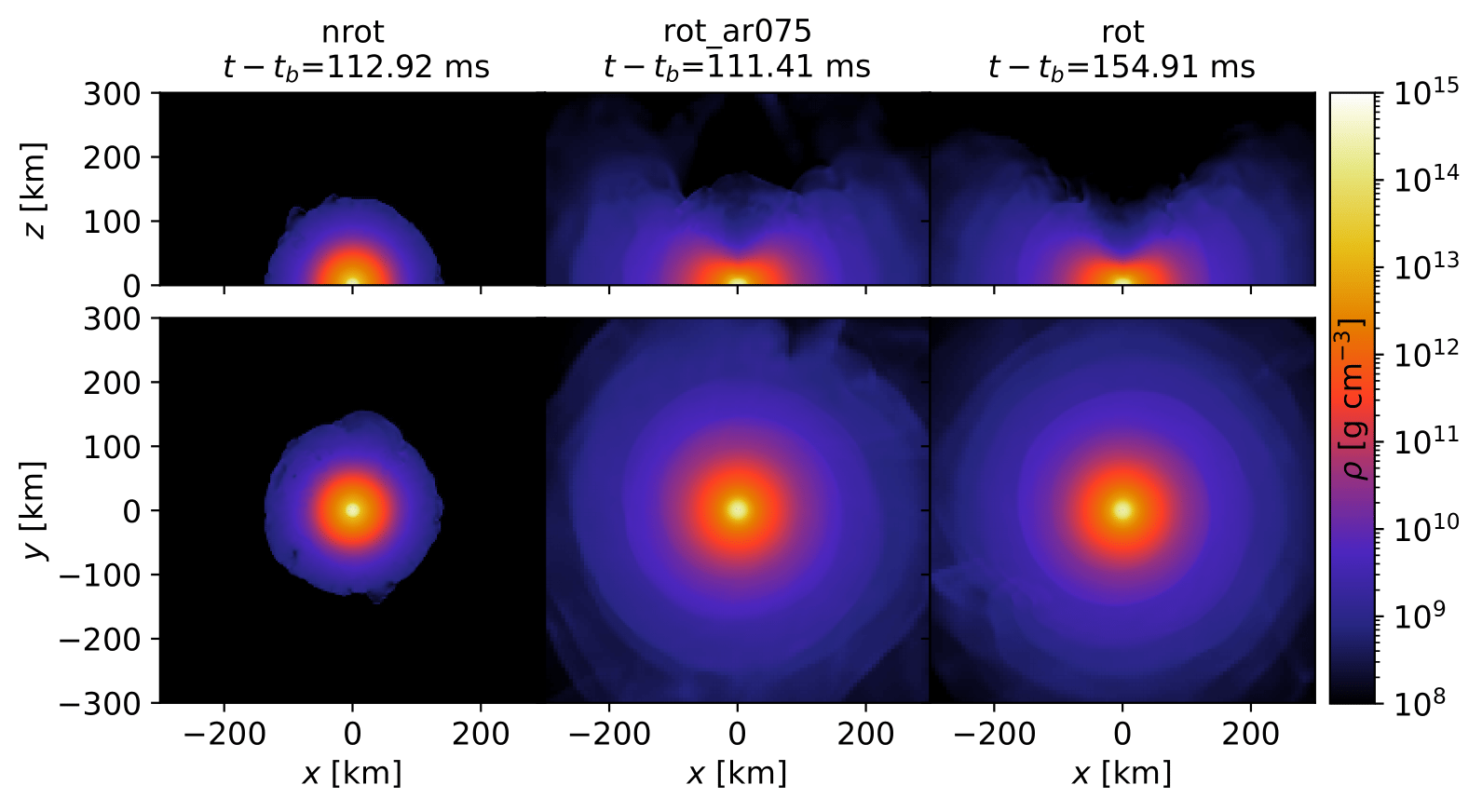}
\end{subfigure}
\end{center}
\caption{Density in the equatorial and meridional planes. The \texttt{rot\_ar075} and \texttt{rot} models present a successful explosion, whereas the shock is stalled for the \texttt{nrot} model at $r\approx 100$ km. A clear torus-like (expanding) structure appears for \texttt{rot} and \texttt{rot\_ar075}. }
\label{fig:dens}
\end{figure}

\begin{figure}

\includegraphics[width=0.5\textwidth]{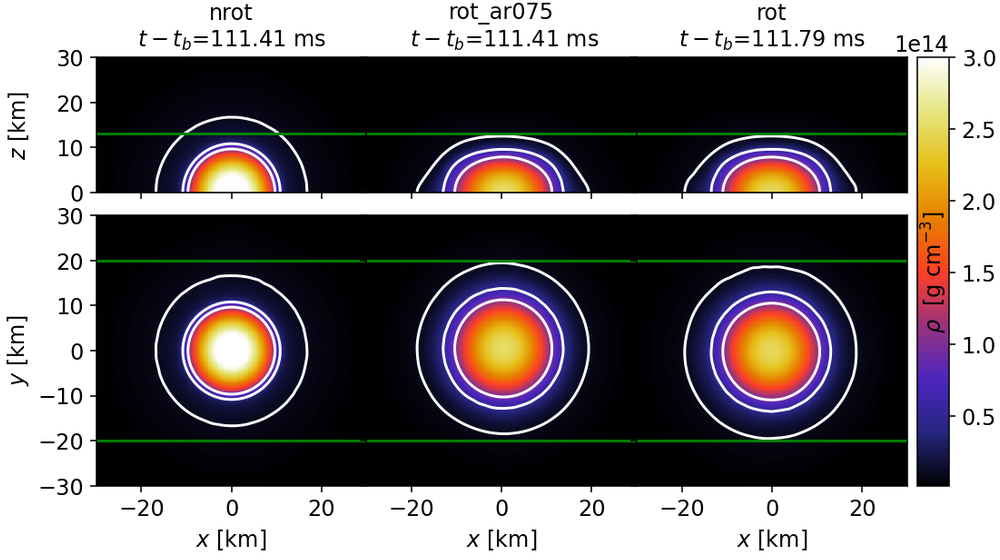}

\caption{Density profile of the central part of the collapsing WD, the PNS. Here we display the final snapshot of the density field with overlapping iso-density surfaces highlighted in white, representing 1, 5 and 10 $\times10^{13}\,{\rm g}/{\rm cm}^3$. The profile of the PNS has already started to settle into a more symmetric configuration. The horizontal green lines are added to aid a visual size comparison: models with initial angular momentum produce larger PNSs.}
    \label{fig:denscentral}
\end{figure}

The electron fraction inside a collapsing WD decreases as electron capture takes place,  driven by the inverse $\beta$-decay: $p+e^{-}\rightarrow n+\nu_{e}$. In our code, we describe this process with the effective deleptonization scheme developed by \citet{Liebendorfer2005}. In Fig.~\ref{fig:Ye}, we show the time evolution of the electron fraction. In the beginning of the collapse stage (not shown in Fig.~\ref{fig:Ye}), we have $Y_{e}\sim0.5$ everywhere inside the star. The deleptonization process starts in the stellar core ($\rho\sim 10^{10}$ g/cm${}^{3}$ in the beginning of our simulations) and the electron degeneracy pressure, that initially supports the star against gravitational collapse, diminishes drastically. Consequently, the star enters the \textit{collapse} stage and becomes increasingly denser, as seen in Figs.~\ref{fig:Central_rho_evolve} and \ref{fig:densityaverage}. 

The collapse phase is mainly characterized by purely ingoing fluid motion. After achieving central densities comparable to the nuclear saturation density ($\rho_{c}\sim 10^{14}$g/cm$^{3}$), the stellar core stiffens significantly. The central density achieves its first local maximum (see Fig.~\ref{fig:Central_rho_evolve}) followed by an abrupt relaxation, marking the time of \textit{bounce} ($t_{b}$) of the core. At this time, $Y_{e}$ achieves values of $\approx 0.25$ in the core. In this context, the collapse phase refers to $t-t_{b}<0$, whereas the \textit{post-bounce} phase refers to $t-t_{b}>0$. When the core bounces, it injects an enormous amount of kinetic energy in the surrounding stellar fluid, which starts an outgoing motion. Fig.~\ref{fig:vel} shows the velocity profile of the stellar fluid for $t-t_{b}>0$. Besides the larger scale outgoing motion in the post-bounce phase, a smaller amount of fluid steadily accretes to the core as the AIC evolves. During the early times of the post-bounce phase it is possible to see oscillations in the density profile of the stellar core (see Fig.~\ref{fig:Central_rho_evolve}), which are associated with the quasi-normal mode oscillations of the remnant PNS.

The rotation of the progenitor WD has a striking effect in the late evolution of the AIC, shown in Fig.~\ref{fig:vel}. Shortly after the bounce, some of the infalling material starts traveling in the opposite direction, achieving mildly relativistic velocities ($v_{r}\lesssim 0.1 c $) for the rotating progenitors. However, the shock of the \texttt{nrot} model stalls at $t-t_{b}\approx$ 40 ms. In this case, the shock ceases to propagate radially at a distance of $\approx 100$ km from the center, showing no sign of a shock revival for the remaining duration of the simulation. The stronger shocks seen in the \texttt{rot} and \texttt{nrot\_ar075} models are enhanced (compared with the stalled shock in \texttt{nrot}) by the kinetic energy of the rotating stars.

While the AIC from a non-rotating progenitor maintains approximate spherical symmetry at all times, the models with rotating progenitors display spiral arms that propagate outwards. These spiral arms maintain a torus-like morphology throughout the post-bounce phase. The spiral motion of the fluid pushes the low $Y_{e}$ material away from the center, and as the material expands the electron fraction becomes nearly symmetric again due to $e^{+}$ capture associated and $\nu_{e}$ absorption. The material with low electron fraction ($Y_{e}\lesssim 0.3$) is constrained to the central 100 (200) km for the \texttt{nrot} model (\texttt{rot\_ar075} and \texttt{rot} models). 

\begin{figure}

\begin{subfigure}{0.5\textwidth}
\includegraphics[width=\textwidth]{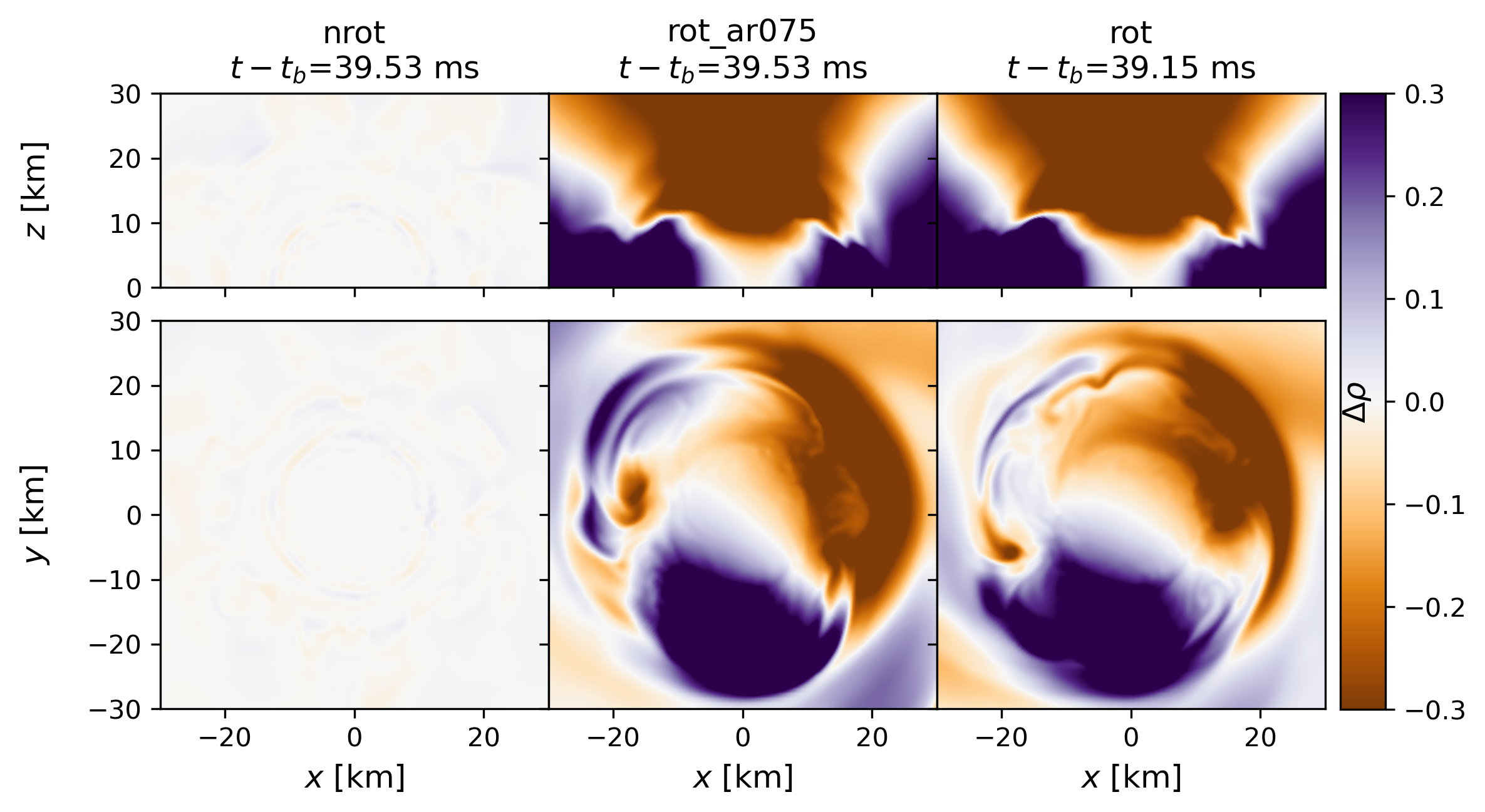}
\end{subfigure}     
\begin{subfigure}{0.5\textwidth}
\includegraphics[width=\textwidth]{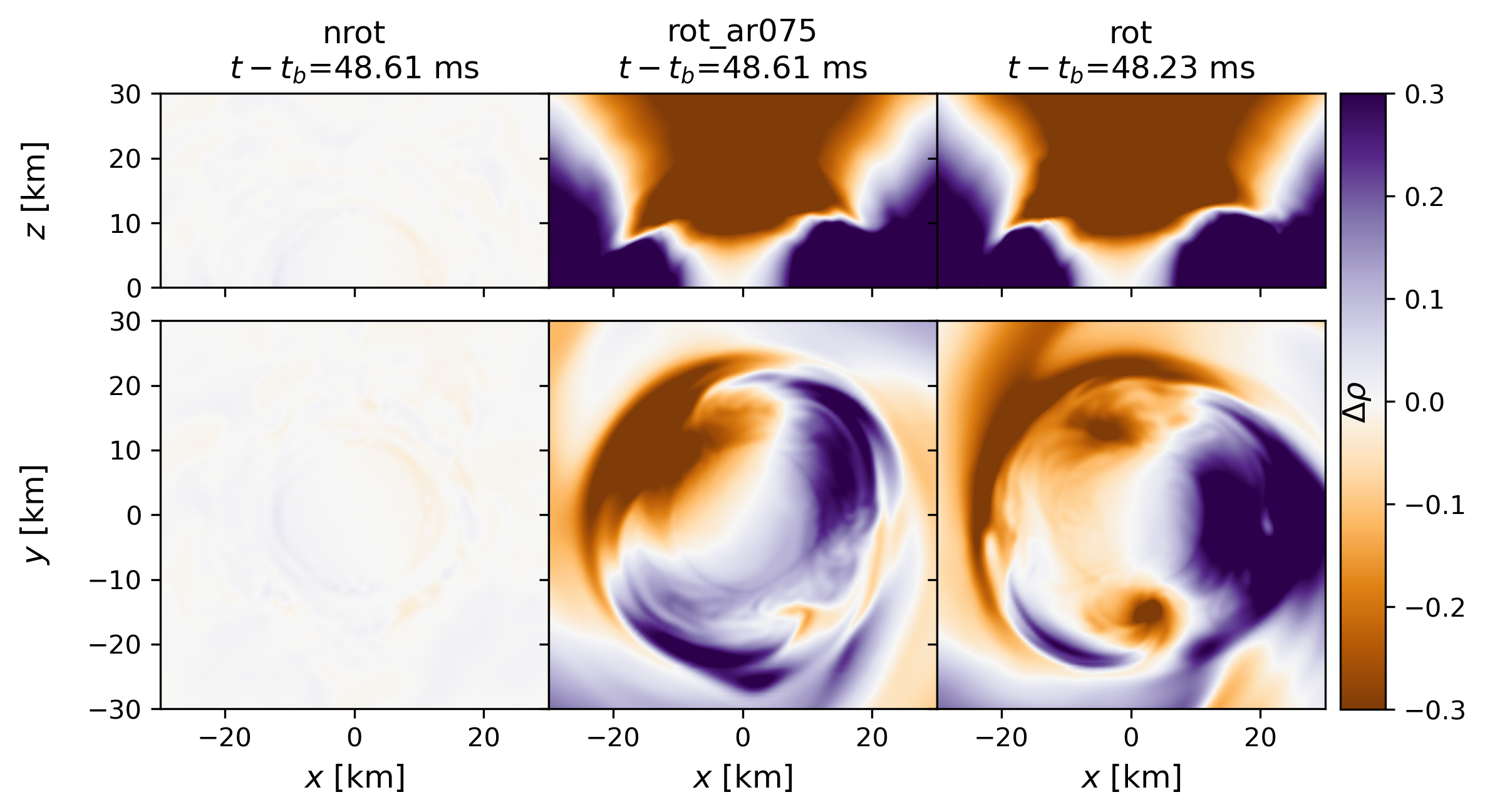}
\end{subfigure}
\begin{subfigure}{0.5\textwidth}
\includegraphics[width=\textwidth]{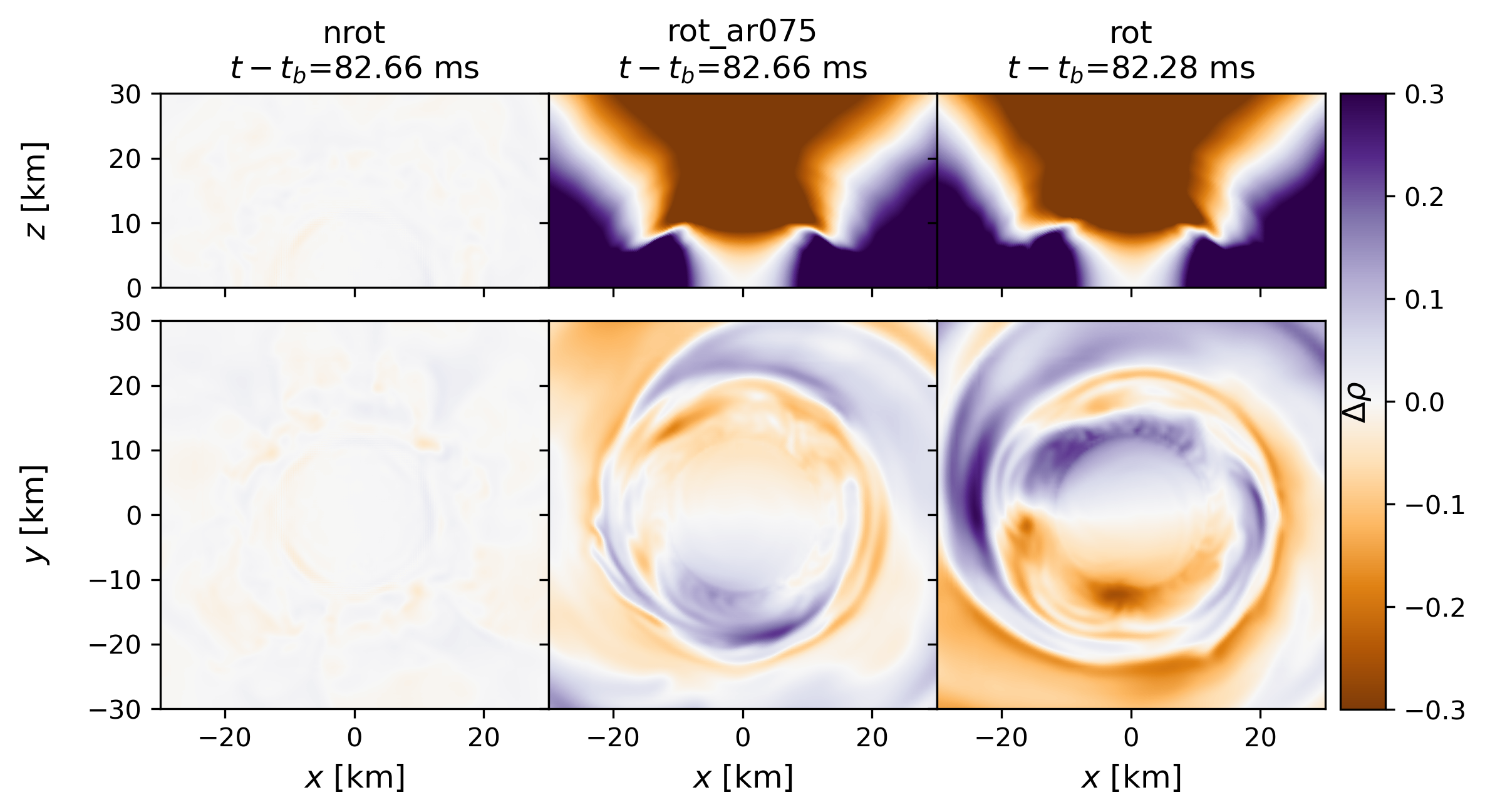}
\end{subfigure}
\begin{subfigure}{0.5\textwidth}
\includegraphics[width=\textwidth]{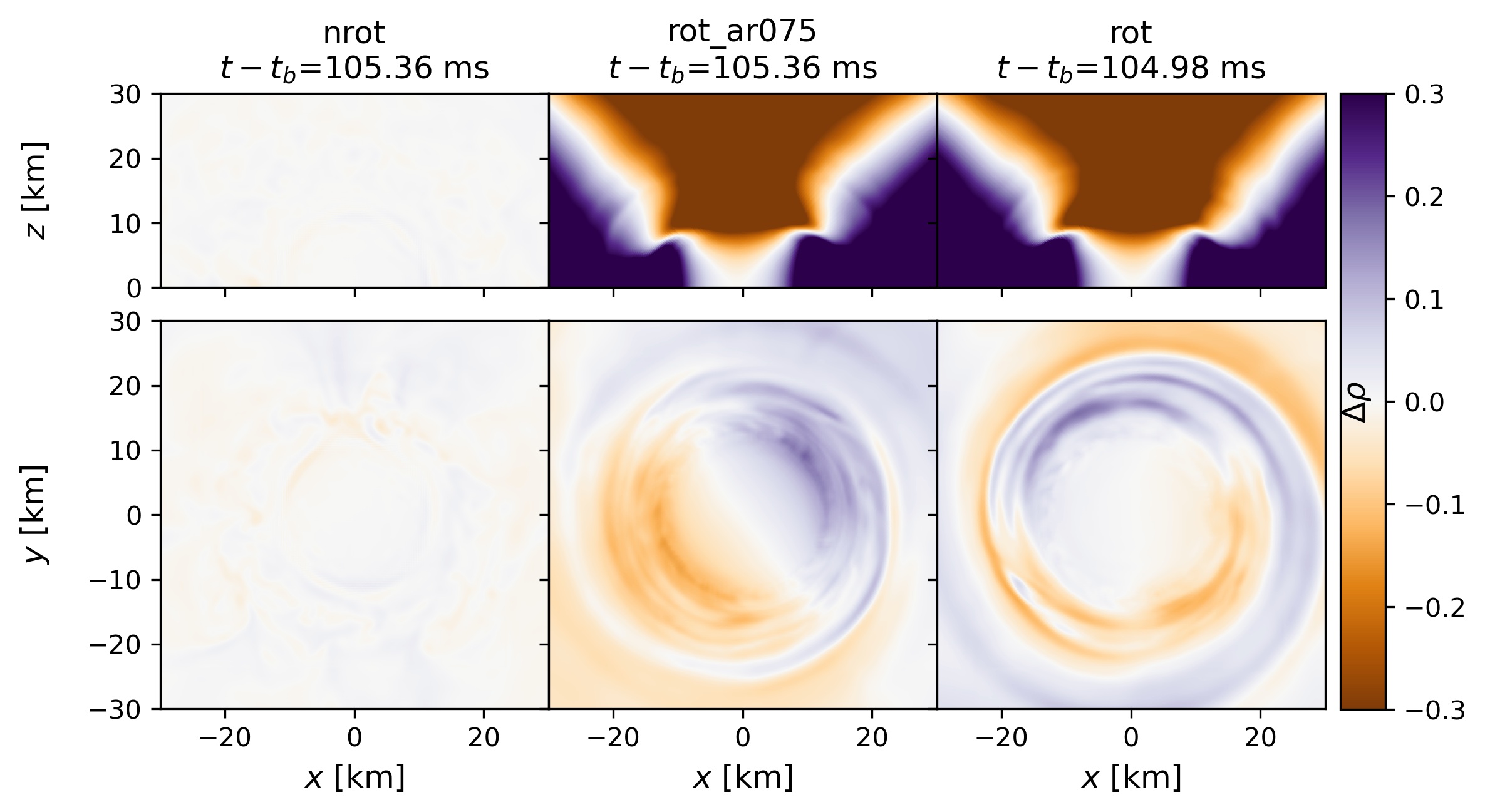}
\end{subfigure}
\caption{Angular-averaged profile of relative density fluctuations ($\Delta\rho$) in the central part of the collapsing WD, the PNS, shown in the equatorial and meridional planes. Density fluctuations for the \texttt{nrot} model are negligible compared with the cases with a rotating progenitor. For the \texttt{rot\_ar075} and \texttt{rot} models, at early times there is a strong $m=2$ mode with two overdense regions (in blue) intercalated between two underdense (in gold) regions. At later times the most prominent mass distribution mode is the $m=1$ where one overdense (in blue) region is diametrically opposed to an underdense (in gold) region. The equatorial plane view shows 
overdense thick disks in the \texttt{rot} and \texttt{rot\_ar075} models. }
    \label{fig:flucdens}
\end{figure}

Comparing Figs.~\ref{fig:vel} and \ref{fig:entropy}, we can see that the shock fronts display relatively large values of entropy ($S \approx 10 k_{\rm{B}}$/baryon). This effect is even more visible in the polar region, where higher values of entropy ($S \approx 15 k_{\rm{B}}$/baryon) can be seen. There is a cone-like structure in the polar region of models \texttt{rot} and \texttt{rot\_ar075}, containing low-density material (see Fig.~\ref{fig:dens}) with  symmetric $Y_{e}$ (Fig.~\ref{fig:Ye}),  mildly relativistic outgoing velocity ($v \approx 0.08c$, see Fig.~\ref{fig:vel}) and high entropy (Fig.~\ref{fig:entropy}), being expelled from the poles. This material is pushed away from the central part of the star by neutrino absorption (see also Section \ref{sec:Neutrinos}). This seems to indicate that ejection is favored at the poles in a collimated fashion. In the presence of magnetic fields, these ejections at the poles could facilitate the breaking of field lines, possibly making these regions hot spots for the formation of jets.

Further examination of Fig.~\ref{fig:densityaverage} confirms that a PNS is left behind a remnant of an AIC. Setting $\rho\sim 10^{11}$g/cm${}^{3}$ as the surface of the PNS, we can see that the remnants are contracting (more rapidly for the \texttt{nrot} model). These remnants have extremely dense cores, achieving central densities $\rho_0 \gtrsim 10^{14}$ g cm${}^{-3}$. Another important property of the PNSs is their electron fraction, which is always smaller than $\lesssim 0.3$ (see Fig.~\ref{fig:Ye}).

\section{Gravitational Wave emission}\label{sec:GW}

The emission of GWs requires the accelerated motion of an asymmetrical (a quadrupole mode or higher) mass distribution. To verify the asymmetry in the mass distribution of our models, we use two independent methods: a visual inspection of the density fluctuations and a mode decomposition of the stellar density profile. First, we compute the angular-averaged relative perturbation of the density profile, defined as:

\begin{equation}
    \Delta \rho \equiv \dfrac{\rho - \left.\langle\rho\rangle\right|_{\phi,\theta}}{\left.\langle\rho\rangle\right|_{\phi,\theta}}, 
\end{equation}

where $\left. \langle \rho \rangle\right|_{\phi,\theta}$ is the angular average in $\phi \equiv \arctan(y/x)$ (when plotting in the xy plane) or $\theta \equiv \arctan(z/r)$ (when plotting the yz plane) of the density $\rho$. Fig.~\ref{fig:flucdens} shows $\Delta \rho$ in the central part of the PNS remnant (as it is the main responsible for the GW emission). The nonrotating progenitor can be very well approximated by a spherically symmetric mass distribution,  even for $t>t_{b}$. In fact, the density fluctuations in the \texttt{nrot} model are at least one order of magnitude smaller than in the models with rotating progenitors, therefore GW signals will be enhanced for the rotating models. Another effect due to rotation is the thick disk on the equatorial plane. For the rotating cases, in the equatorial plane, we can see that at early times there is a noticeable quadrupolar mass distribution, followed by a somewhat longer lasting dipolar distribution.

\begin{figure*}
     \includegraphics[width=\textwidth]{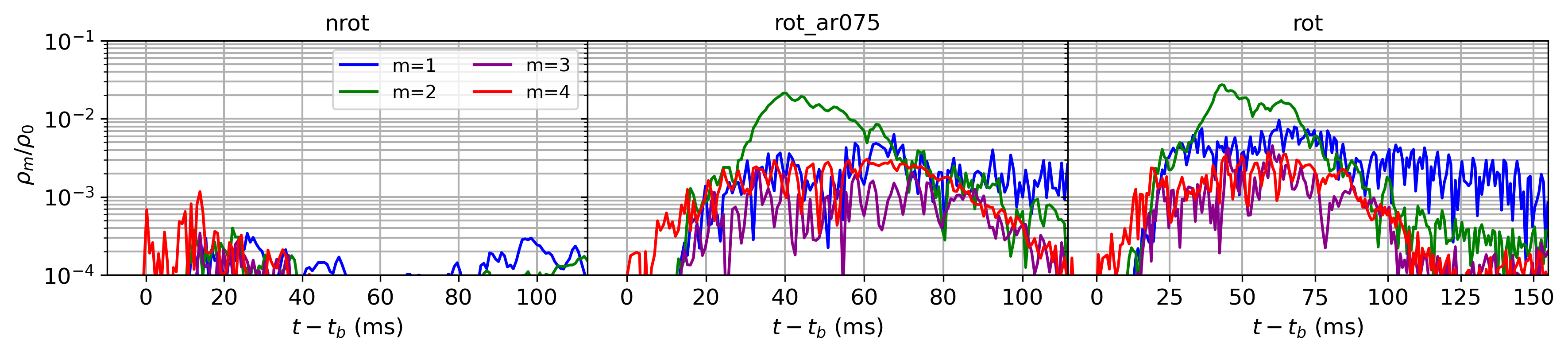}
    \caption{Mode decomposition of the density in the  equatorial plane, see Equation (\ref{rhom}). The \texttt{nrot} model maintains an approximately spherical  mass distribution, as all modes have relative amplitude  lower than 0.1\% at all times. For the models with rotating progenitors, the $m=2$ mode is dominant from $\approx$ 20 ms to $\approx$ 80 ms after the bounce, peaking at a relative amplitude of 2-3\%. At later times $t-t_b \gtrsim 80$ ms, the $m=1$ mode becomes dominant, maintaining a relative amplitude approximately 10 times higher than the other modes.}
    \label{fig:densitymodes}
\end{figure*}

\begin{figure*}
     \centering
     \includegraphics[width=\textwidth]{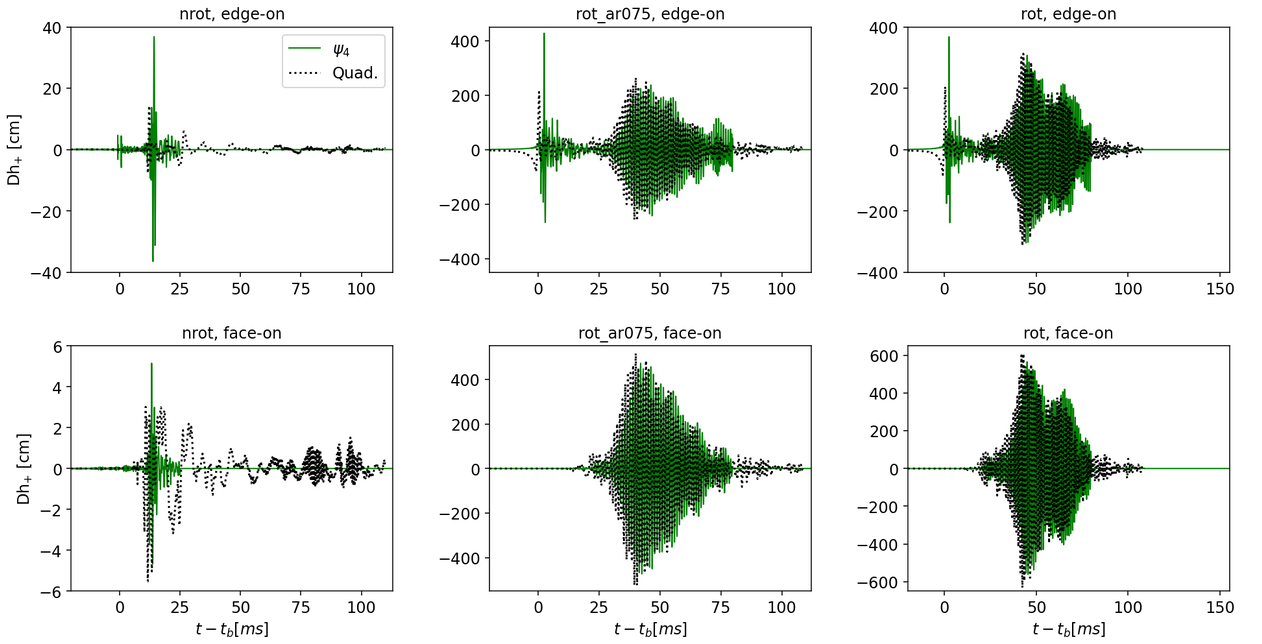}
     \caption{GW emission in the time domain. Here we compare the GW strain ($l=2$ mode) obtained via the quadrupole approximation and with the waveform reconstruction from $\Psi_{4}$. In the edge-on configuration, we can see a prompt $m=0$ signal, followed by a very loud $|m|>0$ pulse coincident with the $\rho_{2}$ perturbation shown in Fig.~\ref{fig:densitymodes}. (We are able to identify the $m=0$ contribution based on an analysis of $\Psi_{4}$, which is originally obtained in the spin-weighted spherical harmonics basis. The signal shown here represents the full $l=2$ mode.) Note that some features vanish when moving from the edge-on to the face-on orientation. This is due the different angular dependence of each one of the $m$ modes. The prompt emission is morphologically consistent with the results from \citet{Abdikamalov2010}. The overall amplitude of the GWs emitted by our non-rotating model is consistent with simulations of GWs from nonrotating CCSNe \citep{Radice_2019}. On the other hand, initial angular momentum seems to amplify the signals by up to two orders of magnitude.}
        \label{fig:GW}
\end{figure*}

A more quantitative description of the results shown in Fig.~\ref{fig:flucdens} is obtained by the mode decomposition of the density in the equatorial plane. The mode coefficients $\rho_{m}$ are defined as

\begin{equation}\label{rhom}
    \rho_{m} \equiv \bigintsss \sqrt{\gamma}\, \rho\, W\, e^{-i m \phi}\, dx \,dy,
\end{equation}

where $W$ is the Lorentz factor, $\gamma$ is the 3-metric determinant and $\rho \, dx \, dy$ is the surface density in the equatorial plane. These results are shown in Fig.~\ref{fig:densitymodes}. In the nonrotating case, all modes have relative amplitude $(\rho_{m}/\rho_{0})$ below $2\times10^{-3}$ for all times. The cases with non-zero initial spin exhibit a prominent peak of the $m=2$ mode at approximately 40 milliseconds after the bounce. For both rotating cases, the $m=2$ mode decays rapidly after the peak, after which the $m=1$ mode becomes the dominant mode. While all other modes tend to decay after 80 ms, the $m=1$ mode lingers on. This $m=1$ feature is related to the one-armed instability \citep{Low3,Low4,Kashyap_2017}.

The gravitational waveforms of our three models are presented in Fig.~\ref{fig:GW}. The GW signals were obtained with two different methods. In the first one, we perform a fixed-frequency double time integration of the Weyl scalar $\psi_{4}$ for all $l=2$ modes. In this method, we use a high-pass filter to keep only frequencies higher than 120 Hz, in order to mitigate the numerical noise at lower frequencies. The second method for obtaining the GW emission used the quadrupole formula \citep{Maggiorebook}, where no post-processing noise mitigation was employed. The resulting GWs from both methods are shown in Fig.~\ref{fig:GW}. Although there is good qualitative agreement between the two ways of reconstructing the waveform, there are clear quantitative disagreements between them, specially in the \texttt{nrot} case, due to the different treatment of numerical noise in each method. Consistent with this interpretation, the \texttt{nrot} model is more affected as its low GW amplitude is comparable to  the numerical noise found in our numerical evolution. One should also have in mind that while the quadrupole formula is approximate (low velocity, weak gravity, and long GW wavelength approximations), while $\psi_4$ is exact, but potentially more sensitive to numerical noise.

As expected, Fig.~\ref{fig:GW} shows that rotating progenitors should emit GWs with higher amplitudes. The nonrotating \texttt{nrot} model yields a GW signal with similar amplitude to a CCSNe with a nonrotating progenitor \citep{PhysRevD.105.063018, PhysRevD.92.084040, Kuroda_2017, Dolence_2015, Murphy_2009, Muller_2013, refId0, CCSN_GW_nrot1, CCSN_GW_nrot2, Morozova_2018, Radice_2019, Mezzacappa_2023, Vartanyan_2023}. The perturbations of the PNS perturbation are amplified by the initial angular momentum (see Figs.~\ref{fig:flucdens} and \ref{fig:densitymodes}), strengthening the GW signals. Comparing our models, we see that the \texttt{rot} model radiates a strain that is up to two orders of magnitude stronger than the \texttt{nrot} model. Our \texttt{rot\_ar075} and \texttt{rot} models have initial angular velocities of $5.09$ Hz and $5.28$ Hz, emit GWs with amplitudes between $Dh \sim 200$ cm and $Dh \sim 600$ cm and exhibit a peak frequency of $\sim 1000$ Hz. As AICs are expect to rotate faster than CCSNe, we argue that AICs are more likely to produce loud high-frequency signals than CCSNe. \citet{CCSNerot1} and \citet{CCSNerot2} studied the GW signal emitted by CCSNe with rotating progenitors ($1\ {\rm Hz} <\Omega < 2\ {\rm Hz}$). In their analysis they found that these events display a GW strain amplitude between $Dh\sim60$ cm and $Dh \sim300$ cm, with peak frequency varying from $\sim$400 to $\sim$900 Hz. The amplitudes and frequencies of these signals increase systematically with rotation.

Our rotating models exhibit a prominent $m=0$ pulse early in the post-bounce phase, identified by the modal decomposition of $\Psi_{4}$ (not shown here). The fact that the initial $m=0$ GW mode disappears in the face-on inclination is a reflection of the different weights, coming from the spin-weighted spherical harmonics basis that each mode has for a given observation angle. This early $m=0$ GW emission is consistent with results from the 2D simulations of \citet{Abdikamalov2010}.

\begin{table}
\centering
\begin{tabular}{lccc}

    Model & LIGO  &  CE & ET \\
    \hline
    \texttt{nrot} ($\Psi_{4}$ edge-on) & 1.7$\times10^{1}$ & 1.5$\times10^{2}$ & 7.7$\times10^{1}$ \\
    \texttt{nrot} (Quad. edge-on)      & 1.1$\times10^{1}$ & 7.7$\times10^{1}$  & 5.5$\times10^{1}$ \\
    \hline
    \texttt{nrot} ($\Psi_{4}$ face-on) & 2.1$\times10^{0}$ & 1.8$\times10^{1}$ & 9.5$\times10^{0}$ \\
    \texttt{nrot} (Quad. face-on)      & 6.7$\times10^{0}$ & 4.9 $\times10^{1}$ & 3.4 $\times10^{1}$ \\
     \hline
     \hline
    \texttt{rot\_ar075} ($\Psi_{4}$ edge-on) & 4.3$\times10^{2}$ & 3.7$\times10^{3}$& 1.9$\times10^{3}$ \\
    \texttt{rot\_ar075} (Quad. edge-on)      & 4.3 $\times10^{2}$ & 3.6$\times10^{3}$ & 2.0$\times10^{3}$ \\
    \hline
    \texttt{rot\_ar075} ($\Psi_{4}$ face-on) & 7.8$\times10^{2}$ & 6.7$\times10^{3}$ & 3.5$\times10^{3}$ \\
    \texttt{rot\_ar075} (Quad. face-on)      & 8.1$\times10^{2}$ & 6.9$\times10^{3}$ & 3.7$\times10^{3}$ \\
     \hline
     \hline
    \texttt{rot} ($\Psi_{4}$ edge-on) & 5.1$\times10^{2}$ & 4.4$\times10^{3}$ & 2.3$\times10^{3}$\\
    \texttt{rot} (Quad. edge-on)      & 5.1$\times10^{2}$ & 4.3$\times10^{3}$ & 2.3$\times10^{3}$\\
    \hline
    \texttt{rot} ($\Psi_{4}$ face-on) &  9.5$\times10^{2}$ & 8.2$\times10^{3}$ & 4.3$\times10^{3}$\\
    \texttt{rot} (Quad. face-on)      &  9.7$\times10^{2}$ &  8.3$\times10^{3}$ & 4.4$\times10^{3}$ \\
     \hline

\end{tabular}
    
    \caption{ Signal to noise ratio (SNR) of our models for different WG detectors: advanced LIGO, Cosmic Explorer (CE), and Einstein Telescope (ET). All SNRs shown in this table were calculated for the full $l=2$ mode at 10kpc, see Fig.~\ref{fig:GWFspace}. We show values obtained both by using the quadrupole approximation and performing strain reconstruction via $\Psi_{4}$. The discrepancies in SNR obtained through different methods can be understood by the fact that they have very distinct noise mitigation schemes. Therefore the final strain obtained in each method is subjected to different levels of noise. In particular, the \texttt{nrot} model is more affected by these uncertainties as its strain amplitude is comparable to the numerical noise present in our simulations, see Fig.~\ref{fig:GW}.}
    \label{tab:SNR_m1_m2}
\end{table}

As \citet{Abdikamalov2010} were limited to 2D simulations, they could only access the $l=2,\ m=0$ mode. Here we moved to 3D models, allowing us to look into higher modes. Given that the quadrupolar contribution is expected to be dominant, we focused our study on the full $l=2$ mode. 
After the prompt (lower frequency) GW emission close to $t-t_{b}=0$, we see a higher frequency signal starting $\approx 20$ ms after the bounce and lasting up to $t-t_{b}\approx80$ ms, as can be seen in Fig.~\ref{fig:GW}. The initial low-frequency signal is produced by the bounce of the PNS core, whereas the later high-frequency emission is related to the  highly perturbed PNS. The asymmetry in the matter distribution of the PNS grows in amplitude until it generates a GW signal comparable to the prompt emission (see Figs.~\ref{fig:flucdens} and \ref{fig:densitymodes}).

\begin{figure*}
    \centering
    \includegraphics[width=0.9\textwidth]{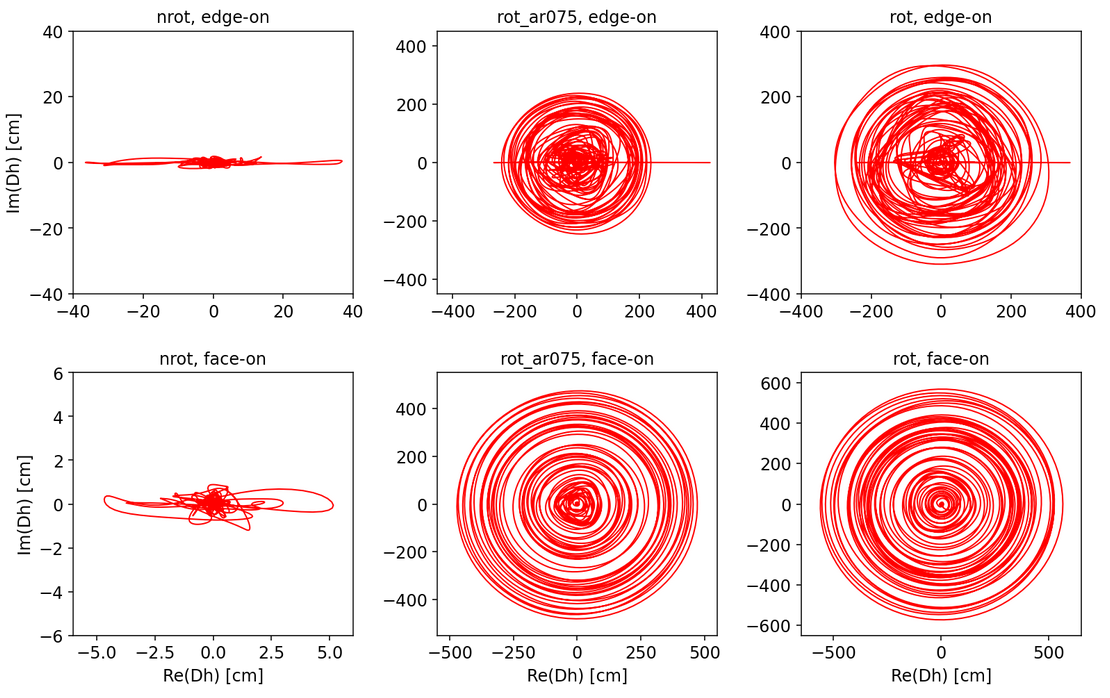}
    \caption{GW polarization patterns using the strain calculated via the Weyl scalar $\Psi_{4}$. For the face-on configuration, the signal follows a very clear circular polarization. The polarization is more complex for the edge-on case, due to the prompt $m=0$ mode which holds a linear polarization. The \texttt{nrot} model is mostly linear polarized but is notably affected by numerical errors.}
    \label{fig:GW_pol_1}
\end{figure*}

\begin{figure}
    \centering
    \includegraphics[width=0.4\textwidth]{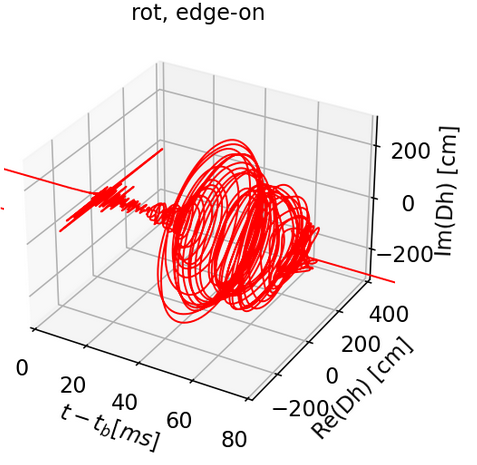}
    \caption{Temporal evolution of the GW polarization pattern for \texttt{rot} in the edge-on alignment. Note that at early times the signal displays a linear polarization associated with the prompt $m=0$ GW emission. The circular polarization corresponds to the $m=2$ GW mode. The strain used is the one obtained via the Weyl scalar $\Psi_{4}$.}
    \label{fig:GW_pol_2}
\end{figure}

\citet{CicularPolarization} showed that the polarization of the GWs of a CCSN can carry information about the EOS of the remnant PNS. The standing accretion shock instability (SASI), present or absent depending on EOS, produces GWs that are circularly polarized, but the sign of the polarization can change abruptly during the CCSN evolution. In contrast to this polarized signal, the GWs emitted by the fundamental mode of oscillation of the PNS in their models are randomly polarized. This separation can be clearly seen in Fig.~1 of \citet{CicularPolarization}. In Fig.~\ref{fig:GW_pol_1} we show the polarization plots for our models, obtained with the quadrupole formula. In our rotating models, we find that the $l=2$ polarization is circular with no sign change in the face-on configuration. The observed polarization is more complex in the edge-on case, due to the presence of the linearly polarized $m=0$ mode at early times, see Fig.~\ref{fig:GW_pol_2}.

\begin{figure*}
    \centering
    \includegraphics[width=1\textwidth]{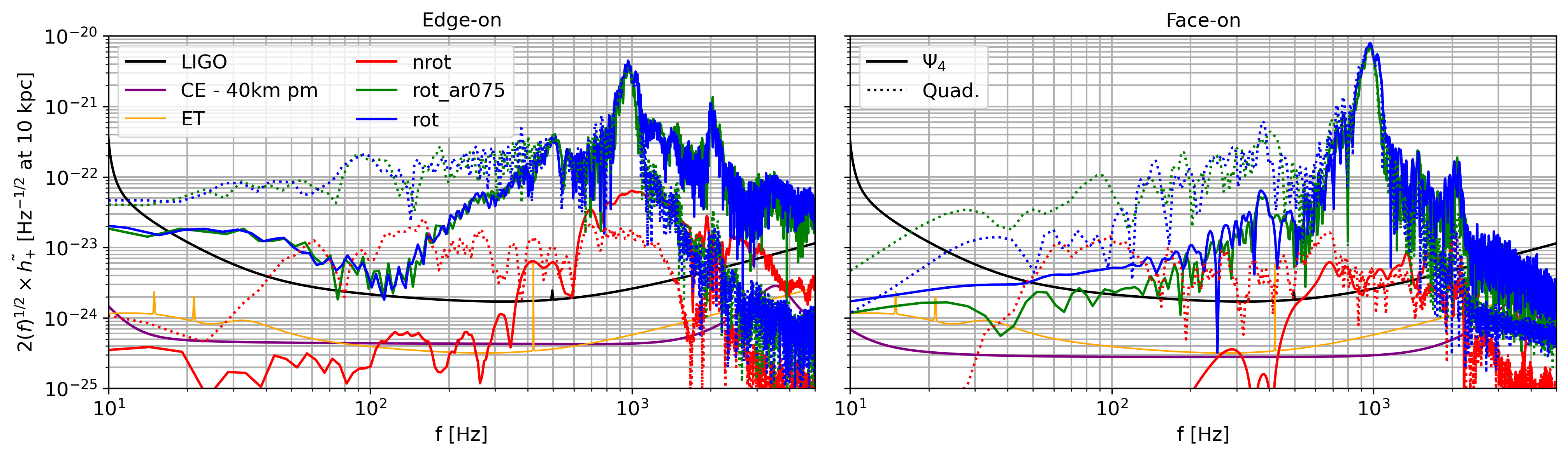}
    \caption{Amplitude spectral density for our models. Here we compare the strain ($l=2$ mode) as obtained via the quadrupole approximation and the waveform reconstruction from $\Psi_{4}$. We show the result for the cases of face-on and edge-on orientation of the source. We compare the signal with the design sensitivity of LIGO, a post-merger (pm) optimized 40 km CE, and ET. The associated SNR for each case is shown in Table \ref{tab:SNR_m1_m2}. Comparing our results to those of CCSNe, we see that the overall amplitude of our non-rotating model is similar to non-rotating CCSNe. On the other hand, initial angular momentum seems to amplify the signals by two orders of magnitude.}
    \label{fig:GWFspace}
\end{figure*}

In order to assess the detectability of GWs emitted by AIC events, we plot in Fig.~\ref{fig:GWFspace} our simulated signals at a fiducial distance of 10 kpc together with the sensitivity curves of LIGO and future GW detectors Cosmic Explorer (CE; \citealt{CE2}) and Einstein Telescope (ET; \citealt{Punturo_2010}). All noise curves were retrieved from \citet{CE_noise}. We define the signal to noise ratio (SNR) as
\begin{equation}
    \left(\dfrac{S}{N}\right) \equiv \bigintsss_{0}^{\infty} df \dfrac{|\tilde{h}(f)|^{2}}{S_{n}(f)},
\end{equation}
where $\tilde{h}(f)$ and $S_{n}(f)$ are respectively the GW strain in the frequency domain and the single-sided noise spectral density. If we require a single-detector SNR of 8 for detectability, then the GW detection horizon for the most optimistic scenario, i.e. a maximally rotating AIC observed face-on, ranges from $\sim$1.2 Mpc (for LIGO) to $\sim$10 Mpc (for a 40 km CE). The \texttt{nrot} model would only be detectable by CE with ${\rm SNR}= 8$ at a maximum distance of $\approx$ 19 kpc. See Table \ref{tab:SNR_m1_m2} for more details. 

The rate of AICs in the galaxy is still very uncertain. In order to estimate the AIC detection rate by future GW detectors, we assume the AIC rate to be of the order of 10\% of the SNIa rate, which is estimated as approximately $3\times 10^{-4} \rm{yr}^{-1} \rm{Mpc}^{-3}$ \citep{SNR_rates}. Our detection horizons for a maximally rotating AIC (best case scenario), yield detection rates of approximately $(x/10\%)\times 0.14\,{\rm yr}^{-1}$ (for CE) and $(x/10\%)\times 2.2\times 10^{-4}\,{\rm yr}^{-1}$ (for LIGO), where $x$ is the (still highly uncertain) true ratio between AIC and SNIa event rates in percentage form.

Our estimated rates are consistent with those made by \citet{Fryer_1999} using nucleosynthesis arguments, which imply an upper limit of 0.08 yr$^{-1}$ within our estimated GW detection horizon for CE. This upper limit, obtained by attributing the formation of all $r$-process elements to AICs, may need to be revised after more 3D simulations become available. For example, our simulations show that the ejecta material has higher $Y_e$ than reported by \citet{Fryer_1999}, which would push their reported upper limits for the AIC rates to higher values.

The numbers in Table \ref{tab:SNR_m1_m2} implicitly assume that the signals were identified through a match-filtering search with a large waveform template bank for these events. In reality, the first detection of an AIC will likely be achieved by a burst analysis \citep{klimenko_sergey_2021_4419902}, which is less sensitive than a match-filtering search. In this case, we can expect a decrease by a factor of order unity in the detection horizon \citep{PycbcCWBcomparison}, nevertheless, the resulting CE observation volume would still include (at least) the Local Group of galaxies representing an improvement compared to LIGO.

\section{ Neutrino emission}\label{chem}

\subsection{Electron capture}

\begin{figure}

\begin{subfigure}{0.5\textwidth}
\includegraphics[width=\textwidth]{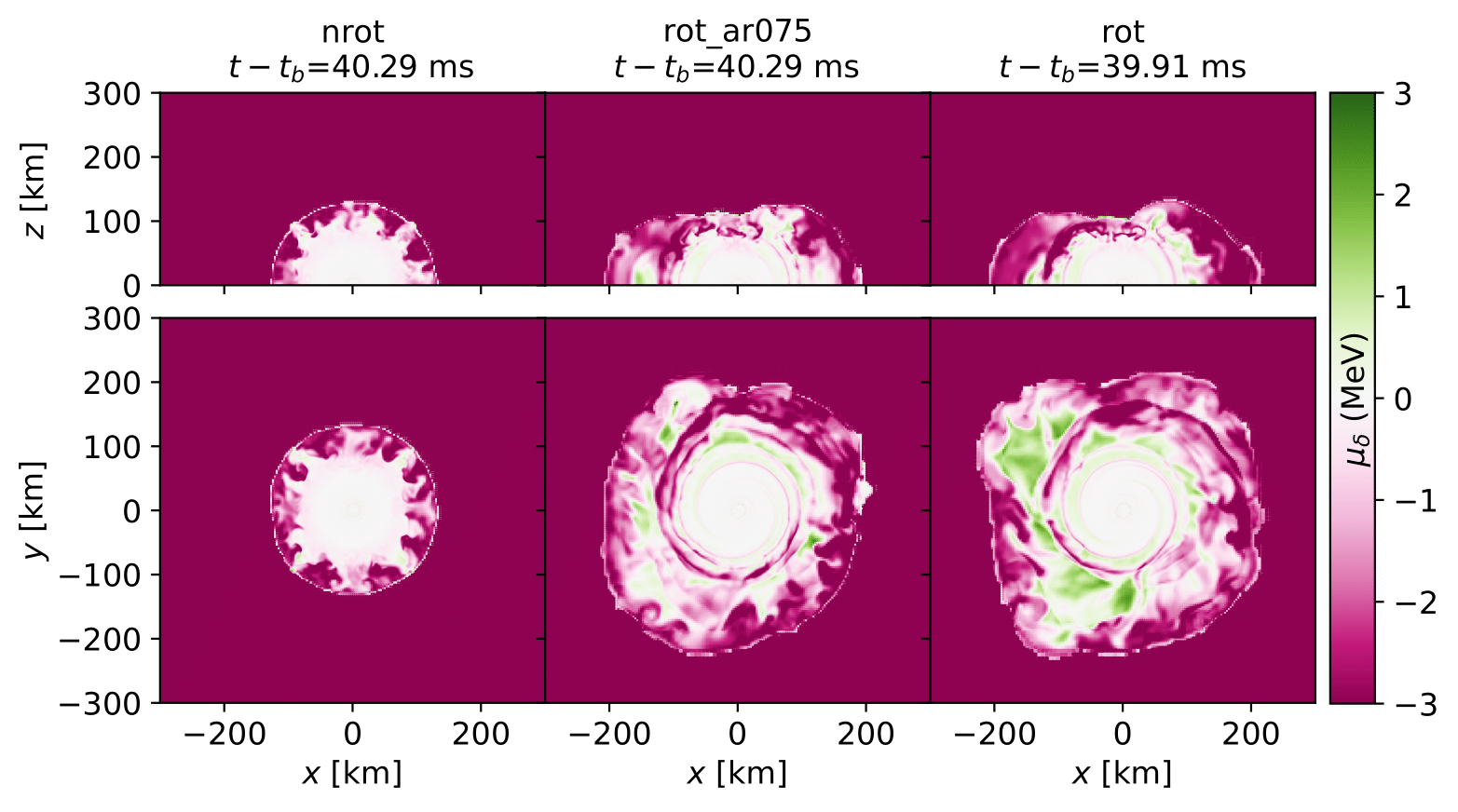}
\end{subfigure}
\begin{subfigure}{0.5\textwidth}
\includegraphics[width=\textwidth]{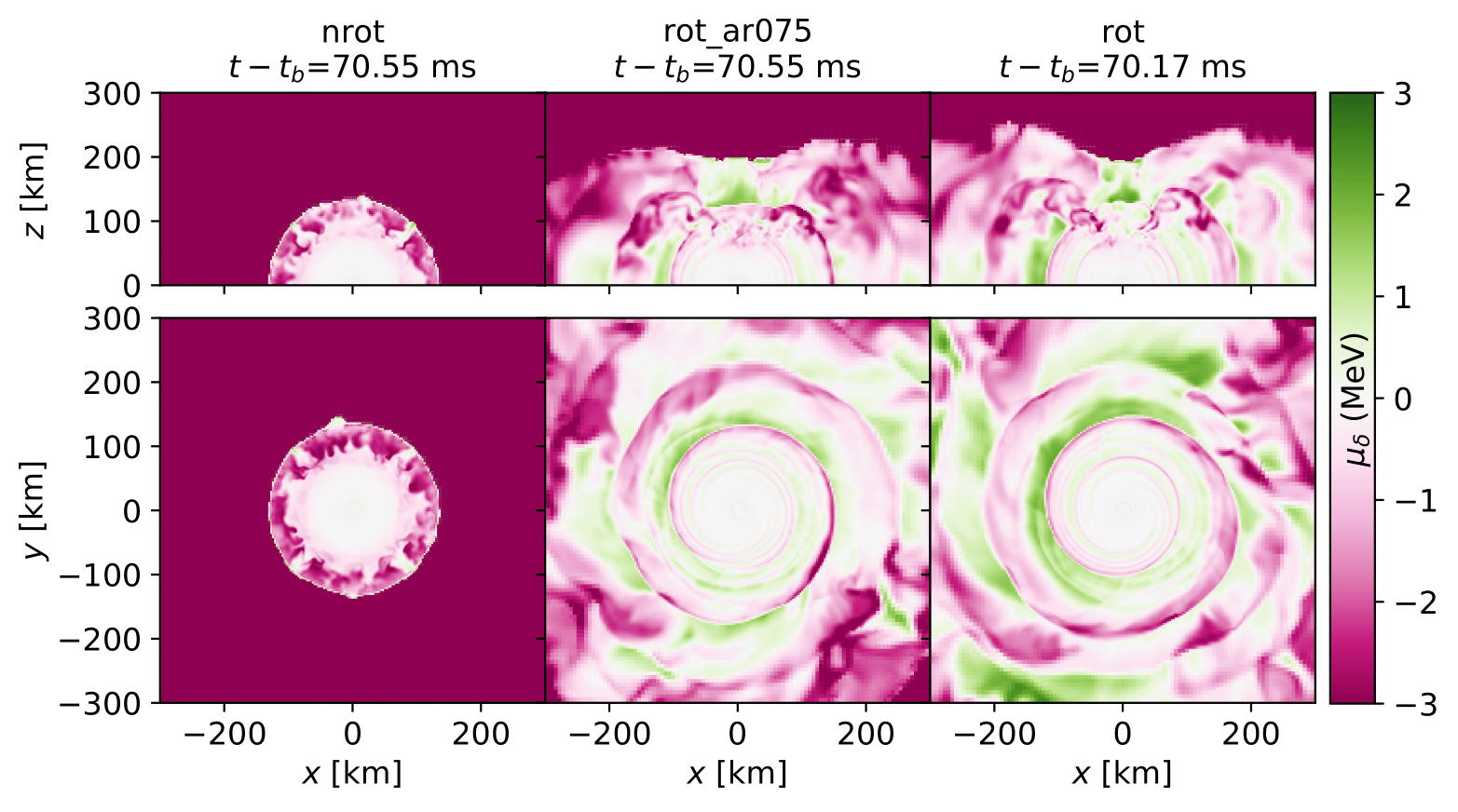}
\end{subfigure}
\begin{subfigure}{0.5\textwidth}
\includegraphics[width=\textwidth]{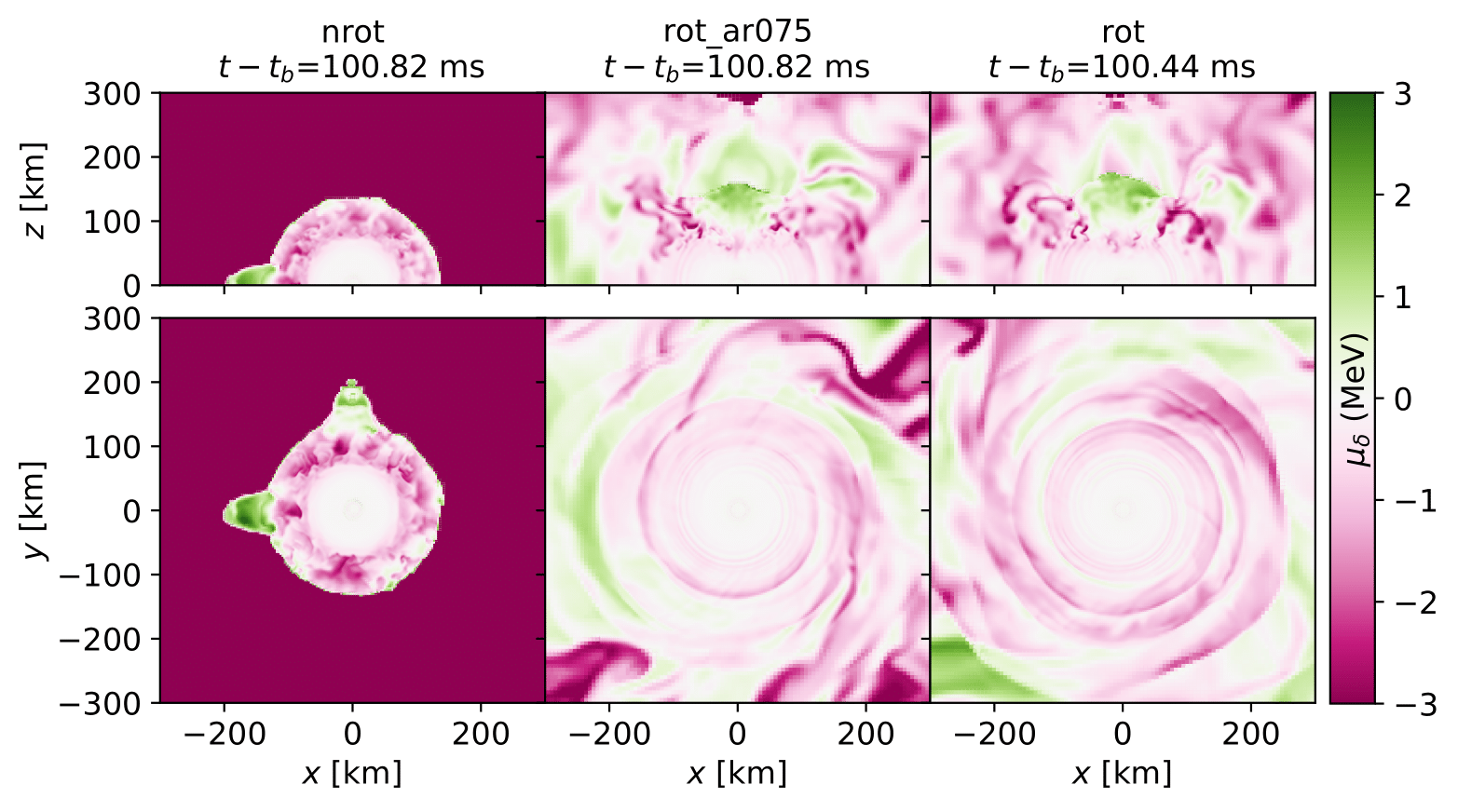}
\end{subfigure}
\begin{subfigure}{0.5\textwidth}
\includegraphics[width=\textwidth]{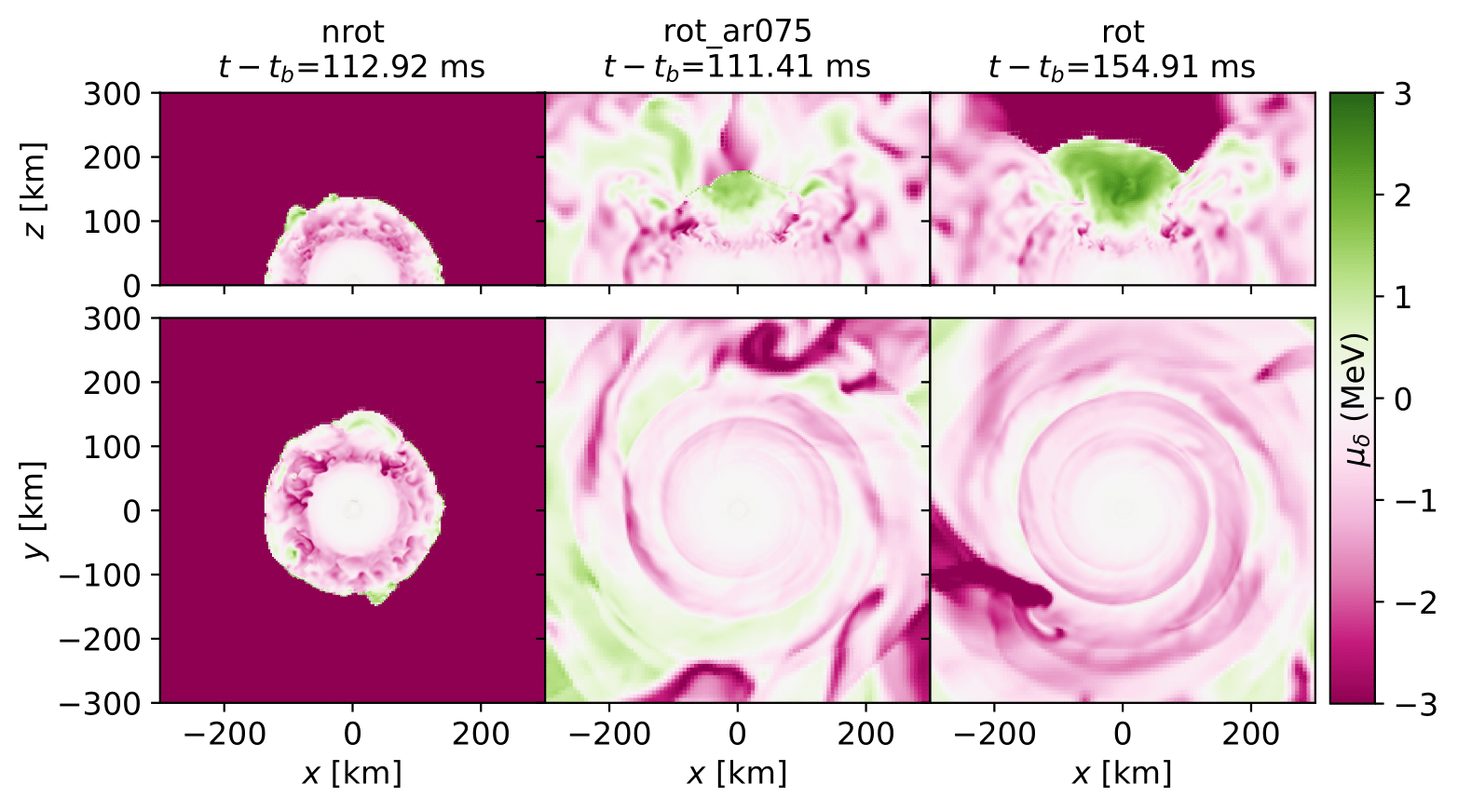}
\end{subfigure}

\caption{Chemical potential imbalance, defined in eq. (\ref{mudelta}). for the inverse $\beta$-decay process. Negative values indicate a tendency to neutronize the material, by emitting neutrinos via inverse $\beta$-decay, see eq. (\ref{betadecay}). The polar regions with positive $\mu_{\delta}$ are more affected by the neutrino wind and can be related to the high entropy bubble being ejected near the pole (see Fig.~\ref{fig:entropy})}

\label{fig:muDelta}

\end{figure}

During the collapse, the increasing density drives up the electron Fermi energy and consequently leads to electron capture inside the WD: 
\begin{equation}\label{betadecay}
    (A,Z) + e^{-} \longrightarrow (A,Z-1) + \nu_{e}\,,
\end{equation}
where $(A,Z)$ is a nucleus of atomic number $Z$ and atomic mass $A$. In our simulations, the nucleon number densities are computed via the prescription for $Y_{e}(\rho)$ described by \citet{Liebendorfer2005} and no neutrino emission is included for $t<t_{b}$. After the bounce, we treat neutrinos using the M1 grey scheme of \citet{M1}. It is useful to define the chemical potential imbalance \citep{Hammond_nu} as
\begin{equation}\label{mudelta}
    \mu_{\delta} \equiv \mu_{n} +\mu_{\nu_{e}} -\mu_{p}-\mu_{e}
\end{equation}
where $\mu_{n}$, $\mu_{p}$, $\mu_{e}$, and $\mu_{\nu_{e}}$ are the chemical potentials of neutrons, protons, electrons, and neutrinos, respectively. The neutrino chemical potential is implicitly defined by   
\begin{equation}
    Y_{\nu_{i}} \equiv  \dfrac{n_{\nu_{i}}}{n_{p}+n_{n} } = \dfrac{4\pi m_{b}}{\rho \left(hc\right)^{3}}\left(k_{B}T\right)^{3}F_{2}\left(\dfrac{\mu_{\nu_{e}}}{k_{B}T}\right),
\end{equation}
where $ Y_{\nu_{i}}$, $n_{\nu_{i}}$, $T$, and $m_{b}$ are the neutrino fraction and the neutrino number density, the temperature, and the baryon effective mass, respectively,  and $F_{2}$ is the Fermi integral of the second order, which can be obtained for cases of physical interest by 

\begin{equation}
F_{k}\left(x\right) \equiv  \bigintss_{0}^{\infty}\dfrac{y^{k} dy}{e^{y-x}+1} =  -\Gamma\left(1 + k \right) {\rm Li}_{k+1}(- e^{x} )\,,
\quad \rm{Re}\left(k\right) > -1\,,
\end{equation}
where $\rm{Li}_{k}$ is the polylogarithm function of the $k$-th order \citep{Perego2019}. The chemical potential imbalance in our models can be seen in Fig.~\ref{fig:muDelta}. Comparing Fig.~\ref{fig:muDelta} and Fig.~\ref{fig:vel}, we can see that the infalling material has $\mu_{\delta}<0$. This implies that the electron capture process is preferred over neutrino absorption. On the other hand, the central part of the star has $\mu_{\delta}\approx 0$, implying a weak equilibrium of the fluid.

\subsection{Neutrino emission}\label{sec:Neutrinos}

\begin{figure}
\begin{subfigure}{0.5\textwidth}
\includegraphics[width=\textwidth]{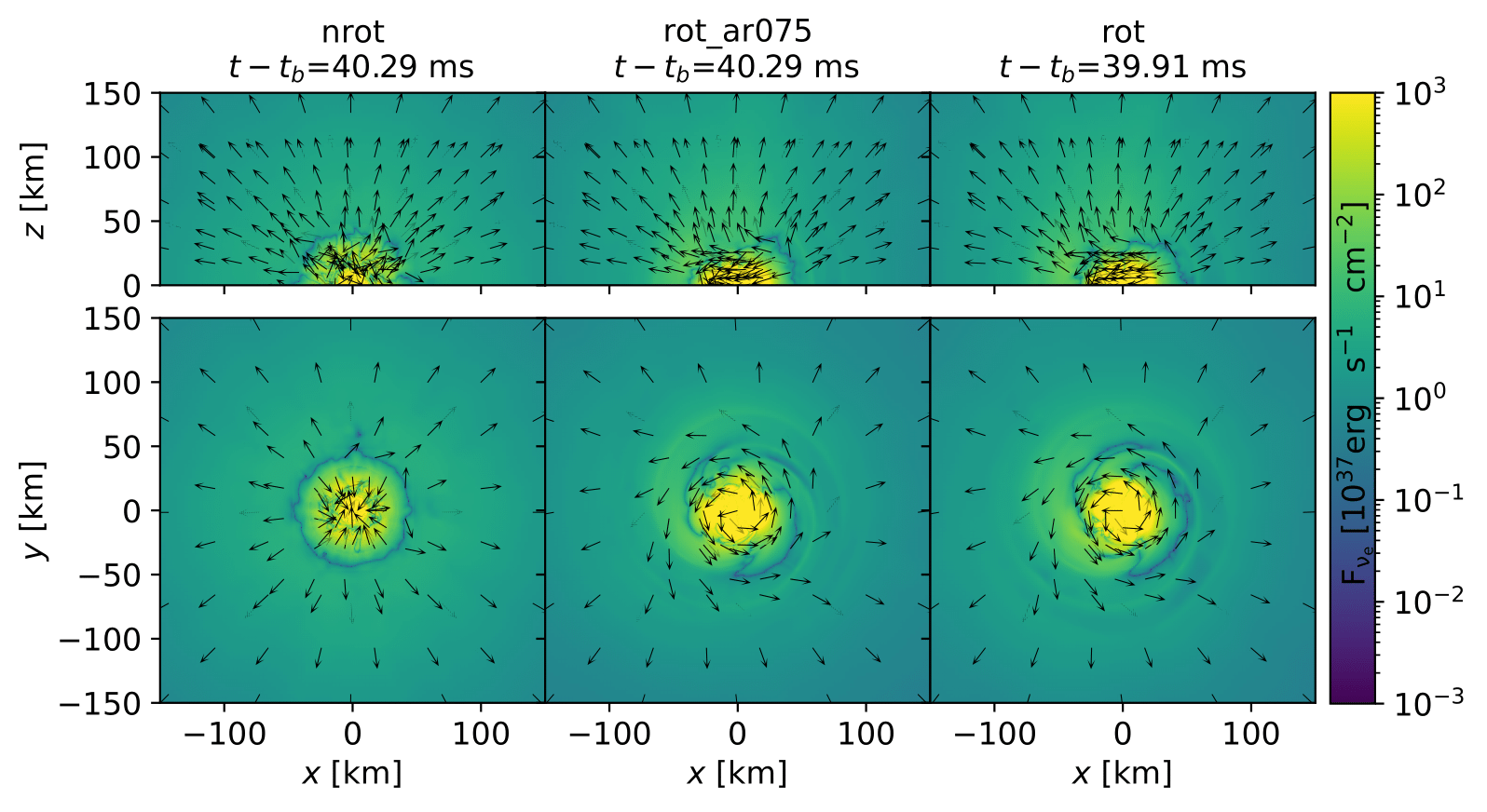}
\end{subfigure}
\begin{subfigure}{0.5\textwidth}
\includegraphics[width=\textwidth]{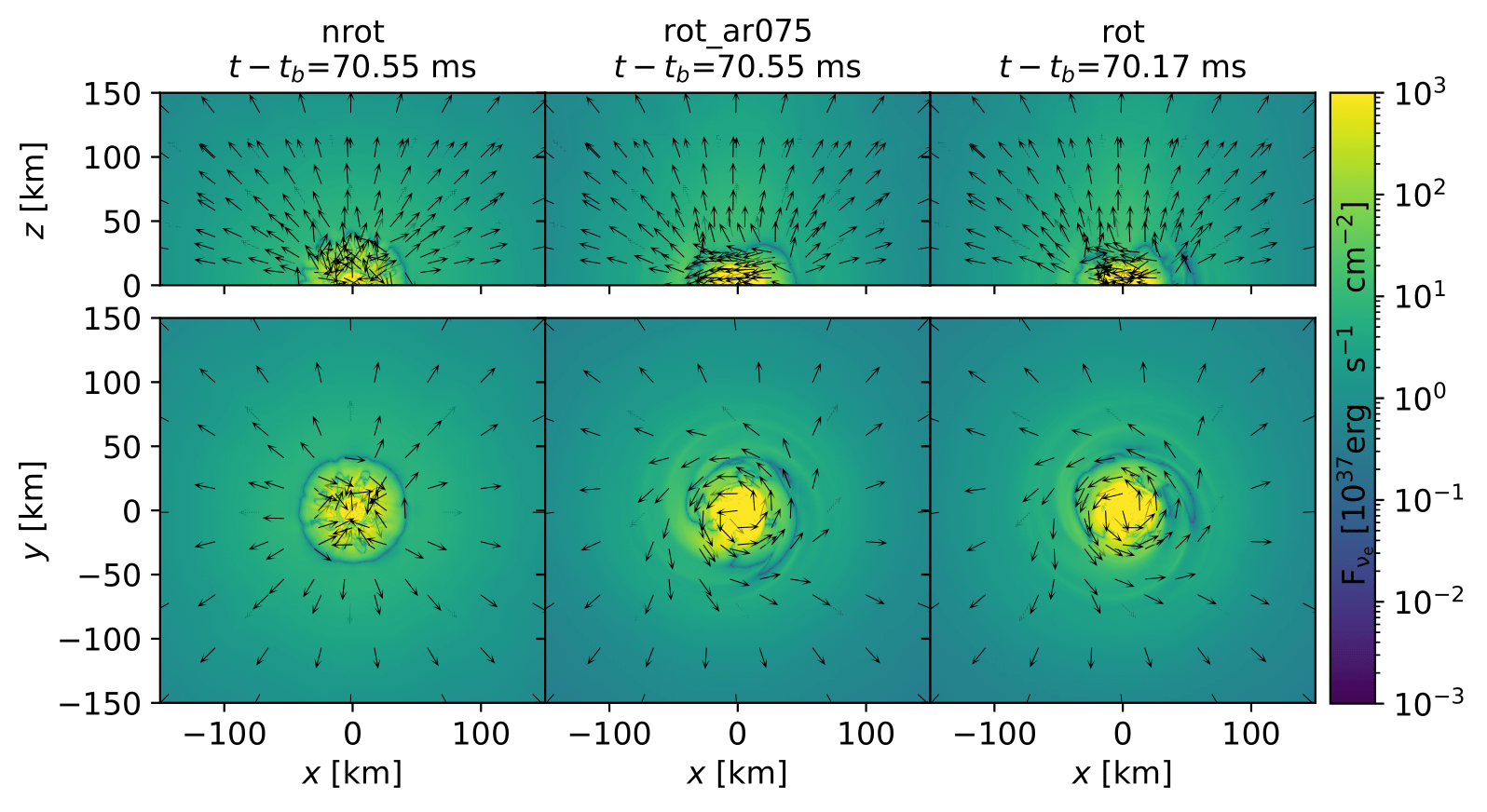}
\end{subfigure}
\begin{subfigure}{0.5\textwidth}
\includegraphics[width=\textwidth]{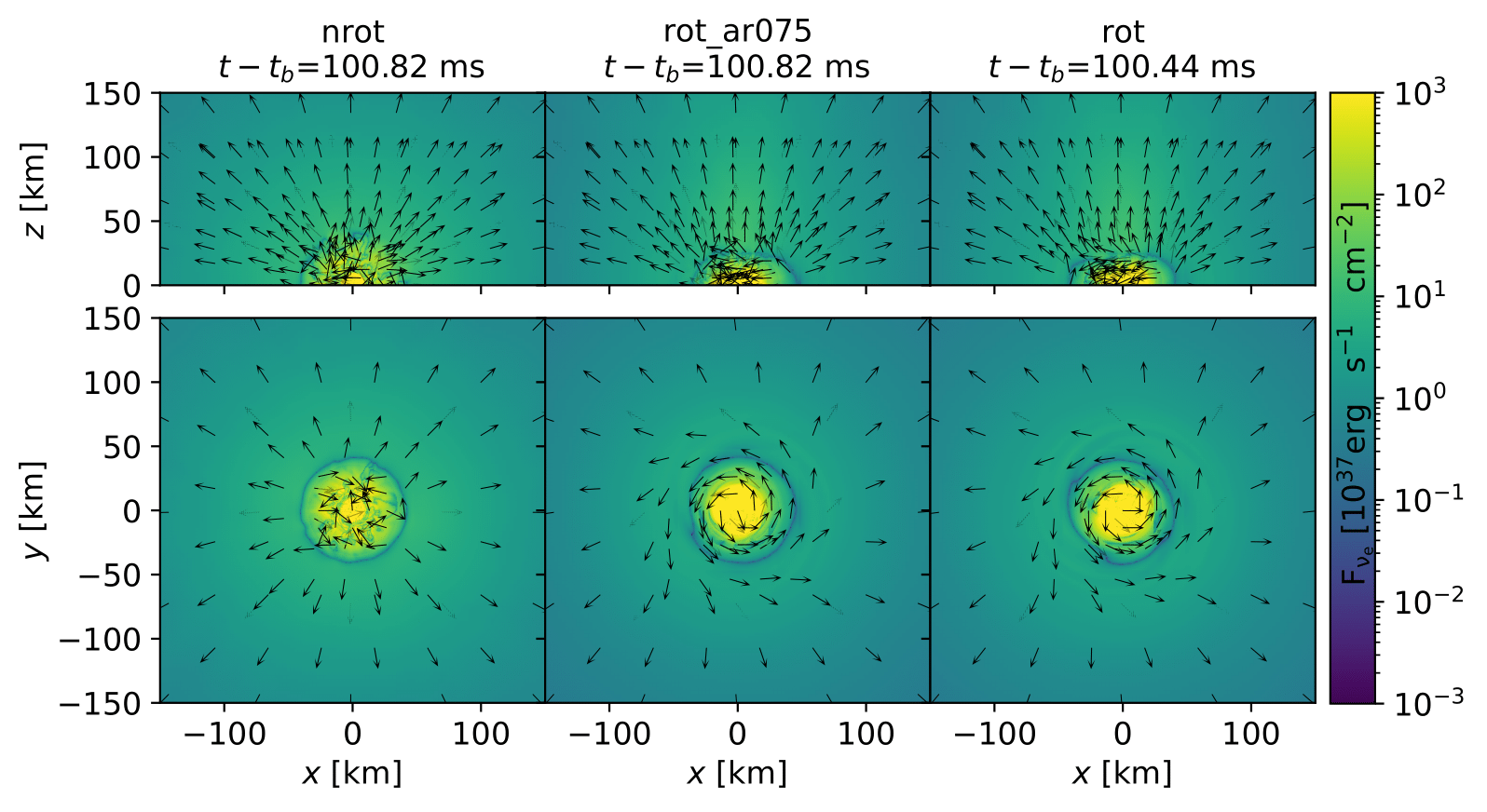}
\end{subfigure}
\begin{subfigure}{0.5\textwidth}
\includegraphics[width=\textwidth]{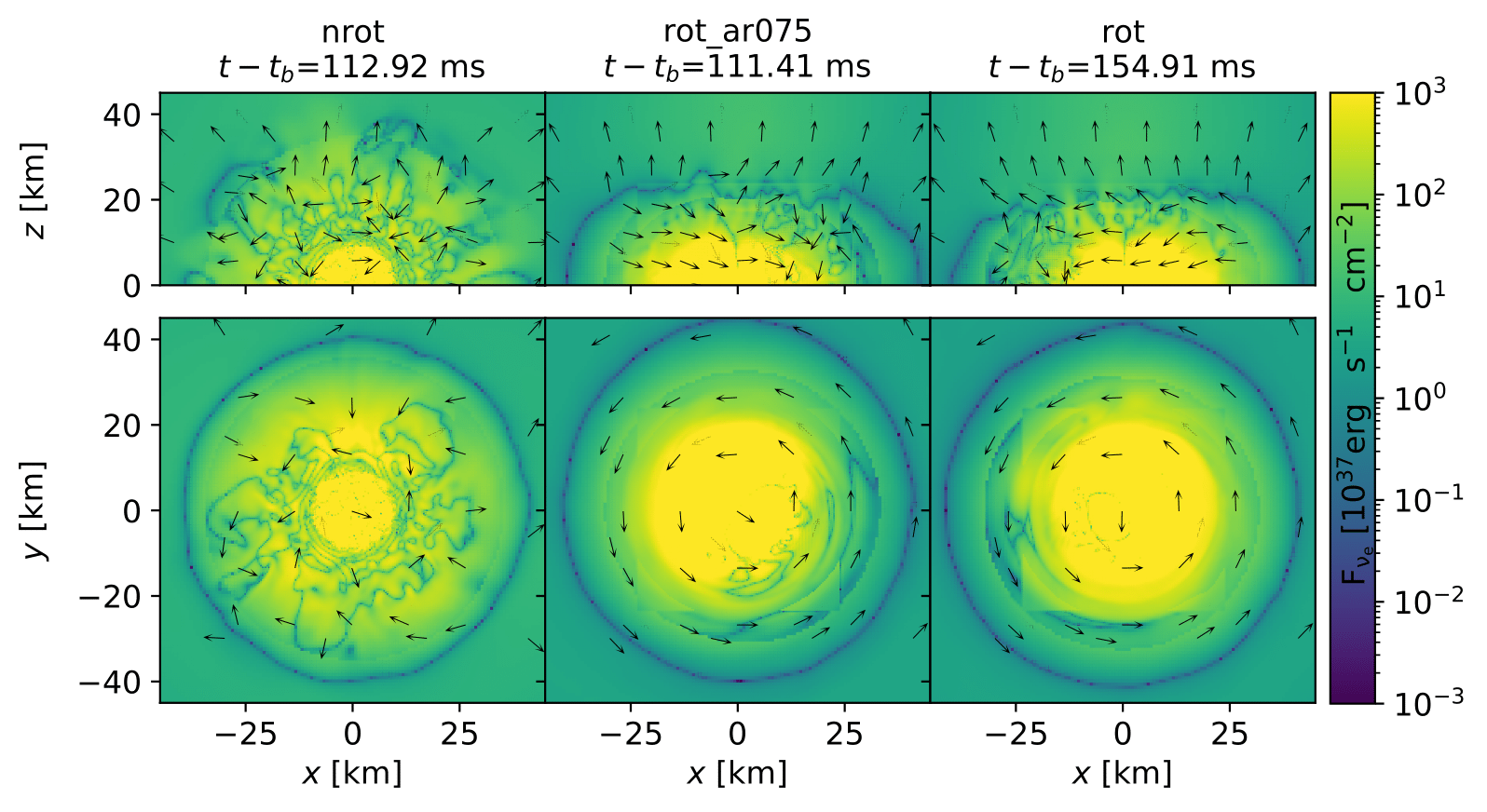}
\end{subfigure}

\caption{Energy flux of $\nu_{e}$. The color scheme represents the intensity of the radial flux component. Arrows represent the local direction of the energy flux. The bottom panel is zoomed-in to show smaller scale details: there is a clear radially outgoing flux at large radii, but not in the core. This fact can be seen more easily in the xy (yz) plane for \texttt{nrot} (\texttt{rot\_ar075} and \texttt{rot}). The surface of zero radial energy flux appears as a blue ring in the equatorial plane and it is related to the surface of thermal decoupling. Note the effect of rotation on this surface: it appears at larger radii for the rotating progenitors. The emission becomes more axisymmetric with rotation.}
\label{fig:Fnue}
\end{figure}

\begin{figure}

\begin{subfigure}{0.5\textwidth}
\includegraphics[width=\textwidth]{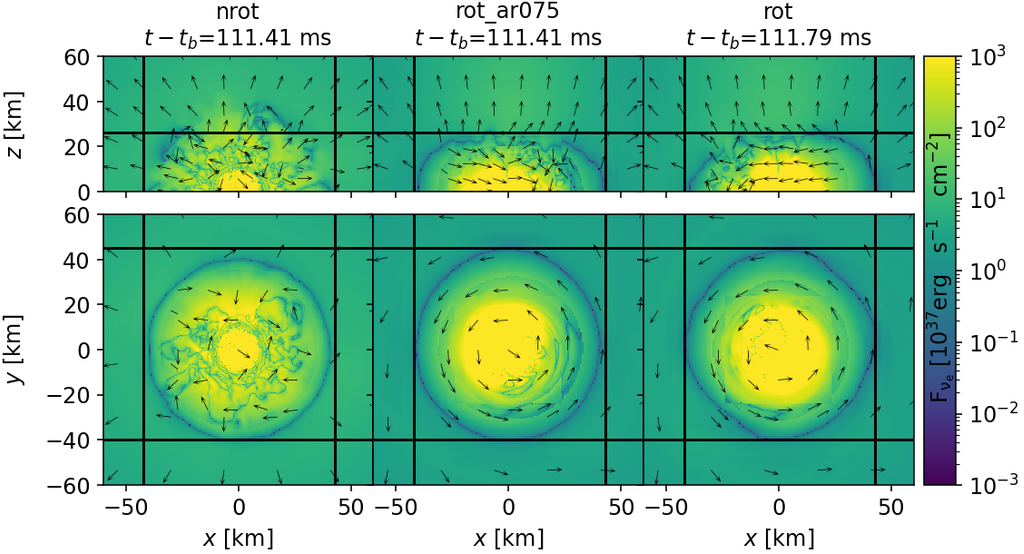}
\end{subfigure}
\begin{subfigure}{0.5\textwidth}
\includegraphics[width=\textwidth]{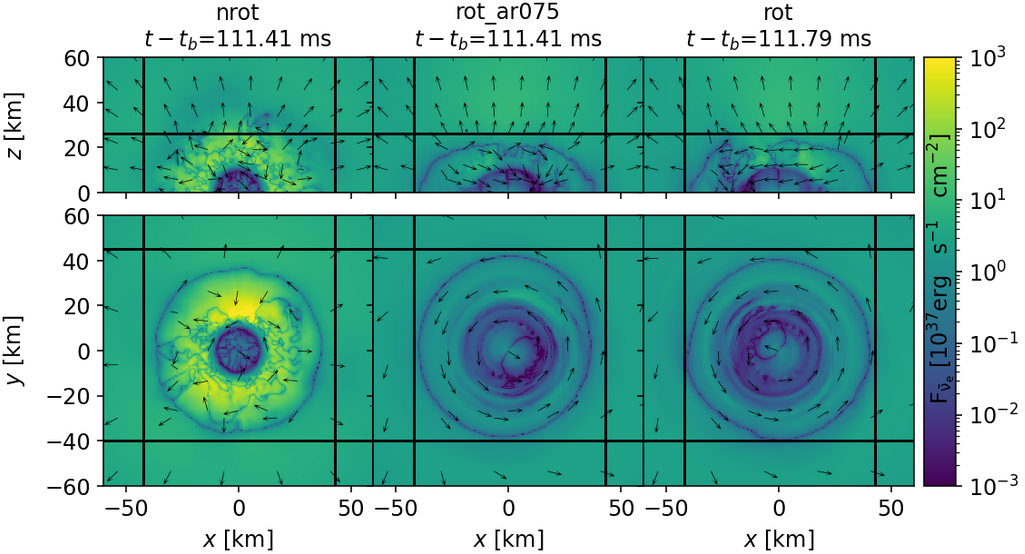}
\end{subfigure}
\begin{subfigure}{0.5\textwidth}
\includegraphics[width=\textwidth]{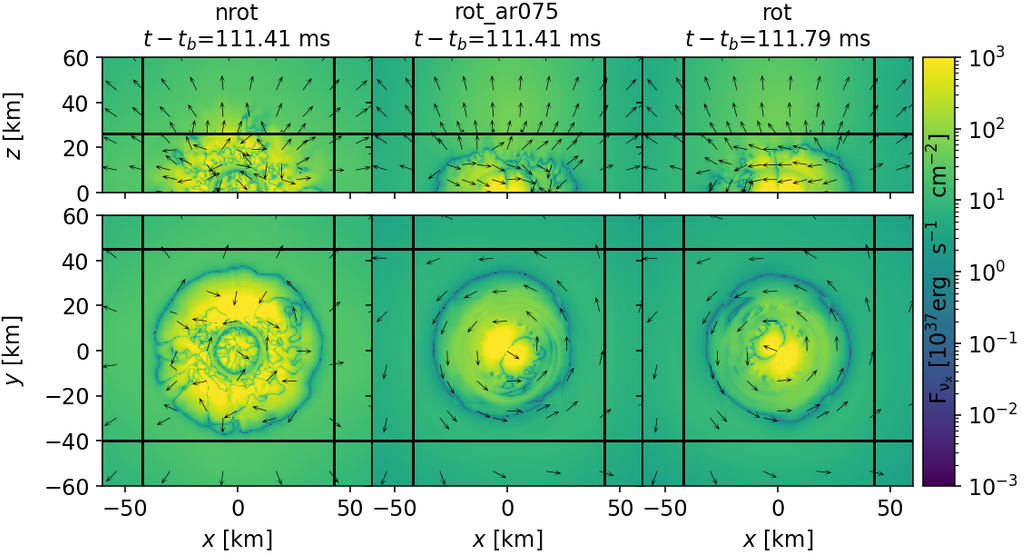}
\end{subfigure}

\caption{Energy flux of neutrinos (from top to bottom: $\nu_{e}$, $\bar{\nu}_{e}$, and $\nu_{x}$ representing heavy lepton neutrinos), similar to the bottom panel of Fig.~\ref{fig:Fnue}. The black lines represent the size of the zero radial energy flux for $\nu_{e}$ in the \texttt{rot} model, and are included to facilitate a visual size comparison between different neutrino flavors and models. The radius of the surface of zero radial energy flux is different for each neutrino flavor (neutrinos can stream outwards freely when outside the largest surface with zero flux, if more than one is present, as in the $\bar{\nu}_{e}$ case). This has a visible imprint in the average neutrino energy, see Fig.~\ref{fig:Enu}. As the inner core of the star is hotter, neutrinos emitted from more central regions will be more energetic. }
\label{fig:Fnufinal}
\end{figure}

The temporal evolution of the $\nu_{e}$ energy flux ($F_{\nu_{e}}$) is depicted in Fig.~\ref{fig:Fnue}. For $r\gtrsim40$ km $F_{\nu_{e}}$ has a visible outgoing radial component that becomes more dominant at larger distances from the center of the star. In the last snapshot of Fig.~\ref{fig:Fnue} (bottom panel), it is possible to see a blue surface around the core of the remnant PNS. This surface separates the exterior region where the radial component of the energy flux is positive (outgoing) from the interior region presenting an energy flux with an overall negative (ingoing) radial component, as indicated by the direction of the arrows. The existence of a central region from which the neutrinos are not escaping implies the existence of trapped neutrinos. In practice, this means that the intense flux in the center will not be released immediately. It will take some time ($\mathcal{O}(1$  $\rm{min})$; \citealt{Roberts2017}) for the PNS to cool down in order to become transparent to these neutrinos.

Similar to the bottom panel of Fig.~\ref{fig:Fnue}, Fig.~\ref{fig:Fnufinal} depicts the neutrino energy flux in the central part of the remnant PNS for different neutrino flavors. In this figure, we insert black lines (fixed at the surface of zero radial energy flux of $\nu_{e}$ in the \texttt{rot} model) to aid a visual size comparison between the different models. From this figure, we can see that the different neutrino flavors start to stream outwards at different radii. The surface of zero radial energy flux is largest for $\nu_{e}$, followed by $\bar{\nu}_{e}$, and $\nu_{x}$. As we argue in the next few paragraphs, the surface of zero radial energy flux is related to the surface of thermal decoupling, and its location leaves imprints on the average neutrino energy.

\begin{figure*}
    \centering
    \includegraphics[width=\textwidth]{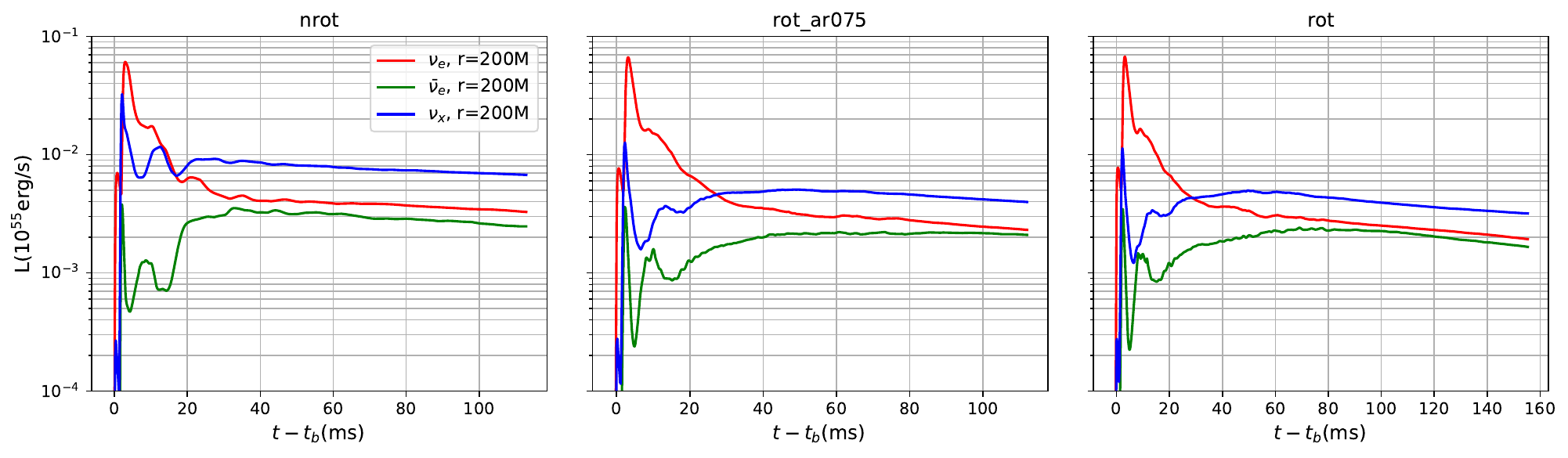}
    \caption{Neutrinos luminosities. Electron neutrinos show an intense prompt emission followed by a sharp decay. Heavy lepton neutrinos dominate the late-time emission. Models with rotating progenitors have lower neutrino emission, see also Fig.~\ref{fig:Enu}.}
    \label{fig:Lnu}
\end{figure*}

\begin{figure}
\begin{subfigure}{0.5\textwidth}
\includegraphics[width=\textwidth]{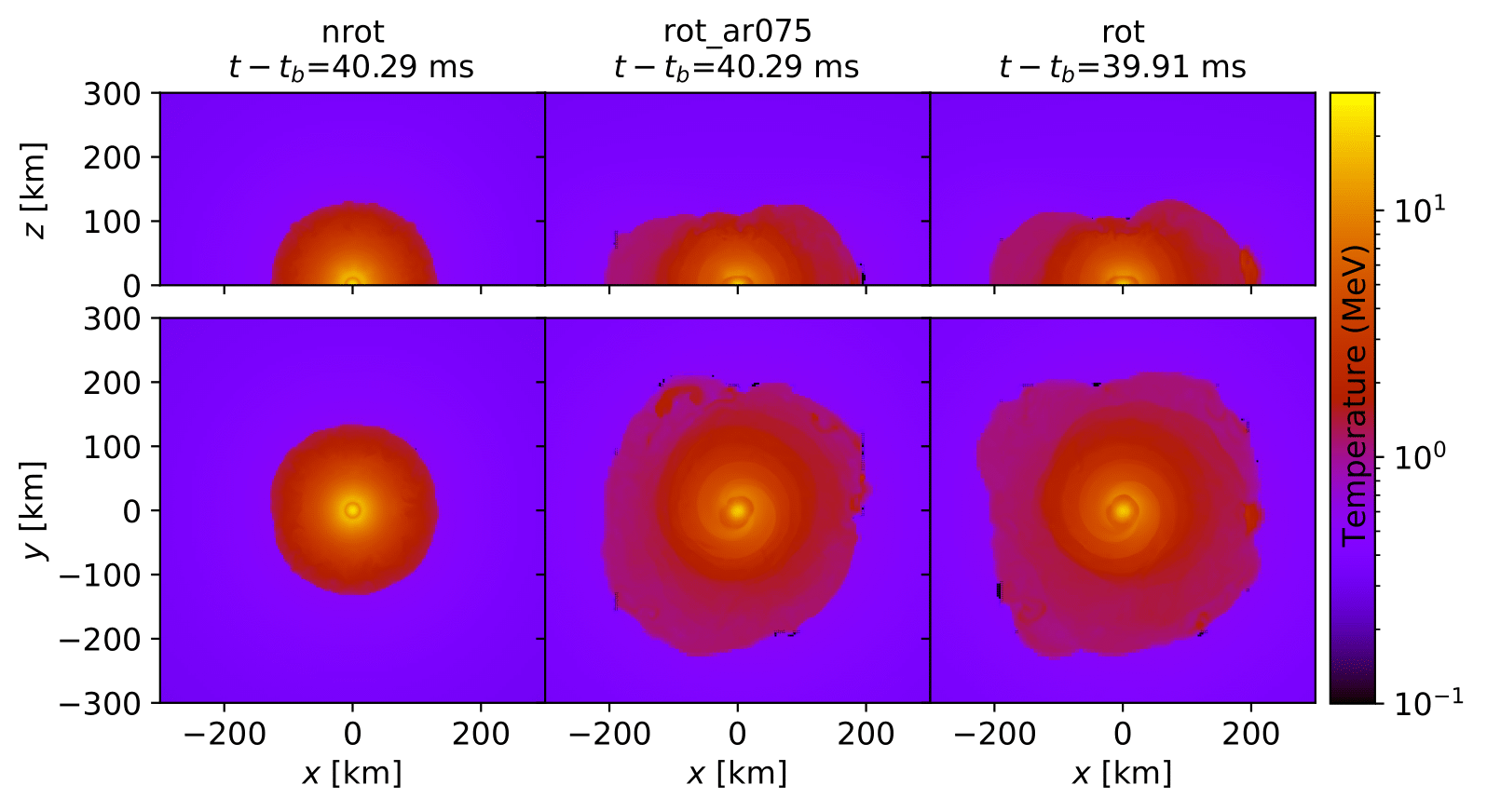}
\end{subfigure}
\begin{subfigure}{0.5\textwidth}
\includegraphics[width=\textwidth]{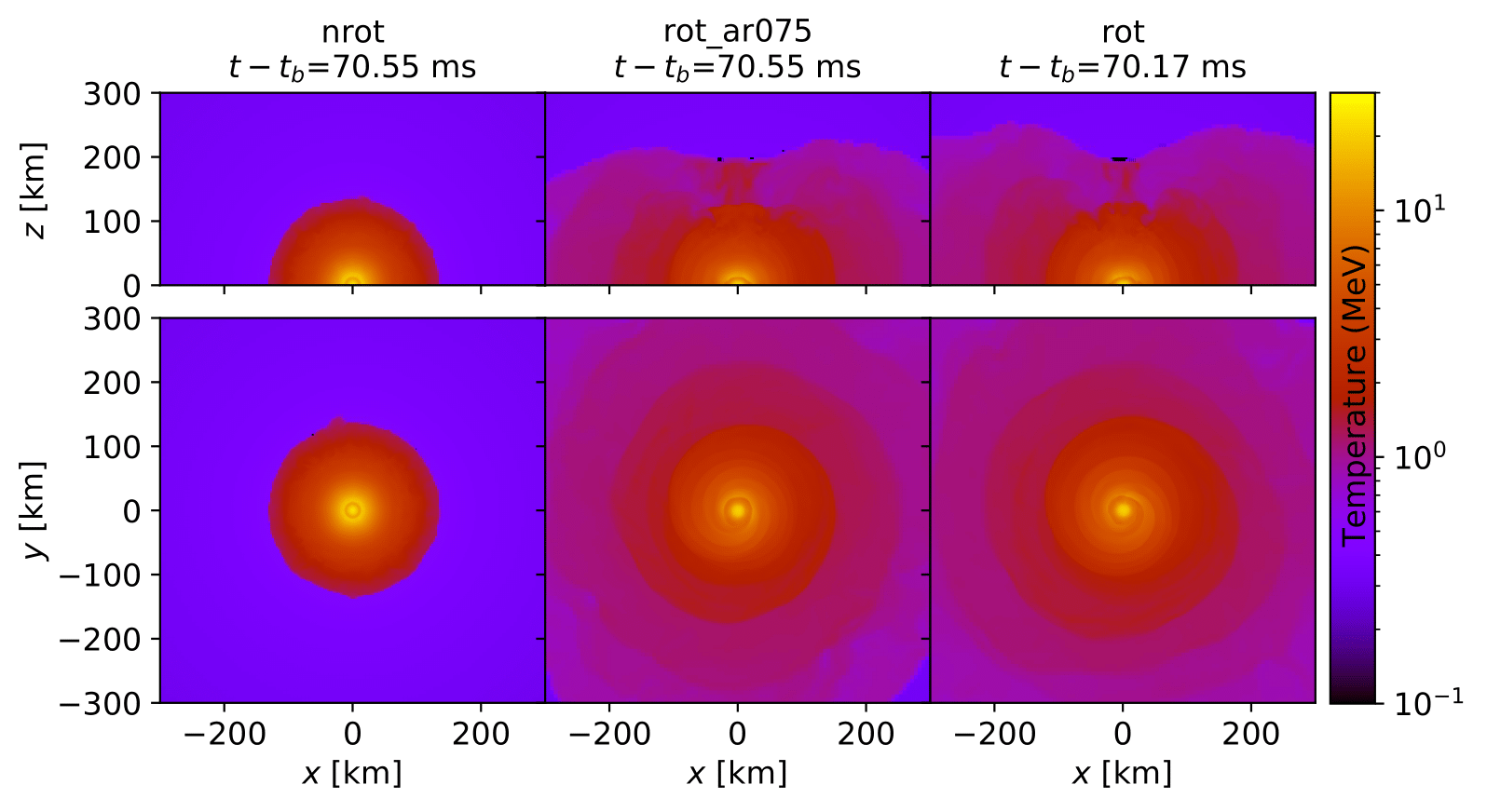}
\end{subfigure}
\begin{subfigure}{0.5\textwidth}
\includegraphics[width=\textwidth]{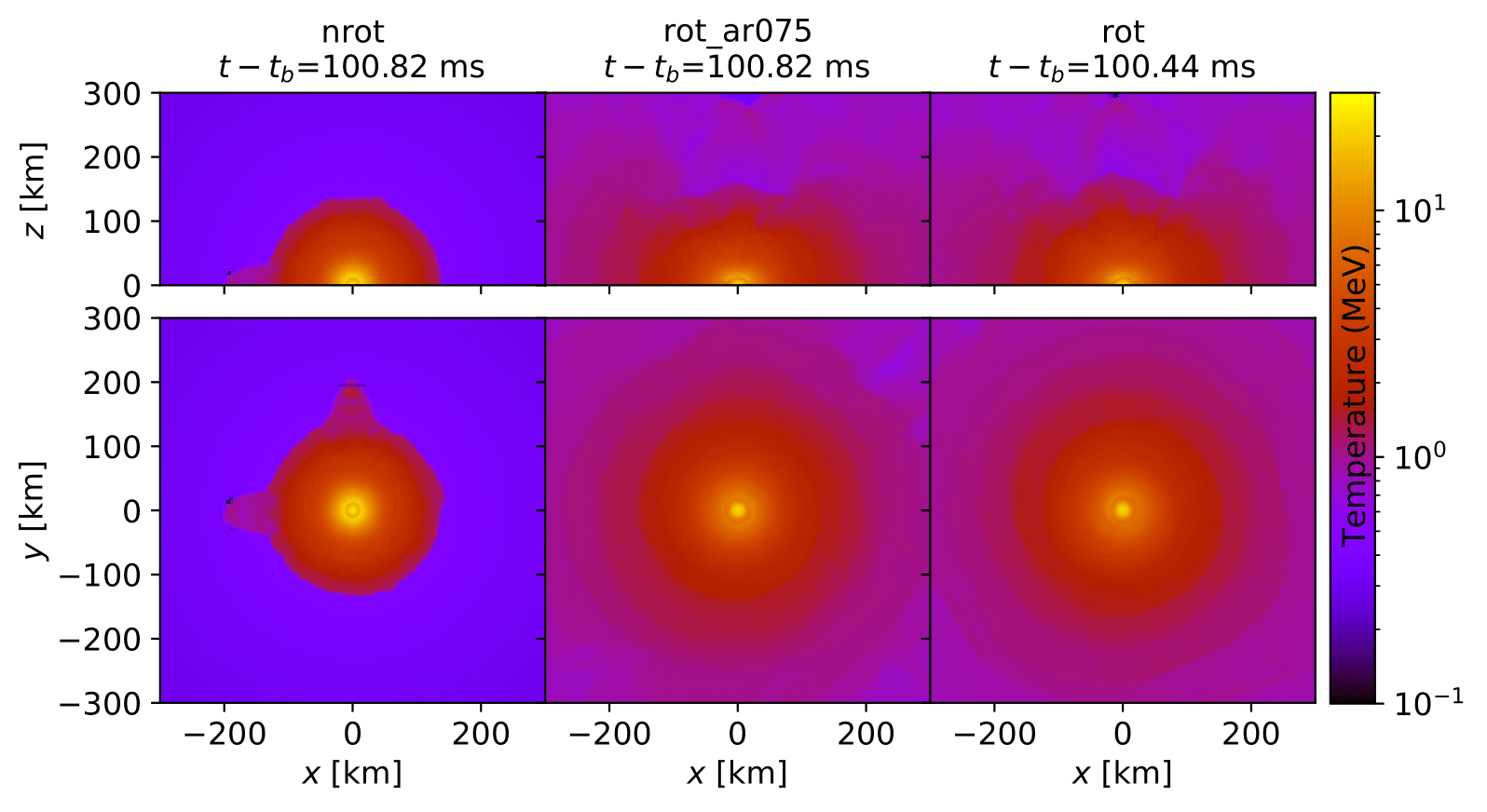}
\end{subfigure}
\begin{subfigure}{0.5\textwidth}
\includegraphics[width=\textwidth]{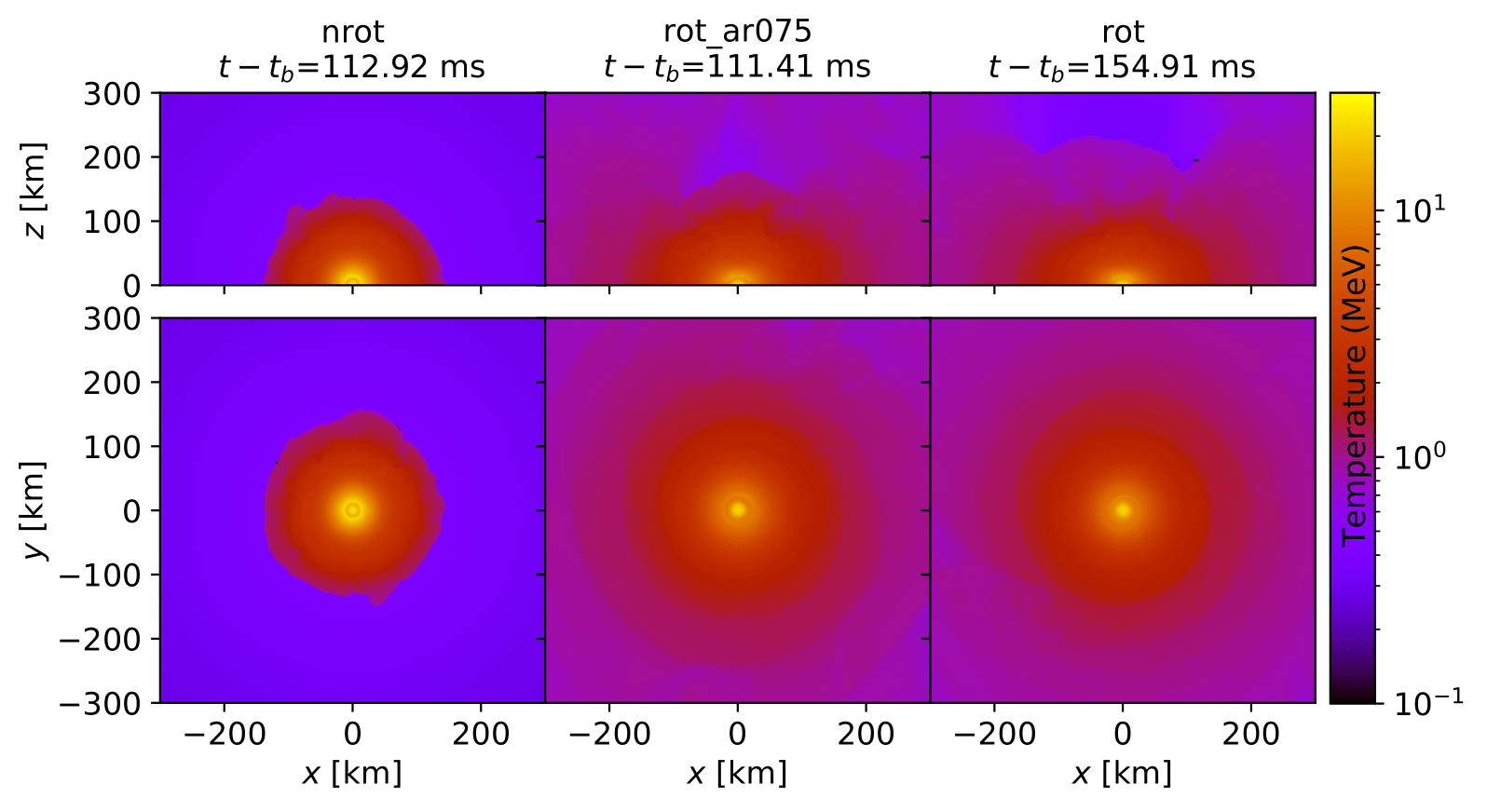}
\end{subfigure}
\caption{Temperature profiles for all our three models. Temperature is an overall decreasing function of the radius. The high-temperature material reaches larger radii due to rotation and there is a low-temperature bubble extending from the poles of the \texttt{rot} and \texttt{rot\_ar075} models.}
\label{fig:temperature}
\end{figure}

To compute the neutrino fluxes as seen by a distant observer, we performed the surface integral of the proper flux defined as $\mathscr{F}^{i}_{\nu} \equiv \sqrt{\gamma}\left(\alpha F^{i}_{\nu} - \beta^{i} E_{\nu} \right)$, where $F^{i}_{\nu}$ and $E_{\nu}$ are the neutrino flux and  energy in the Eulerian frame, $\alpha$ is the lapse function and $\beta^{i}$ is the shift vector. Accordingly, the neutrino luminosity is given by

\begin{equation}\label{eq:Lnu}
    L_{\nu} = \int_{S_{r}} \hat{r}_{i}\, \mathscr{F}^{i}_{\nu} \, dS,
\end{equation}
where $\hat{r}_{i}$ is the unit covector normal to a coordinate sphere $S$ of radius $r$. The neutrino luminosities we obtain in our simulations are summarized in Fig.~\ref{fig:Lnu}. There is a large peak in the luminosity near the time of bounce, which is present in all neutrino flavors for our three models. The total luminosity of each neutrino flavor varies over time from $\sim$$10^{51}$ to $\sim$$10^{54}$ erg/s. At early times $t-t_{b}\lesssim 20$ ms, we have  $L_{\bar{\nu}_{e}}<L_{\nu_{x}}<L_{\nu_{e}}$, but at later times the electron neutrino luminosity drops below the heavy lepton neutrino luminosity, while the electron antineutrino emission remains subdominant, i.e. $L_{\bar{\nu}_{e}}<L_{\nu_{e}}<L_{\nu_{x}} $. This sharp decrease in $L_{\nu_{e}}$ is related to the behavior of $Y_{e}$ after the bounce. Although not shown in Fig.~\ref{fig:Ye}, during the collapse phase we have $Y_{e} \gtrsim 0.3$. Shortly after the bounce ($t-t_{b}\approx 7$ ms) the PNS reaches $Y_{e} \lesssim 0.2$ for the first time in a thin ring (in the xy plane) that expands for the next tens of ms. The further decrease in $Y_{e}$ naturally suppresses the inverse $\beta$-decay as there will be less electrons available to be captured. After $t-t_{b}\approx 20$ ms, the low-$Y_{e}(\lesssim0.2)$ region remains within a radius of $r\approx 100$ km and $Y_e$ evolves more slowly, consistently with the slow evolution of $L_{\nu_{e}}$ at late times. Increasing the rotation of the WD progenitor suppresses the $\nu$-emission of all flavors because the centrifugal forces pushes the material to larger radii resulting in lower temperatures, see Fig.~\ref{fig:temperature}.

\begin{figure*}
    \includegraphics[width=\textwidth]{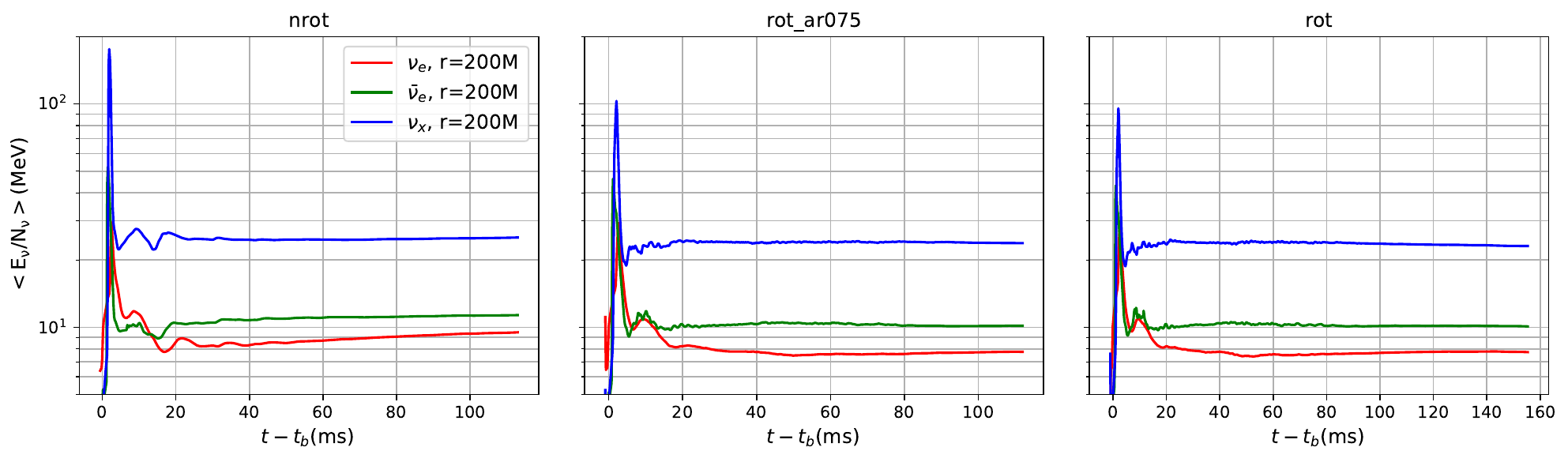}
    \caption{Average neutrino energy. Heavy lepton neutrinos are the most energetic neutrinos at all times; electron neutrinos are the least energetic. This reflects the temperature of the surface at which each neutrino flavor becomes thermally decoupled, see also Fig.~\ref{fig:Fnufinal}. Rapidly rotating progenitors show lower average neutrino energies as the material is pushed to larger radii where the temperature is lower.}
    \label{fig:Enu}
\end{figure*}

Another observable of interest for the neutrino emission process is the average neutrino energy $\langle E_{\nu}/N_{\nu}\rangle$, shown in Fig.~\ref{fig:Enu}. Although the electron neutrinos have the highest luminosity, they have the lowest average energy because they also have the highest number density. The average neutrino energy is related to the temperature of the surface of the thermal decoupling of each neutrino type. Heavy-lepton neutrinos are the most energetic and become thermally decoupled in innermost regions of the PNS, whereas electron neutrinos are the least energetic and decouple further away from the PNS center, as can be seen by comparing the temperature maps shown in Fig.~\ref{fig:temperature} and the surfaces of zero radial flux in Fig.~\ref{fig:Fnufinal}. 

Our neutrino luminosity and average neutrino energy are in good qualitative and quantitative agreement with those reported in the CCSN literature \citep{NeutrinoCCSN, SNnu1, SNnu2, SNnu3, SNnu4}. Our results agree quantitatively with these earlier papers, where the neutrino luminosities and average neutrino energies for all flavors were found to be between $\sim 2-8 \times 10^{52}$ erg/s and $\sim 9-16$ MeV, respectively, for similar times scales as those considered here. As in the case for the neutrino luminosities, the addition of angular momentum suppresses the average neutrino energy irrespective of their flavors.

\section{Ejecta and Electromagnetic Counterpart}\label{EMcounter}

In order to estimate the ejecta mass we use the Bernoulli criterion \citep{Bernoulli_criterion,Haddadi_2023}, according to which fluid elements are unbound if they satisfy
\begin{equation}\label{BernCrit}
   h u_{t} \leq -h_{min} \,,
\end{equation}
where $h$ is the enthalpy, $u_{t}$ is the time component of the fluid 4-velocity and $h_{min}$ is the minimum enthalpy value allowed by our EOS table. The Bernoulli criterion assumes that the fluid is in a steady-state configuration, but our models are highly dynamic.  Thus Equation \ref{BernCrit} should be used only as an approximation/upper limit: 
the amount of ejecta can be overestimated because this criterion assumes that the internal energy of the fluid is completely transformed into kinetic energy. 
We also require that fluid elements satisfying the Bernoulli criterion should move outward ($v^{r}>0$) in order to be considered unbound. Finally, we only analyze material inside a radius of 1200 km, in order to avoid any contamination from spurious outflow material on the edge of the star. The shock front did not exceed this radius during our simulations. 

\begin{figure}
    \centering
\includegraphics[width=0.5\textwidth]{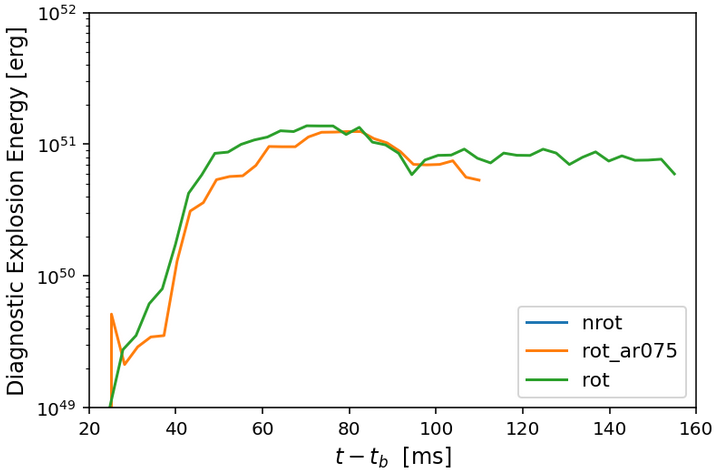}
    \caption{Diagnostic explosion energy for our three models. For the \texttt{nrot} model we find the explosion energy to be numerically consistent with zero at all times. This is consistent with the interpretation that the shock has stalled. Both models with initial angular momentum (\texttt{rot\_ar075} and \texttt{rot}) have diagnostic explosion energy in the range 5-6 $\times 10^{50}$ erg at the of our simulations.}
    \label{fig:explosion}
\end{figure}

The first quantity of interest for the ejecta is the diagnostic explosion energy, presented in Fig.~\ref{fig:explosion}. We calculate the diagnostic explosion energy as: 

\begin{equation}\label{diagE}
    E_{\rm exp} = \int_{V} \left(\rho \varepsilon W^{2} + \rho W(W-1) + P(W^2-1) \right) \sqrt\gamma dV,
\end{equation}
where the first term represents the internal energy of the fluid ($\varepsilon$ is the specific internal energy), the second term is the kinetic energy and the last term accounts for the pressure ($P$) contribution. The volume $V$, over which the integral in Equation (\ref{diagE}) is evaluated, is the volume occupied by outgoing fluid elements satisfying the Bernoulli criterion, given by Equation (\ref{BernCrit}). We note that the diagnostic explosion energy overestimates how much internal energy is transformed into kinetic energy during the explosion, in addition to using the overestimated ejecta amount based on the Bernoulli criterion.

The diagnostic explosion energy for the \texttt{nrot} model is found to be numerically consistent with zero. For the two rotating models, the diagnostic explosion energies are approximately $5 \times 10^{50}$ and $6 \times 10^{50}$ ergs for the \texttt{rot\_ar075} and \texttt{rot} models, respectively. Rotation seems to favor mass ejection: \texttt{nrot} does not present any ejecta, whereas both \texttt{rot\_ar075} and \texttt{rot} eject $M_{\rm{ej}} \approx 3\times 10 ^{-2} M_{\odot}$, in good agreement with the ``naked AIC" model proposed by  \citet{NakedAIC}.  From explosion energies and ejecta masses we can estimate the asymptotic velocities at radial infinity as $v^{r}_{\infty} \approx (2 E_{\rm exp}/M_{\rm{ej.}})^{1/2} \approx 0.14 c$ for both \texttt{rot\_ar075} and \texttt{rot}.

\begin{figure}
    \includegraphics[width=0.5\textwidth]{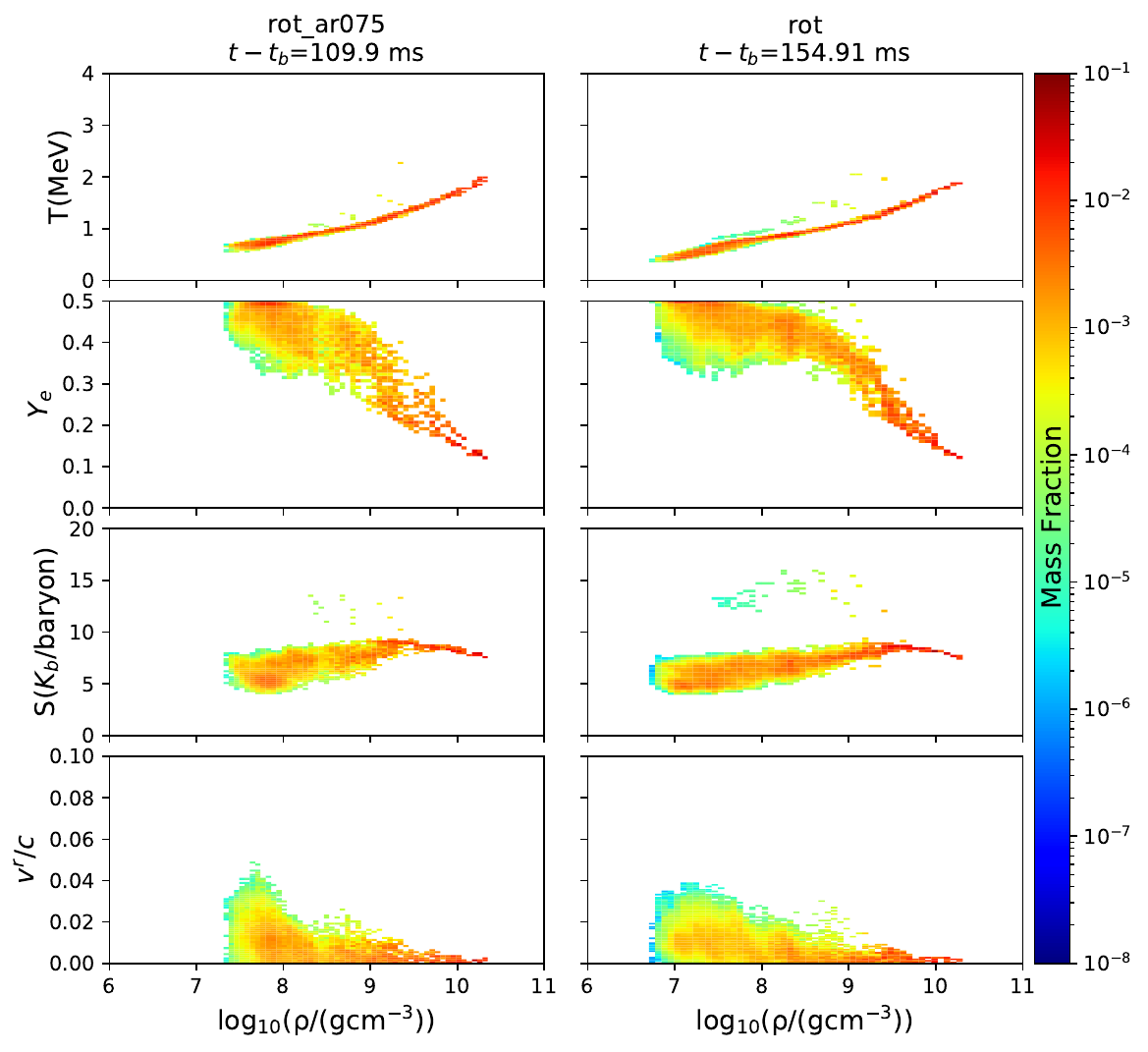}
    \caption{Snapshots of the ejecta properties as a function of density for the rotating models. (The \texttt{nrot} model presented a stalled shock and, for that reason, no ejecta was found.) The color scale shows the mass fraction normalized to the total ejecta mass, $M_{\rm ej.} \approx 3 \times 10^{-2}\,M_{\odot}$.
    First row: the tight correlation between temperature and density indicates that the ejecta material cools while it expands. Second row: the electron fraction $Y_e$ shows a strong correlation with the density, implying that $Y_{e}$ evolves as the fluid expands. For densities lower than $10^{8}{\rm g}/{\rm cm}^{3}$, the value of $Y_{e}$ is $\sim 0.4 - 0.5$. Third row: the entropy is almost constant over a density variation of three orders of magnitude, showing that the expansion is nearly adiabatic. Fourth row: higher velocities are found for lower density material, indicating an acceleration of the fluid during the expansion due to the explosion.}
    \label{fig:ejecta}
\end{figure}

The properties of the ejecta are presented in Fig.~\ref{fig:ejecta}, which displays the temperature, electron fraction, entropy and radial velocity as a function of density. Notably, the radial velocities observed in the ejecta are only a fraction of those obtained from the explosion energy estimates. However, this apparent contradiction can be reconciled by understanding that Fig.~\ref{fig:ejecta} represents only a snapshot of the time evolution of the system. Therefore, the velocities observed in the plot are not yet the asymptotic values. 

Furthermore, lower density material in the ejecta shows higher velocities. At the same time, the electron fraction distribution ranges from $\approx 0.1$ to $\approx 0.5$. We estimate $Y_{e} \approx 0.46$ for the lowest density ejecta material ($\rho \le 10^8\, {\rm g}/{\rm cm}^3$), using a mass-weighted average. Therefore, it is essential to consider the time evolution of the system and the behavior of low-density material to draw accurate conclusions about the properties of the ejecta.

\quad We estimate the peak luminosity of the light curve following \citet{ArnettLaw} and \citet{Metzger2019}
\begin{align}\label{eq:ArnettL}
    L_{\rm{peak}} &\approx 10^{41} {\rm{erg \quad s^{-1}}} \left( \dfrac{M_{\textrm{ej.}}}{10^{-2}M_{\odot}} \right)^{0.35}\left( \dfrac{v}{0.1 c} \dfrac{1 \rm{cm}^{2}\rm{g}^{-1}}{\kappa}\right)^{0.65}& \end{align}
and time of peak luminosity $t_{\rm{peak}}$ is estimated as
\begin{align}\label{eq:Arnettt}
    t_{\rm{peak}} 
    &\approx 1.6 \quad  {\rm{days}}\left( \dfrac{M_{\textrm{ej.}}}{10^{-2} M_{\odot} } \right)^{1/2} \left( \dfrac{v}{0.1 c} \dfrac{1 \rm{cm}^{2} \rm{g}^{-1}}{\kappa} \right)^{-1/2},&
\end{align}
where $\kappa$ is the fluid opacity, $v$ is the ejecta material typical velocity, and $M_{\textrm{ej.}}$ is the total ejecta mass. Our analysis reveals that the ejecta mass has a nearly symmetric composition. To approximate the opacity of the ejecta material, we follow \citet{Kasen_2013} and assume it to be composed by elements of the iron group (a mixture of 10\% Ni, 80\% Co, 10\% Fe), which has a wavelength-dependent opacity ranging from ${\sim}0.6\ {\rm cm}^{2}/{\rm g}$ to ${\sim}0.01\ {\rm cm}^{2}/{\rm g}$ in the optical band ($4000\, \AA$ to $10000\, \AA$). For our analysis, we use a representative value $\kappa = 0.2\ {\rm cm}^{2} / {\rm g}$. By applying Arnett's law (Equations \ref{eq:ArnettL} and \ref{eq:Arnettt}), we obtain estimates for $L_{\rm peak}$  and $t_{\rm peak}$: ${\sim} 5.1 \times 10^{41}$~erg/s and 1 day for the \texttt{rot\_ar075} model, and ${\sim}5.5 \times 10^{41}$~erg/s and 1.3 days for the \texttt{rot} models, similar to the predictions for the ``naked AIC" model presented in \citep{NakedAIC}.

Our estimates for the peak luminosity of our AIC models are 2 orders of magnitude mower than for  CCSNe and SNIa, see for example \citet{Khokhlov_1993,ArnettBook,Woosley_2007, Morozova_2020}. This difference is largely influenced by the fact that our models predict very low mass ejecta compared to CCSNe and SNIa. Also, the evolution timescale of the AIC light curves (less than two days for our models) is about a tenth of the  timescale for CCSNe and SNeIa (typically a few weeks). These differences are a consequence of the low ejecta mass and rapid rotation achieved in our models.

\section{Conclusions and Future Work}\label{conclusions}
Accretion-induced collapse (AIC) of WDs is a theorized possible source for multimessenger astronomy and, in particular, of GWs for next-generation ground-based detectors. We perform numerical simulations of three AIC models, starting with WD progenitors that are nonrotating, fast rotating or maximally rotating. Our main results are as follows: 
\begin{enumerate}
    \item The non-rotating progenitor leads to a fairly spherical symmetric AIC, with no spiral arms forming as the shock seems to stall. Our rotating progenitors form spiral arms and a thick disk on the equatorial plane surrounding the PNS remnant. Rotation also induces the excitation of higher order modes in the mass distribution. A transient $m=2$ mode is connected to the GW emission; a long-lived $m=1$ mode with approximately constant amplitude may be related to the one-armed instability \citep{Abdikamalov2010}.
    \item GW radiation of AICs is strongly enhanced in our models with rotating progenitors, with amplitudes 1-2 orders of magnitude stronger than predictions for CCSNe with slowly rotating progenitors. CCSNe with fast-rotating ($\Omega=$2 Hz) progenitors display GW signals similar to our models with rotation although still $\sim30\%$ fainter and with $\sim10\%$ lower frequencies. Our results imply that the additional rotational energy is the decisive feature for the stronger GW emission of AICs when compared to CCSNe. The AIC detection horizon (for the maximally rotating model in a face-on configuration) is approximately 10 Mpc for CE and 1.2 Mpc for LIGO. Estimated detection rates are $(x/10\%) \times 0.14\,{\rm yr}^{-1}$ for CE and $(x/10\%) \times2.2\times10^{-4}\,{\rm yr}^{-1}$ for LIGO, where $x$ is the ratio (in percentage form) between AIC and SNIa event rates. 
    \item Neutrino emission from all of our AIC models is comparable with that of CCSNe. 
    \item Electromagnetic emission of our rotating AIC models is estimated to be approximately $5\times 10^{41}$erg/s, i.e. two orders of magnitude dimmer than  SNIa models \citep{Khokhlov_1993,ArnettBook,Woosley_2007,Morozova_2020}, with peak times between one and two days.
    Our predictions are in agreement with the previous estimates of \citet{Metzger_2009} and \citet{Darbha_2010}, although their mass ejection mechanism is different from the one considered here.
    At the distances accessible to GW detectors, however, this would be a very bright transient.
\end{enumerate}
The estimated light curve properties of AICs indicate that an AIC detection based on their optical transient alone (without a GW counterpart) would be rather challenging. WD ZTF J190132.9+145808.7 \citep{Caiazzo2021} is a possible candidate for a future AIC progenitor. This object is very close to the Chandrasekhar mass limit ($M\approx 1.327-1.365 M_{\odot}$) with a rotation period of 6.94 minutes and seems to be cooling down via electron capture on sodium.

As usual, the inclusion of magnetic fields would be desirable for a better description of these astrophysical events. Because AIC progenitors are expected to be rapidly rotating due to their accretion history, existing magnetic fields are expected to be amplified during the collapse but details are still uncertain. \citet{Yi_GRB}, \citet{AICasSGRBEE}, and \citet{Perley2009} suggested that AICs could power GRBs; we see that some conditions for the formation of jets are already present in our simulations, but conclusions regarding jet formation should wait until the full inclusion of magnetic fields in the simulation. We are currently studying the implementation of magnetic field effects via the general-relativistic magnetohydrodynamical \texttt{Gmunu} code \citep{Gmunu1,Gmunu2,Gmunu3}, which recently has been updated in order to include a two-moment based multi-frequency neutrino radiation scheme \citep{Gmunu4}. Including magnetic fields will allow us to assess effects such as jet formation and evolution of the light curves. These two effects are essential for comparing our AIC models with a GRB event such as GRB211211A \citep{Rastinejad2022}. It will also be interesting to investigate whether the oscillations of a PNS formed in an AIC could explain the pulsations observed in some short GRB light curves \citep{Chirenti2023}. Relaxing the assumption of the deleptonization scheme by \citet{Liebendorfer2005} and including possible environmental effects (such as a remaining accretion disk) also constitute an interesting avenue of investigation.

\section*{Acknowledgements}
It is a pleasure to acknowledge Viktoriya Morozova for performing the collapse portion of the simulations and Ernazar Abdikamalov for providing the deleptonization table. We are grateful to Brian Metzger for his input on a previous version of this manuscript. LFLM also acknowledges useful discussions with Cole Miller, Igor Andreoni, John Baker, and Scott Noble. LFLM thanks the financial support of the S\~ao Paulo Research Foundation (FAPESP) grant 2021/09531-5 under the BEPE program, and thanks Pennsylvania State University for the hospitality during the realization of this project. DR acknowledges funding from the U.S. Department of Energy, Office of Science, Division of Nuclear Physics under Award Number(s) DE-SC0021177 and from the National Science Foundation under Grants No. PHY-2011725, PHY-2020275, PHY-2116686, and AST-2108467. CC acknowledges support by NASA under award numbers 80GSFC17M0002 and TCAN80NSSC18K1488, and of the Aspen Center for Physics, which is supported by National Science Foundation grant PHY-2210452. Simulations were performed on PSC Bridges2 and SDSC Expanse (NSF XSEDE allocation TG-PHY160025). This research used resources of the National Energy Research Scientific Computing Center, a DOE Office of Science User Facility supported by the Office of Science of the U.S.~Department of Energy under Contract No.~DE-AC02-05CH11231. Computations for this research were performed on the Pennsylvania State University’s Institute for Computational and Data Sciences’ Roar supercomputer.

\section{Data Availability}
Simulation data will be made available upon reasonable request to the corresponding author.



\typeout{}
\bibliographystyle{mnras}
\bibliography{mnras_template} 




\appendix

\section{Error Control}

It is common practice in the Numerical Relativity (NR) community to present convergence tests of the simulations by showing results obtained with three different resolutions. However, numerical studies of supernova-like events usually do not present such tests. In our case, a factor 2 increase in the resolution would result in a total computational cost increase from 2.1 million to 33.6 million CPU hours. Furthermore, events such as AICs, CCSNe and SNIa are known to be turbulent and chaotic, implying that these systems are not fully deterministic. In this case, even slight changes in the grid setup may lead to changes in the evolution of the system and arbitrarily fine (and expensive) resolution schemes would be required for the convergence. This stochastic feature of supernova-like simulations has been discussed in the literature by following the shock radius for models differing by small random density perturbations, see for example Fig.~8 of \citet{OConnor:2015rwy} and Fig.~17 of \citet{Summa:2015nyk}. 

\begin{figure*}
    \centering
    \includegraphics[width=0.5\textwidth]{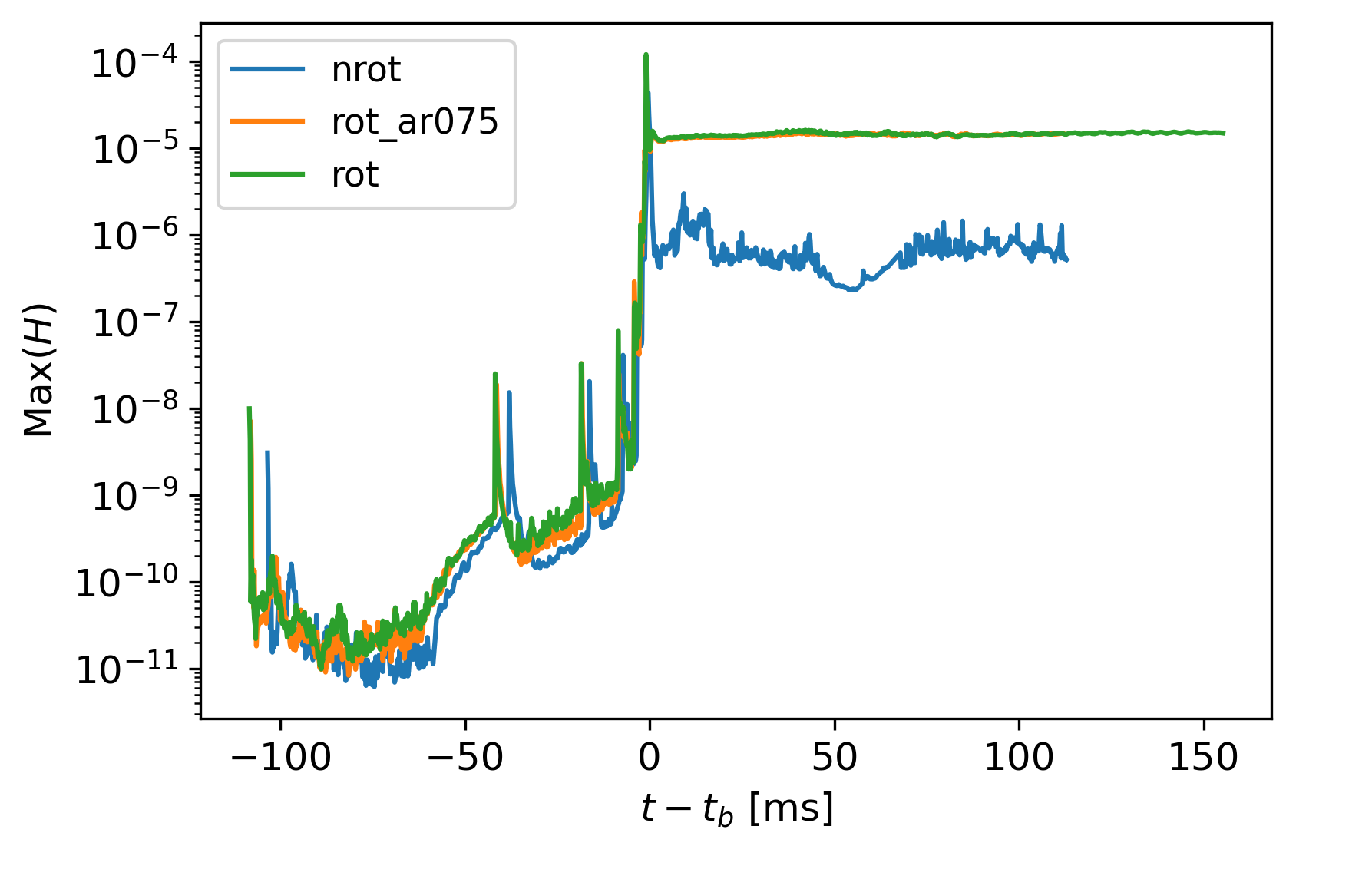}
    \caption{Maximum Hamiltonian constraint violation in our simulations. Throughout the time evolution of our models, we kept track of the Hamiltonian constrain violation. The smaller the violation, the more accurate the simulation is. Nevertheless, there is a lack of a natural scale to which these violations should be compared in order to be considered small. We use the \texttt{nrot} model as a reference and note that the violations do not grow further after the bounce, indicating the numerical stability of our code. The Hamiltonian constraint violation reaches a maximum (${\sim}10^{-4}$) near the time of bounce ($t=t_{b}$) due to the transition between the neutrino treatment provided by \citet{Liebendorfer2005} and \texttt{WhiskyTHC}.}
    \label{fig:errors}
\end{figure*}

Alternatively to the standard NR convergence tests, we follow the evolution of the Hamiltonian constraint to monitor the error in our simulations. The Hamiltonian constraint violation is defined as \citep{3p1NR}
\begin{equation}
    H \equiv {}^{\textrm{3D}}R + K^{2} +K_{ij}K^{ij} - 16 \pi E, 
\end{equation}
where $^{\textrm{3D}}R$, $K_{ij}$, and $E$ are, respectively, the 3D Ricci scalar, the 3D extrinsic curvature tensor, and the matter energy density as measured by an Eulerian observer. It represents the lack of conservation of energy in our numerical evolution. Analytically, this quantity should always be zero. Numerically, the Hamiltonian constraint violation will not be trivially equal to zero and can be used as a measurement of the robustness of the numerical evolution: the smaller the violation, the more accurate the simulation is. 

Our simulations presented an average Hamiltonian constraint violation below $2\times10^{-5}$ (with a maximum value of $10^{-4}$ at $t=t_{b}$), see Fig.~\ref{fig:errors}. The peak of the Hamiltonian constraint violation coincides (not by chance) with the time of bounce, when a shock wave is formed and the dynamics becomes more turbulent. The Hamiltonian constraint violation remains at higher values after the bounce due to the more complex neutrino treatment and turbulent motion in the post-bounce phase. 

As previously discussed, our \texttt{nrot} model is consistent with previous works in the literature. Therefore, we use its Hamiltonian constrain violation as a reference value for comparison with our rotating models. All models show very similar behavior before the bounce. After the bounce, the rotating models have a higher constraint violation which does not grow further until the end of the simulations. This behavior is similar to the \texttt{nrot} model. These results make us confident that our simulations are numerically robust and physically meaningful and that all numerical errors are under control. 


\bsp	
\label{lastpage}
\end{document}